\documentclass[ba, preprint]{imsart}

\RequirePackage{amsthm,amsmath,amsfonts,amssymb}
\RequirePackage[authoryear]{natbib}%% uncomment this for author-year citations
\RequirePackage[colorlinks,citecolor=blue,urlcolor=blue,backref=page,backref=page]{hyperref}
\RequirePackage{graphicx}

\usepackage[misc]{ifsym}
\usepackage{xcolor}
\usepackage{bm}
\usepackage{booktabs}
\usepackage{subcaption}
\usepackage{amsmath}
\usepackage{amsfonts}
\usepackage{hyperref}
\usepackage{comment}
\usepackage{algorithm}
\usepackage{algpseudocode}
\usepackage{threeparttable}

\pubyear{2024}
\volume{TBA}
\issue{TBA}
\firstpage{1}
\lastpage{1}

\startlocaldefs
\theoremstyle{plain}

\theoremstyle{definition}

\theoremstyle{remark}

\endlocaldefs

\begin{document}

\begin{frontmatter}
\title{Model Uncertainty in Latent Gaussian Models with Univariate Link Function}
%\title{A sample article title with some additional note\thanksref{T1}}
\runtitle{Model Uncertainty in ULLGMs}
%\thankstext{T1}{A sample of additional note to the title.}

\begin{aug}
\author[A]{\fnms{Mark F.J.}~\snm{Steel}\ead[label=e1]{m.steel@warwick.ac.uk}}
%\author[B]{\fnms{???}~\snm{???}\ead[label=e2]{???@???}}
\and
\author[C]{\fnms{Gregor}~\snm{Zens}\ead[label=e3]{zens@iiasa.ac.at}}
%%%%%%%%%%%%%%%%%%%%%%%%%%%%%%%%%%%%%%%%%%%%%%
%% Addresses                                %%
%%%%%%%%%%%%%%%%%%%%%%%%%%%%%%%%%%%%%%%%%%%%%%
\address[A]{Department of Statistics, Warwick University\printead[presep={\ }]{e1}.}
%\address[B]{???\printead[presep={\ }]{e2}.}
\address[C]{International Institute for Applied Systems Analysis\printead[presep={\ }]{e3}.}
\end{aug}

\begin{abstract}
We consider a class of latent Gaussian models with a univariate link function (ULLGMs). These are based on standard likelihood specifications (such as Poisson, Binomial, Bernoulli, Erlang, etc.) but incorporate a latent normal linear regression framework on a transformation of a key scalar parameter. We allow for model uncertainty regarding the covariates included in the regression. The ULLGM class typically accommodates extra dispersion in the data and has clear advantages for deriving theoretical properties and designing computational procedures. We formally characterize posterior existence under a convenient and popular improper prior and show that ULLGMs inherit the consistency properties from the latent Gaussian model. We propose a simple and general Markov chain Monte Carlo algorithm for Bayesian model averaging in ULLGMs. Simulation results suggest that the framework provides accurate results that are robust to some degree of misspecification. The methodology is successfully applied to measles vaccination coverage data from Ethiopia and to data on bilateral migration flows between OECD countries.
\end{abstract}


\end{frontmatter}

\section{Introduction}
\label{sec:intro}

Non-Gaussian regression models are extensively applied across numerous disciplines. The emergence of large datasets, coupled with significant uncertainty regarding the relevant variables for explaining an outcome of interest, has highlighted the importance of variable selection and model averaging techniques in non-Gaussian settings. The Bayesian approach to addressing model uncertainty involves placing a prior probability on each model, typically defined by a subset of predictors, as well as a prior on the corresponding parameters. This approach yields a joint posterior distribution of models and parameters, offering insights into the importance of specific variables within the regression model and making it particularly well-suited for predictive inference.

Bayesian model averaging (BMA) for non-Gaussian data encounters two primary challenges. First, in the presence of $p$ covariates, the model space is of size $2^p$, making it infeasible to enumerate in many cases. Second, the weights used to construct model-averaged estimates are typically based on marginal likelihoods, which are often unavailable analytically in non-Gaussian frameworks. To address these challenges, several procedures for variable selection and model averaging under non-Gaussian likelihoods have been proposed. Well-known  approaches rely on approximate marginal likelihoods (\citealp{volinsky1997bayesian}; \citealp{rossell2021approximate}) or reversible jump Markov chain Monte Carlo (MCMC) algorithms (\citealp{dellaportas2002bayesian}; \citealp{lamnisos2009transdimensional}) to calculate posterior model probabilities. More recently, the increasing availability of data augmentation schemes for non-Gaussian regression models (\citealp{fruhwirth2006auxiliary}; \citealp{polson2013bayesian}) has led to the development of specialized augmented MCMC algorithms to address model uncertainty in Poisson (\citealp{dvorzak2016sparse}), negative binomial (\citealp{jankowiak2023nb}), and logistic (\citealp{wan2021adaptive}) models.

We extend this literature by proposing a general and exact framework for formal BMA in a wide class of non-Gaussian regression models. Specifically, we focus on models that combine a standard likelihood specification (such as Poisson or Binomial) with a latent Gaussian linear regression framework applied to a transformation of a key scalar parameter. This model class accommodates a broad spectrum of outcome data types, including non-negative real-valued data and %overdispersed 
count and rate data. Members of this model class typically contain a dispersion parameter, which allows the model to mimic potentially large amounts of overdispersion, as commonly found in real data sets. Modeling overdispersed, non-Gaussian outcomes is a frequently encountered and challenging problem across various scientific disciplines including epidemiology, public health, demography, ecology, economics, insurance modeling, and environmental sciences. 

From a methodological perspective, the model class offers clear advantages for deriving theoretical properties and designing computational procedures. Importantly, we demonstrate the existence of the posterior distribution under a convenient and popular uninformative prior setting. This is crucial as BMA is typically sensitive to prior choices, making the theoretical justification of available uninformative benchmark priors highly relevant in practice. 
The modeling framework also allows us to prove that known results on model selection consistency carry over from the simple Gaussian case to the entire, much wider, model class used here. For posterior simulation, we introduce a simple and general MCMC algorithm for parameter estimation under model uncertainty.

We study two members of the model class in more detail, both used for overdispersed count data regression. These models are applied to simulated data and further illustrated using real-world datasets on early childhood measles vaccination coverage rates in Ethiopia and bilateral migration flows between OECD countries. In addition, we conduct an extensive out-of-sample cross-validation exercise with the real-world datasets to examine the comparative predictive performance of the models. Our results demonstrate the accuracy and predictive quality of the proposed framework, as well as its robustness under misspecification. A software implementation of the algorithms used is provided in the \texttt{R} package \texttt{LatentBMA}, available from \texttt{CRAN}.

The remainder of this article is organized as follows. Section~\ref{sec:intro_z} introduces the model class we consider and discusses two members of the model class in detail. Section~\ref{sec:prior} discusses prior specifications. Section~\ref{sec:posterior_results} summarizes our formal results on posterior existence and model selection consistency, while also providing details on the  posterior distributions. Section~\ref{sec:computation} develops the computational framework for posterior simulation. Section~\ref{sec:simulations} reports the results from a simulation study, while Section~\ref{sec:real_data_apps} examines real-world applications. Section \ref{sec:conclusion} concludes the paper and suggests directions for future research. Proofs as well as additional details and results are provided in the supplementary material.

\section{Univariate Link Latent Gaussian  Models}\label{sec:intro_z}

Consider the following general class of models for observations $y_i, i=1,\dots,n$
\begin{align}
y_i | z_i,r &\stackrel{ind}{\sim}  F_{h(z_i),r} \label{eq:Gen_y}\\
    z_{i} &= \alpha + \bm{x}'_{i}\bm{\beta} + \varepsilon_{i} \quad 
    \text{with}
    \quad \varepsilon_{i}\sim \mathcal{N}(0, \sigma^2),
\label{eq:Gen_z}
\end{align}
where, given $z_i$ and $r$,  the $y_i$ are independently drawn from some (continuous or discrete) distribution $F$ with support $\cal Y$ and which is indexed by a scalar parameter $h(z_i)$ and possibly another (low-dimensional) parameter vector $r$. The index $h(z_i)$ is 
%built on a normal linear  regression 
constructed on the basis of a latent variable $z_i$ using an invertible and continuously differentiable link function $h(\cdot)$ which takes values in some univariate space. Assuming a Gaussian distribution in (\ref{eq:Gen_z}) to model unobserved heterogeneity can be motivated as capturing a large number of independent heterogeneity terms, using a central limit theorem. The latent $z_i$ is modelled through the normal linear  regression model in (\ref{eq:Gen_z}), where $\alpha$ is an intercept term, $\bm{\beta}$ is a $p \times 1$ regression coefficient vector, $\sigma^2$ is (usually) an overdispersion parameter and $\bm{x}_{i}$ groups $p$ observable covariates for observation $i$.  We consider a Gaussian prior for $\bm{\beta}$ and an improper flat prior on $\alpha$ and on $\ln(\sigma^2)$, see Section \ref{sec:prior} for details.

The class of models formed by (\ref{eq:Gen_y}) and (\ref{eq:Gen_z}) are covered by the definition of ``Latent Gaussian Models with a Univariate Link Function''  in \cite{Hrafnkelsson2023}. In particular, the class of models we consider is a subclass of the model family considered in \cite{Hrafnkelsson2023}, who also consider settings with a larger  number of random effects, possibly modeled jointly to account for latent dependency structures in the data. We will refer to models defined via (\ref{eq:Gen_y}) and (\ref{eq:Gen_z}) as \textit{ULLGMs (Univariate Link Latent Gaussian Models)}. Approximate Bayesian inference for latent Gaussian Models %(LGMs)
was discussed in \cite{rue_etal_09}. In contrast to most of the existing literature, $F$ for our ULLGMs does not need to belong to the exponential family\footnote{For example, the Negative Binomial distribution with $r$ a free parameter is not in the exponential family.} and  $h(z_i)$ is not necessarily equal to the mean of $y_i$ (the latter need not even exist). In addition and more importantly, we will formally deal with  model uncertainty regarding the choice of regressors in (\ref{eq:Gen_z}); see Subsection \ref{sec:uncertainty}. 

\begin{table*}[t]
\centering
\resizebox{\textwidth}{!}
{\begin{tabular}{lcccc}
\toprule
 Model & $\cal Y$ & $F$ & $h(z)$ & Proper \\
 \midrule
 \addlinespace
Poisson Log-Normal (PLN) & $\{0,1,2,\dots\}$ & Poisson($\lambda$) & $\lambda=\exp(z)$ & yes \\
Binomial Logistic (BiL) & $\{0,1,2,\dots,N\}$ & Bin($N,\pi$), $N=2,3,\dots$ & $\pi=\frac{\exp(z)}{1+\exp(z)}$ & yes\\
Negative Binomial Logistic (NBL) & $\{0,1,2,\dots\}$ & Neg Bin($r,\pi$), $r=1,2,\dots$ & $\pi=\frac{\exp(z)}{1+\exp(z)}$ & yes\\
Erlang Log-Normal (ErLN) & $\Re_+$ & Erlang($r,\lambda$), $r=1,2,\dots$ & $\lambda=\exp(z)$ & yes \\
Log-Normal Normal (LNN) & $\Re_+$ & log-Normal($\mu, 1$) & $\mu=z$ & yes \\
Log-Normal Log-Normal (LNLN) & $\Re_+$ & log-Normal($r, \lambda$), $r\in\Re$ & $\lambda=\exp(z)$ & yes, for fixed $r$ \\
Bernoulli Cdf (BeC) & $\{0,1\}$ & Bernoulli($\pi$) & $\pi=Q(z)$ & no\\
\bottomrule
\end{tabular}}
\caption{Examples of Univariate Link Latent Gaussian Models (ULLGMs). $\lambda$ indicates a parameter in $\Re_+$. $\mu$ takes values in $\Re$. $\pi$ is a parameter on the unit interval $(0,1)$ and $Q(\cdot)$ denotes a known continuous cumulative distribution function (cdf) defined on $\Re$. The last column indicates posterior propriety (discussed in Section \ref{sec:exist}) under a convenient improper prior introduced in Section \ref{sec:prior}.}
\label{tab:RRSPM}
\end{table*}
 
The members of the ULLGM class are mapped out by choosing different $F$ and $h(\cdot)$. Table \ref{tab:RRSPM} lists some examples. Some models in the table have an additional parameter $r$, which allows for more flexibility and is considered fixed for now (until Subsection \ref{sec:add_param}). Certain choices for $F$ can generate more than one member of the ULLGM class, depending on which of the parameters we model through the latent Gaussian variable $z_i$, one example being the case where $F$ is log-normal. The LNN model in Table \ref{tab:RRSPM} can be shown to be equivalent to the usual log-Normal regression model (where $y_i\sim$~log-Normal$(\alpha+\bm{x}_i'\bm{\beta}, \omega^2)$ with $\omega^2=\sigma^2+1$), and it tends to this standard model with $\omega^2=1$ as $\sigma^2\to 0$. Advantages of expressing this model as a member of the ULLGM class include the ease of deriving theoretical results on posterior existence and model selection consistency and the simple treatment of model  uncertainty (see Section~\ref{sec:advantages}).  The LNLN model introduces the Gaussian regression for the scale parameter of the log-Normal and treats the location parameter as an additional parameter $r$. 

PLN, NBL and ErLN models converge to the usual Poisson, negative Binomial and Erlang 
regression models as $\sigma^2$ tends to zero. The Erlang distribution is a Gamma distribution with integer shape parameter and reduces to the Exponential distribution for $r=1$. The negative Binomial distribution with $r=1$ is also called the geometric distribution. These standard regression models (Generalised Linear Models or GLMs) are often found to be unable to account for overdispersion in observed data. For nonzero $\sigma^2$, the random nature of the latent Gaussian component in ULLGMs will allow for such extra variation or dispersion.

The subclass of models based on Bernoulli sampling is a special case of Binomial sampling models when $N_i = 1$ and is defined by the choice of the link cdf $Q(\cdot)$. For example, if the cdf of a standard normal distribution is chosen for $Q(\cdot)$, the BeC model becomes equivalent to a probit model with an additional unidentified parameter $\sigma^2$. For other choices of $Q(\cdot)$, the BeC model can be shown to interpolate between the corresponding binary regression model (where $\sigma^2 = 0$) and the probit model, with the value of $\sigma^2$ indicating its proximity to these extremes. Further theoretical details and empirical examples are provided in Supplement A1. Nevertheless, since $\sigma^2$ is typically unidentified in BeC models, this subclass is expected to be mainly of theoretical interest and is unlikely to have major %practical
empirical utility.

\begin{figure*}[t]
\centering
\begin{subfigure}{0.47\textwidth}
    \includegraphics[width=\textwidth]{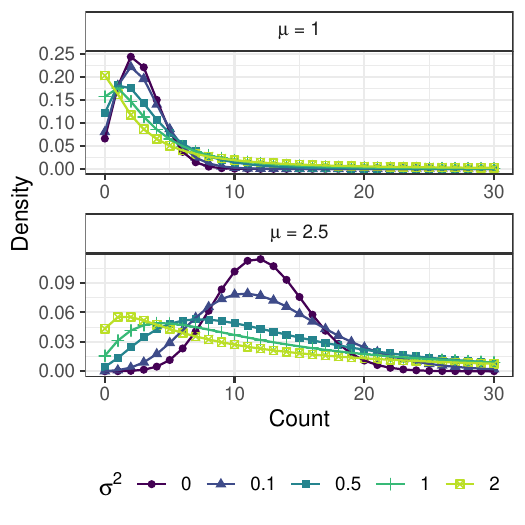}
    \caption{Poisson Log-Normal.}
    \label{fig:pmf_pln}
\end{subfigure}
\hfill
\begin{subfigure}{0.47\textwidth}
    \includegraphics[width=\textwidth]{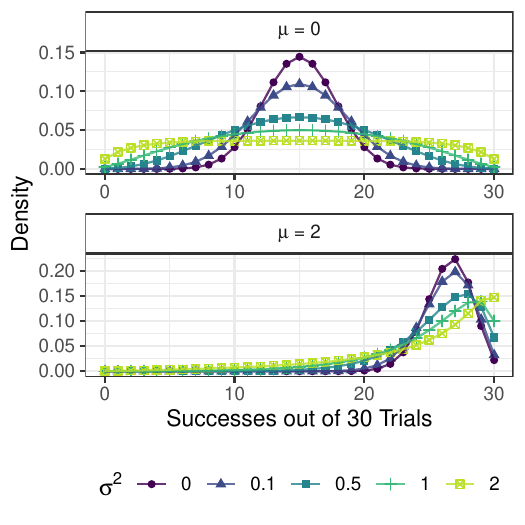}
    \caption{Binomial Logistic Normal.}
    \label{fig:pmf_bil}
\end{subfigure}
\caption{Probability mass functions for random variables arising from a Poisson Log-Normal distribution $y \sim \mathcal{P}(e^z), z \sim \mathcal{N}(\mu, \sigma^2)$ (left) and a Binomial Logistic Normal distribution $y \sim \text{Bin}(30, [1+e^{-z}]^{-1}), z \sim \mathcal{N}(\mu, \sigma^2)$ (right).}
\end{figure*}

\subsection{Selected ULLGMs for Count Data Regression}

Consider the PLN model, which applies to count-valued data and is based on a Poisson likelihood.
%\begin{equation}\label{eq:y_Poi}
% y_{i} \sim \mathcal{P}({\lambda_{i}}),
%\end{equation}
The observed counts  $y_i$ ($i = 1, \dots, n$) are assumed to be Poisson distributed with an intensity parameter $\lambda_{i}$. In the standard, equi-dispersed,  Poisson regression framework $\lambda_i$ is a deterministic function of observed covariates. In the presence of unobserved heterogeneity and overdispersion, it makes sense to assume that $\lambda_i$ is random, arising from an appropriate mixing distribution. %Several distributional choices for $\lambda_i$ are possible, as long as the model implies $\lambda_i > 0$. %(\cite{karlis2005mixed}).
Commonly considered mixing distributions include the Gamma distribution, which %in combination with the Poisson likelihood
results in a negative binomial model (\citealp{greenwood1920inquiry}), or an inverse Gaussian distribution which was used in \cite{dean1989mixed}. A log-normal mixing distribution model has appeared as such in the literature:  \cite{bulmer1974fitting} uses this mixture model in a location-scale context, which was extended to a multivariate setting in  \cite{aitchison1989multivariate}. The regression structure as used here was % already 
mentioned in \cite{Hinde1982} and used in \cite{tsionas2010bayesian} in a Bayesian setting. 

For the PLN model as in Table \ref{tab:RRSPM}, we can show that
\begin{equation}
    \begin{split}
        \mathbb{E}(y_i|\bm{x}_i) &= e^{\alpha+\bm{x}_i'\bm{\beta} + 0.5\sigma^2}\\
        \mathbb{V}(y_i|\bm{x}_i) &= \mathbb{E}(y_i|\bm{x}_i) + \mathbb{E}^2(y_i|\bm{x}_i)(e^{\sigma^2} - 1),
    \end{split}
\end{equation}
allowing for overdispersion since $\mathbb{E}(y_i|\bm{x}_i) < \mathbb{V}(y_i|\bm{x}_i)$. The expression for the expected value further shows that the PLN model maintains a simple and intuitive interpretation of the regression parameters $\bm{\beta}$, similar to a Poisson regression model.\footnote{Note that both in ULLGMs and in GLMs, the precise interpretation of regression coefficients depends on the chosen likelihood specification. In general, the marginal effect of any $x_k$ on $y$ will differ from $\beta_k$ and depend on the value of the other regressors, coefficients, $\sigma^2$ and the chosen transformation $h(\cdot)$.} Note that the usual dispersion index 
\begin{equation}
\mathbb{D}(y_i|\bm{x}_i) = 1 + \mathbb{E}(y_i|\bm{x}_i)(e^{\sigma^2} - 1)
\end{equation}
is a monotonous function of $\sigma^2$ taking values on all of $\Re_+$. With the exception of the BeC class, similar results hold for the other models in Table \ref{tab:RRSPM}, which gives $\sigma^2$ the interpretation of a dispersion parameter, controlling excess dispersion beyond the one implied by $F$.  Figure~\ref{fig:pmf_pln} shows example probability mass functions (pmfs) for the PLN model. 

The BiL model represents another important member of the ULLGM family, generalising a Binomial model. To address overdispersion, it employs a logistic-normal distribution for the success probabilities of individual observations (\citealp{atchison1980logistic}). Illustrations of pmfs %probability mass functions 
of Binomial logistic normal distributions are provided in Figure~\ref{fig:pmf_bil}, underscoring the role of $\sigma^2$ as overdispersion parameter. BiL regression constitutes a flexible alternative to Beta-Binomial regression models for the analysis of overdispersed binomial outcomes. Although analytical expressions for the moments  $\mathbb{E}(y_i|\bm{x}_i, N_i)$ and $\mathbb{V}(y_i|\bm{x}_i, N_i)$ are not known for a logistic link function, %the following
approximate results can be derived, such as
\begin{equation}
\label{eq:binomial_approx_moments}
%\begin{split}
\mathbb{E}(y_i~|~\bm{x}_i, N_i) = N_i~\mathbb{E}[\pi_i] \approx N_i~\Phi \left( \frac{b(\alpha + \bm{x}'_i\bm{\beta})}{\sqrt{1 + b^2\sigma^2}} \right) %\\
%\mathbb{V}(y_i|\mu, \sigma^2, N_i) &= N_i \mathbb{E}[\pi_i] -  N_i\mathbb{E}[\pi_i^2] + N_i^2 \mathbb{V}(\pi_i)\\
%\end{split}
\end{equation}
for a suitable value of $b>0$, where $\Phi(\cdot)$ is the cdf of a standard Gaussian random variable. Full details and an approximation of the variance %term %$\mathbb{V}(\pi_i)$ 
are provided in Supplement A2. %, where also exact results under a probit link are discussed. 
From (\ref{eq:binomial_approx_moments}), 
the interpretation of the coefficients and error variance is intuitive in the BiL model. %Specifically, the sign and statistical significance of $\bm{\beta}$ can be understood directly in relation to the increase or decrease in success probabilities. 
%The role of $\sigma^2$ is to adjust the intensity of the effects suggested by $\bm{\beta}$. 
As the value of $\sigma^2$ increases, the impact of the coefficients $\bm{\beta}$ is more muted. %, and the success probabilities tend towards $\frac{1}{2}$, which leads to the largest possible variance for $y_i$.
Also, $\mathbb{V}(y_i|\mu, \sigma^2, N_i)$ can be shown to approach the usual binomial variance %$N_i \mathbb{E}[\pi_i (1 - \pi_i)]= 
$N_i\pi_i (1 - \pi_i)$ for $\sigma^2 \rightarrow 0$. %, as $\sigma^2 \rightarrow 0$ implies $\mathbb{V}(\pi_i) \rightarrow 0$. 
Similarly, the dispersion index $\mathbb{V}(y_i|\mu, \sigma^2)/\mathbb{E}(y_i|\mu, \sigma^2)$ tends to the binomial dispersion index $(1-\pi_i)$ for $\sigma^2 \rightarrow 0$, but is larger than the binomial dispersion index when $\sigma^2 > 0$, as also shown in Supplementary 
Figure A3; see Supplement A2 for more details.

\subsection{Advantages of the ULLGM class}\label{sec:advantages}

As previously discussed, for most underlying distributions $F$, the ULLGM specification intuitively allows for overdispersion, which is regulated by the extra parameter $\sigma^2$ in (\ref{eq:Gen_z}). In regression analysis, failing to account for overdispersion in the data can lead to an underestimation of the posterior variability. Empirically, this means that credible intervals for ULLGM regression parameters will typically be larger than for their GLM counterpart. In the context of model selection and model averaging, ignoring overdispersion will often lead to a preference for overly complex models in order to compensate for the inability to account for the additional variation in the outcome. This undesirable phenomenon is illustrated for Poisson and Binomial regression models in Supplement A3. 

The particular structure of the models in the ULLGM class in (\ref{eq:Gen_y}) and (\ref{eq:Gen_z}) also has a number of theoretical and practical benefits. First, in the context of model uncertainty, the tractability of the Gaussian distribution lends itself to convenient applications of standard BMA methods. In particular, the parameters $\alpha, \bm{\beta}$ and $\sigma^2$ can be integrated out analytically under a popular and convenient prior, conditional on $z_i$. This greatly simplifies the computational implementation (see Section \ref{sec:computation}), the characterisation of posterior existence under this improper prior (see Section \ref{sec:exist}) and  proofs of model selection consistency (see Section \ref{sec:consistency}).

Second, the computational implementation of ULLGMs is simple, allows for exact inference, and is largely generic in the sense that it can be easily modified to accommodate a broad family of members of the ULLGM class. In terms of model averaging, the computational approach enables simple, efficient  and exact posterior simulation even in settings with large numbers of predictors $p$. This contrasts for example with existing GLM averaging approaches, which on top of lacking the ability to account for overdispersion typically impose upper limits on $p$, rely on complex reversible jump MCMC algorithms for exact inference, or allow only for approximate model averaging. In contrast, ULLGMs leverage simple Gaussian regression updates that allow for exact inference and joint updates of all $p$ coefficients at comparable or reduced computational cost per MCMC iteration relative to RJMCMC algorithms, even if $p$ grows large.

Third, anecdotal empirical evidence suggests that, for specific ULLGMs, relying on Gaussian error terms in the latent linear specification often provides superior model fits compared to allowing for overdispersion using non-Gaussian terms, such as log-Gamma errors in negative binomial regression models (\citealp{winkelmann2008econometric}; \citealp{tsionas2010bayesian}). The Gaussian latent specification also holds theoretical merit, as normal error terms can be justified by a central limit theorem, if they capture a sum of latent shocks to the linear predictor. Finally, ULLGMs possess great potential for relatively straightforward generalization to multivariate settings with correlated observations (\citealp{aitchison1989multivariate}; \citealp{chib2001markov}).

Finally, certain limiting cases of ULLGMs are closely related to Gaussian regression models with transformed outcomes. For example, as \( y_i \rightarrow \infty \), the PLN model converges to a Gaussian regression model with outcome \(\log(y_i)\). Similarly, as \( N_i \rightarrow \infty \), the BiL model converges to a Gaussian regression model with the outcome \(\text{logit}(y_i/N_i)\), see Supplement A4 for further discussion. However, even in those cases where these approximations are satisfactory, the ULLGM class has a number of key benefits compared to such Gaussian approximations often employed by practitioners. For example, ULLGMs provide valid uncertainty quantification and can naturally handle zero outcomes.

\subsection{Model uncertainty}\label{sec:uncertainty}

Given the model class defined in (\ref{eq:Gen_y}) and (\ref{eq:Gen_z}), the goal is to design a theoretical framework and a computational strategy for posterior and predictive inference, in the face of  model uncertainty. Specifically, we are interested in model uncertainty with respect to inclusion and exclusion patterns of the components  of the regression coefficient vector $\bm{\beta}$. Models will thus be characterized by the inclusion or exclusion of any of the columns of $\bm X$, which is the $n \times p$ matrix with $\bm{x}_{i}'$ as its $i$th row. We denote the total number of potential covariates in $\bm X$ by $p$ while $p_k$ indicates the number of covariates from $\bm X$ that are included in model $M_k$. An intercept term is included in all models.
This gives us a model space with $K=2^p$ elements and for model $M_k$ the distribution of $\bm{z}=(z_1,\dots,z_n)'$ now becomes
\begin{equation}
\label{eq:model_k_z}
\bm{z}|\alpha, \bm{\beta}_k, \sigma^2, M_k \sim \mathcal{N}(\alpha \bm{\iota}_n + \bm{X}_k \bm{\beta}_k , \sigma^2 \bm{I}_n),
\end{equation}
where $\bm{\iota}_n$ is a column vector of $n$ ones, $\bm{I}_n$ is the $n$-dimensional identity matrix, $\bm{X}_{k}$ consists of the $p_k$ columns of $\bm{X}$ that correspond to the regressors that are included in $M_k$ and $\bm{\beta}_k$ groups the corresponding regression coefficients. The regressors in $\bm{X}$ are centered by subtracting their means, which makes them orthogonal to the intercept and renders the interpretation of the intercept common to all models.

\subsection{ULLGMs with random $r$}\label{sec:add_param}

Sofar, we have focused on inference on the scalar observation-specific parameter, represented by $h(z_i)$ for observation $i$. We now consider situations where we also want to conduct inference on other parameters added to the model, grouped in $r$ and common to all observations. %so that (\ref{eq:Gen_y}) is replaced by
%\begin{equation}
%y_i | z_i,r\sim F_{h(z_i),r}, %\label{eq:Gen_yr}
%\end{equation}
%where $r$ is a parameter (could be a vector) that is the same for all observations.  
We will assume that $r$ is a priori independent of $\bm{z}$ given a model within the set of models described in Subsection \ref{sec:uncertainty}:
\begin{equation}
r\amalg \bm{z}| M_k, \text{for all } M_k. \label{eq:r_indep}
\end{equation}

Examples are the NBL and ErLN models, where we now allow $r$ to be an unknown parameter on which we conduct inference, rather than simply fixing it. In these models, $r$ is an integer scalar and it would be natural to assume that (\ref{eq:r_indep}) holds. 

\section{Prior Specification}\label{sec:prior}

We will focus on the prior setup that is most often encountered in the context of BMA. For the linear regression model in (\ref{eq:model_k_z}) taken in isolation, this prior satisfies many of the desiderata of \cite{Bayarri_etal_12} for objective priors, such as measurement and group invariance and exact predictive matching. Specifically, we assume an improper, `non-informative' prior on the parameters common to all models 
\begin{equation}
    p(\alpha, \sigma^2) \propto {\sigma^{-2}},\label{PriorAS}
\end{equation}
which is a convenient prior that has the advantage of being invariant with respect to rescaling and translating the $z_{i}$s. For the regression coefficients $\bm{\beta}_k$, we adopt a so-called $g$-prior which is invariant under affine linear transformations of the covariates
\begin{equation}
    \bm{\beta_k} |\sigma^2, M_k \sim \mathcal{N}(0 \bm{\iota}_{p_k}, g\sigma^2(\bm{X}_{k}'\bm{X}_{k})^{-1}),\label{PriorThetaMA}
\end{equation}
where $g>0$. Throughout, we will assume that the matrix formed by adding a column of ones to $\bm{X}_k$ is of full column rank. If the model space contains models for which this is not the case (for example because $p_k\ge n$), we will assign prior probability zero to those models, as shown in (\ref{eq:PM_hyper}) below.\footnote{This can be easily implemented while running the MCMC sampler, without needing to restrict the total number of possible covariates $p$. Alternative approaches to
use $g$-priors in situations where $p\ge n$ can be found in  \cite{MaruyamaGeorge_11} and \cite{Berger_etal_16}, based on different
ways of generalizing the notion of inverse matrices.} In practice, these rank-deficient models are likely to be of very limited relevance. The scalar $g$ can either be deterministic or assigned a hyperprior $p(g)$ as described in, e.g., \cite{liang_etal_08} or \cite{leysteel2012}. We will consider both deterministic and random $g$ when illustrating the framework in later sections. For BMA or model selection we require well-defined pair-wise Bayes factors between all models in the model space. In case a hyperprior is specified on $g$, it is necessary to take into account that $g$ does not appear in the null model (with $p_k=0$). Hence, %$g$ is not common to all models and, if $g$ is considered random, 
a proper $p(g)$ is necessary in order to ensure meaningful model comparisons. For the null model with no regressors and only an intercept, the prior will simply be (\ref{PriorAS}). Components of $\bm\beta$ that correspond to excluded regressors under $M_k$ are assigned a prior point mass at zero for that model. 

As a prior on the model space, we employ the beta-binomial structure of \cite{Brown_etal_98}, \cite{ley2009effect} and \cite{ScottBerger}, which amounts to using a Beta$(a,b)$ prior on the common prior inclusion probability for each covariate and results in
\begin{equation}\label{eq:PM_hyper}
P(M_k)=
\begin{cases}
\frac{\Gamma(a+b)}{\Gamma(a)\Gamma(b)} \frac{\Gamma(a+p_k)\Gamma(b+p-p_k)}{\Gamma(a+b+p)} & \text{if } (\bm{\iota}:\bm{X}_k) \text{ has full column rank} ,\\[1mm]
 0 & \text{else.}
\end{cases}
\end{equation}
This type of prior is less informative in terms of model size than fixing the prior inclusion probability of the covariates. 
Following the suggestions of \cite{ley2009effect}, we choose $a=1$ and $b=(p-m)/m$, where $m$ is the prior expected model size. This means that the user only needs to specify a value for $m$. Typical choices for $m$ include $m=p/2$, implying a uniform prior on model size \citep{ScottBerger}, or $m$ set to a small value to put more prior mass on sparse models. A detailed discussion of the effects of this prior choice is given in \cite{ley2009effect}. 

If there are any additional parameters $r$ as in Subsection \ref{sec:add_param}, we specify a proper prior on $r$, satisfying (\ref{eq:r_indep}). 

\section{Posterior Results}
\label{sec:posterior_results}

If we combine the  $g$-prior setup proposed in Section \ref{sec:prior} with the sampling model in (\ref{eq:Gen_y}) and (\ref{eq:model_k_z}), the conditional posterior distributions and the  marginal likelihoods of the latent data $\bm{z}$ can be easily derived. We summarize the posterior on the model parameters as follows: 

\begin{equation}
\label{eq:beta_post}
\beta_k|\alpha,\sigma^2,\bm{z} ,M_k \sim {\cal{N}}\left(\delta(\bm{X}_k'\bm{X}_k)^{-1} \bm{X}_k'\bm{z}, \delta\sigma^2 (\bm{X}_k'\bm{X}_k)^{-1} \right),
\end{equation}
where $\delta=\frac{g}{1+g}$,
\begin{equation}
\label{eq:alpha_post}
   \alpha|\sigma^2,\bm{z},M_k\sim {\cal{N}}\left(\bar z, \frac{\sigma^2}{n}\right),
\end{equation}
with $\bar z=\frac{1}{n} \sum_{i=1}^n z_{i}$ and $\sigma^{-2}|\bm{z},M_k \sim \mathcal{G}(c_n, C_n)$ (a gamma distribution) with
\begin{equation}
\label{eq:sigma_post}
\begin{aligned}
       c_n &= \frac{n-1}{2}\\
       C_n &= \frac{1}{2}\left[\delta\bm{z}'Q_{(\iota:X_k)}\bm{z}+(1-\delta) (\bm{z}-{\bar z}'\iota)'(\bm{z}-{\bar z}'\iota)\right],
\end{aligned}
\end{equation}
where $Q_A=I_n-\bm{A}(\bm{A}'\bm{A})^{-1}
\bm{A}'$ for any matrix $\bm{A}$ of full column rank.
Finally, the marginal likelihood under deterministic $g$ is
\begin{equation}\label{eq:ml_a}
p(\bm{z}|M_k) \propto (1+g)^{\frac{n-1-p_k}{2}}
\left[\{1+g(1-R_{k}^2)\}(\bm{z}-{\bar z}\iota)'(\bm{z}-{\bar z}\iota)\right]^{-\frac{n-1}{2}},
\end{equation}
where $R_{k}^2$ is the coefficient of determination of $\bm z$ regressed on $\bm{X}_k$ (and an intercept) and the proportionality constant is the same for all models, including the null model for which
$p(\bm{z}|M_0)\propto \left[(\bm{z}-{\bar z}\iota)'(\bm{z}-{\bar z}\iota)\right]^{-\frac{n-1}{2}}$. Under random $g$ with hyperprior $p(g)$, the marginal likelihood is 
\begin{equation}\label{eq:ml_randomg}
    p(\bm{z}|M_k)\propto \int_0^{\infty} (1+g)^{\frac{n-1-p_k}{2}} 
\left[\{1+g(1-R_{k}^2)\}(\bm{z}-{\bar z}\iota)'(\bm{z}-{\bar z}\iota)\right]^{-\frac{n-1}{2}} p(g)dg.
\end{equation}

\subsection{Posterior existence}\label{sec:exist}

The prior for each given model $M_k$ is improper, as can be seen from the prior specification on the common parameters shared by all models in (\ref{PriorAS}). Thus, we need to make sure that the posterior distribution of the parameters in each model is well-defined in the sense that the marginal likelihood is a finite quantity for each possible value of the observations $\bm{y}=(y_1,\dots,y_n)'$. We can state the following for cases where the possible additional parameter $r$ is fixed:

\vskip 0.5cm

\noindent\textit{\textbf{Theorem 1:}}
\noindent\textit{If we combine the sampling model in (\ref{eq:Gen_y}) and (\ref{eq:model_k_z}) (ULLGMs defined in Table \ref{tab:RRSPM}) with the improper prior structure in (\ref{PriorAS}) and (\ref{PriorThetaMA}), then the posterior is well-defined for any model $M_k$ in the model space, if and only if 
the matrix composed of a column of ones and $\bm{X}_{k}$
has full column rank and, in addition, the following condition holds: 
\begin{itemize}
\item for the PLN and NBL models: at least two of the observations are nonzero;
\item for the BiL model: at least two observations are nonzero and smaller than $N_i$, where $N_i$ is the number of trials for observation $i$;
\item for the ErLN, LNN and LNLN models: we have at least two observations.
\end{itemize}
ULLGMs based on Bernoulli sampling (the BeC models) do not allow for a posterior under the prior in (\ref{PriorAS}) and (\ref{PriorThetaMA}).}

\noindent\textit{\textbf{Proof:} See Supplement A.5}.

Theorem 1 provides  necessary and sufficient conditions for all the models in  Table \ref{tab:RRSPM} that are not based on Bernoulli sampling, and thus fully characterizes posterior propriety for these ULLGMs. As the conditions in Theorem 1 are necessary and sufficient, we know that models for which the matrix $(\bm{\iota}:\bm{X}_k)$ is not of full column rank do not admit a posterior under the prior on the model parameters assumed here. This could occur in situations where $p>n$ by considering a model for which $p_k>n$. Such large models would, in most practical settings, be of very little empirical relevance and we simply assign zero prior probability to such models, as stated by (\ref{eq:PM_hyper}) in Section \ref{sec:prior}.

For models where the additional  parameter $r$ is treated as random as in Subsection \ref{sec:add_param}, we can derive the following:

\vskip 0.5cm

\noindent\textit{\textbf{Theorem 2:}}
\noindent\textit{If we combine the sampling model in (\ref{eq:Gen_y}) and (\ref{eq:model_k_z}) with the improper prior structure in (\ref{PriorAS}) and (\ref{PriorThetaMA}), along with a proper prior on $r$, $p(r|M_k)$ satisfying (\ref{eq:r_indep}),  then the posterior is well-defined for any model $M_k$ in the model space, if the corresponding model with $r$ fixed leads to a proper posterior (see Theorem 1) and if, in addition
\begin{equation}
\int f(r) p(r|M_k) dr < \infty ,  \label{eq:Th2}
\end{equation}
where we have defined 
\begin{equation}\label{eq:def_f(r)}
f(r)\equiv \int \prod_{i\in \cal N} P(y_i| z_i, r) dz_i , 
\end{equation}
with $\cal N$ the set of observation indices for which $\int P(y_i|z_i,r) d z_i$ is finite.
}

\noindent\textit{\textbf{Proof:} See Supplement A6}.

An immediate consequence of Theorem 2 is that the NBL model and the ErLN model with random $r$ have proper posteriors under any proper prior on $r$ respecting (\ref{eq:r_indep}), since $f(r)$ is constant in $r$ for these models (see Supplement A5.2 and A5.4).
The situation is quite different for the LNLN model where Supplement A6 shows that we can not conclude that posterior inference on $r$ can be conducted with the overall prior structure assumed here.  

\subsection{Model selection consistency}\label{sec:consistency}

One of the important desiderata in \cite{Bayarri_etal_12} for objective model selection priors is model selection consistency, see \cite{FLS01} and \cite{li2018mixtures}. This requires that if the data have been 
generated by model $M_k$ in the model space, then the posterior probability of $M_k$ should converge to 1 with sample
size $n$. We can show that the consistency results for the underlying Gaussian model in (\ref{eq:model_k_z}) essentially carry over to ULLGMs. More precisely, we have the following results.  
\vskip 0.5cm

\noindent\textit{\textbf{Theorem 3:}}
\noindent\textit{If we combine the sampling model in (\ref{eq:Gen_y}) and (\ref{eq:model_k_z}) with the prior structure in (\ref{PriorAS}), (\ref{PriorThetaMA}) and (\ref{eq:PM_hyper}), and we assume that the relevant conditions in Theorem 1 hold in combination with a choice for $g$ or a hyper-prior on $g$ that leads to consistency in the Gaussian case,  then we achieve model selection consistency, in the sense that if the data were generated by any model $M_k$ in the model space, we have $P(M_k|\bm{y})\to 1$ as $n\to\infty$. %, provided that the matrix $(\iota : \bm{X}_i)$ is of full column rank %({\bf Condition 1} of Theorem 1) 
%for any model $M_i$ in the model space.
}

\noindent\textit{\textbf{Proof:} See Supplement A7}.

If we consider models with an  additional random  parameter $r$ as in Subsection \ref{sec:add_param}, this result extends as follows: 

\noindent\textit{\textbf{Theorem 4:}}
\noindent\textit{If we combine the sampling model in (\ref{eq:Gen_y}) and (\ref{eq:model_k_z}) with the prior in (\ref{PriorAS}), (\ref{PriorThetaMA}) and (\ref{eq:PM_hyper}), in combination with a choice for $g$ or a hyper-prior on $g$ that leads to consistency in the Gaussian case and a proper prior on $r$ satisfying that 
\begin{equation}
r\amalg \bm{z}, M_k, \label{eq:r_indep2}
\end{equation}
then model selection consistency holds under the conditions in Theorems 1 and 2.}

\noindent\textit{\textbf{Proof:} See Supplement A8}.

In conclusion, for all those cases where ULLGMs lead to a well-defined posterior and we have no additional random variables $r$, model selection consistency holds with %the priors assumed in Section \ref{sec:prior} and 
a choice for $g$ that leads to consistency in the Gaussian case. The latter holds for the unit information prior \citep{FLS01} and the hyper-$g/n$ prior \citep{li2018mixtures} while more choices that lead to consistency in the Gaussian model can be found in Table 1 of \cite{leysteel2012}. If we have additional random variables, it is enough to also replace condition (\ref{eq:r_indep}) with (\ref{eq:r_indep2}), which does not seem restrictive in practice. 

\section{Computational Considerations and Implementation}\label{sec:computation}

%In spite of their benefits, ULLGMs can be difficult to estimate. 
% For PLN models, traditional maximum likelihood methods can yield unreliable parameter estimates even in simple scenarios, as highlighted in \cite{tsionas2010bayesian}. As an alternative, \cite{tsionas2010bayesian} focuses on a Bayesian treatment of the univariate PLN regression setting, %eliciting basic theoretical properties and 
% using an expectation-maximization algorithm for model estimation. MCMC estimation for the PLN model and extensions to $t$-distributed noise distributions are discussed in \cite{chib2001markov}. However, these contributions do not account for model uncertainty. Likewise, although estimation of BiL frameworks has been examined previously (\cite{coull2000random}), model uncertainty is typically not addressed.

The computational strategy we propose is based on the observation that, conditional on $z_i$, the posterior distributions of the latent Gaussian regression parameters, along with the marginal likelihoods and Bayes factors, assume a simple and convenient form. Hence, data augmentation, where the observed data is augmented with a posterior sample of $z_i$, is a natural choice for a posterior simulation strategy (\citealp{tanner1987calculation}). %Given $z_i$, the parameters $\alpha$, $\bm{\beta}$, and $\sigma^2$ can then be updated using a simple Bayesian regression update. 

To conduct inference under model uncertainty, we construct a posterior sampling scheme over latent outcomes, regression parameters, and models. In particular, defining $\bm{\theta}=(\alpha, \bm{\beta}, \sigma^2)$, we target the joint posterior density $P(\bm{z}, \bm{\theta}, M_k|y)$. Instead of constructing a Gibbs sampling algorithm based on full conditionals, we will consider a partially collapsed Gibbs sampler (\citealp{van2008partially}) to increase MCMC efficiency. 

The model structure readily allows to iterate between drawing from (i) $P(M_k | \bm{z})$, (ii) $P(\bm{\theta} | M_k, \bm{z})$ and (iii) $P(\bm{z} | \bm{\theta}, \bm{y}, M_k)$. A similar blocking strategy for MCMC in latent Gaussian models is suggested in \cite{geirsson2020lgm}. A related sampling scheme for variable selection in probit models is described in \cite{lee2003gene}. Sampling from $P(\bm{\theta} | M_k, \bm{z})$ is composed of drawing from (ii-a) $P(\sigma^2|\bm{z}, M_k)$ and (ii-b) $p(\alpha, \bm{\beta}|\bm{z}, M_k, \sigma^2)=p(\alpha|\bm{z}, M_k, \sigma^2)p(\bm{\beta}|\bm{z}, M_k, \sigma^2)$. All of these conditional densities are easy to simulate from. Note that, due to parameter blocks being marginalized out in some of these densities, the ordering of the updating steps is not arbitrary. Details of the MCMC algorithm are summarized in Algorithm~\ref{alg:mcmc}. 

\begin{algorithm*}[t!]
\caption{MCMC Sampling Procedure for fixed $r$ and $g$}
\label{alg:mcmc}

 \begin{algorithmic}[1]

\State Initialize model $M_k$, latent outcomes $\bm{z}$ and parameters $\sigma^2$, $\alpha$, $\bm{\beta_k}$

\For{each iteration}

\State \textbf{Sample} $M_k$ from $P(M_k|\bm{z})$
\State \hspace{\algorithmicindent} \textbf{Propose} $M^*$ using an add-delete-swap  proposal as in Supplementary 
\State
\hspace{\algorithmicindent} Section A10
\State \hspace{\algorithmicindent} \textbf{Compute} $p(\bm{z}|M^*)$ and $p(\bm{z}|M)$ using (\ref{eq:ml_a})
\State \hspace{\algorithmicindent} \textbf{Accept or Reject} moving to $M^*$ using the prior in (\ref{eq:PM_hyper}) and the acceptance \State
\hspace{\algorithmicindent}
probability
\State \hspace{\algorithmicindent} \hspace{\algorithmicindent} \hspace{\algorithmicindent} $\zeta = \min\left(1, \frac{p(M^*)}{p(M_k)} \times \frac{p(\bm{z}|M^*)}{p(\bm{z}|M_k)} \times \frac{q(M_k \mid M^*)}{q(M^* \mid M_k)}\right)$, \State
\hspace{\algorithmicindent} where $q(\cdot,\cdot)$ is the proposal pmf
\State \textbf{Sample} $\bm{\theta}$ from $P(\bm{\theta}|\bm{z}, M_k)$
\State \hspace{\algorithmicindent} \textbf{Sample} $\sigma^2$ from $p(\sigma^2|\bm{z}, M_k)$ defined in (\ref{eq:sigma_post})
\State \hspace{\algorithmicindent} \textbf{Sample} $\alpha$ from $p(\alpha|\sigma^2, \bm{z})$ defined in (\ref{eq:alpha_post})
\State \hspace{\algorithmicindent} \textbf{Sample} $\bm{\beta_k}$ from $p(\bm{\beta_k}|\bm{z}, \sigma^2, M_k)$ defined in (\ref{eq:beta_post})
\State \textbf{Sample} $\bm{z}$ from $P(\bm{z}|\bm{\theta}, \bm{y}, M_k)$ as described in Section \ref{sec:computation}
\EndFor
\end{algorithmic}
\end{algorithm*}

To obtain a sample of $z_i$, note that its conditional posterior distribution can be written as the product of a likelihood term defined via (\ref{eq:Gen_y}) and a Gaussian `prior' term (\ref{eq:model_k_z}). This factorization implies that  the $z_{i}$s are all conditionally independent, given the remaining parameters and the data. Consequently, $n$ independent univariate updates can be performed, one for each $z_i$, in each iteration of the Gibbs sampler. To simulate from $p(z_i | \cdot)$, a simple strategy is to employ independent random-walk Metropolis-Hastings updates for all $i$. However, the simple structure of $p(z_i | \cdot)$ renders gradient-based methods a convenient and more efficient alternative. In the ULLGM framework, gradients of the likelihood and priors are typically available analytically and inexpensive to compute. We found that updating $z_i$ using an adaptive version of the \textit{Barker proposal} from \cite{livingstone2022barker} offers a good balance between mixing speed and robustness of the algorithm. Robustness is particularly important in certain ULLGMs, such as those involving the Poisson distribution, where gradient-based methods may exhibit numerical instabilities. The adaptive MCMC scheme we implement is based on diminishing adaptation rates, aiming for an acceptance probability of 0.57 for each $i$ (\citealp{roberts2009examples}). We provide the log posterior gradients of $z_i$ for selected models in Supplement A9.

When $z_i$ is weakly identified by the likelihood, the proposed data augmentation scheme can induce some autocorrelation in the posterior draws.\footnote{For the PLN and BiL models, likelihood identification of a given $z_i$ becomes weaker when either the count $y_i$ is close to zero (PLN; BiL) or the number of trials $N_i$ is close to one (BiL); see Supplement A4 for details.} Nonetheless, the proposed algorithm is straightforward to implement and strikes a favorable balance between computation time and sampling efficiency. Moreover, it integrates effortlessly into the standard BMA framework. In contrast, conventional posterior simulation algorithms for high-dimensional non-Gaussian regression models often encounter significant difficulties, such as low sampling efficiency, expensive repeated likelihood evaluations, or the necessity for complex algorithmic techniques like Hamiltonian Monte Carlo. Extending conventional non-Gaussian regression model algorithms to handle model uncertainty efficiently is also challenging, as gradient-based methods can struggle with discrete sampling spaces, motivating approximate methods or intricate and computationally intensive reversible jump MCMC algorithms. In comparison, the ULLGM framework only requires very basic algorithmic techniques and knowledge of model averaging in linear Gaussian models, making it a simpler and more accessible approach while not relying on approximations.
In addition, it allows for an easy adaptation of the general sampler in Algorithm \ref{alg:mcmc} to accommodate specific members of the ULLGM class, simply by modifying the update of the latent variables $\bm{z}$.

For model proposals, we utilize an add-delete-swap (ADS) algorithm where each iteration involves adding, deleting, or swapping variables to create a new model proposal. This method has proven effective for the scenarios we examined. For very high-dimensional applications, future research might extend adaptive model proposals as suggested in \cite{zanella2020informed}, \cite{griffin2021search}, and \cite{liang2023adaptive} to the ULLGM context. Details on the ADS proposal can be found in Supplement A10. 

Assuming $g$ is random requires only minor modifications to the MCMC scheme outlined above. Given a prior density $p(g)$, we follow \cite{leysteel2012} and construct a Gibbs sampler that jointly explores latent outcomes, regression parameters, models, and values of $g$. This entails, in addition to the `within-model' update steps for $\alpha$, $\bm{\beta}$, and $\sigma^2$, simulating a new value of $g$ after sampling a new model $M_k$. We use a univariate Metropolis-Hastings step with proposal mechanism $\log(g^*) \sim \mathcal{N}\left(\log(g), \tau_g\right)$. The corresponding acceptance probability $\min\left(1, \frac{p(g^*)}{p(g)} \times \frac{p(\bm{z}|M_k, g^*)}{p(\bm{z}|M_k, g)} \times \frac{g^*}{g}\right)$ involves the prior $p(g)$, the marginal likelihood in (\ref{eq:ml_a}), and the appropriate Jacobian, accounting for the proposal on the log-scale. Similar to the adaptive approach for $z_i$, the $g$ update utilizes adaptive MCMC techniques, aiming for an acceptance rate of $0.234$. Consequently, the Gibbs sampler is fully automatic, requiring no manual input beyond the initial prior specification. Finally, if additional auxiliary parameters are involved, the MCMC scheme can be expanded to a Gibbs sampler that jointly explores latent variables, regression parameters, models, $g$, and auxiliary parameters. These auxiliary parameters will typically be univariate or low-dimensional, rendering further (adaptive) Metropolis steps a viable sampling strategy.

\subsection{Assessing the Presence of Overdispersion}

An interesting question is whether the framework can be used to formally assess the need to account for an overdispersion parameter or whether $\sigma^2=0$. While significant posterior mass of $\sigma^2$ on areas with large positive values is suggestive of the presence of overdispersion, deriving formal statements is more challenging. Bayes factors could compare models with $\sigma^2>0$ versus $\sigma^2=0$, but the unavailability of marginal likelihoods leads to various difficulties. Most importantly, computing marginal likelihood estimates becomes computationally very intensive due to the high-dimensional parameter space, as the dimension of the joint posterior density is $n + p + 2$. This makes established methods such as Laplace approximations or bridge sampling computationally highly demanding, which motivated much of our current work.

Even if this challenge was resolved, such Bayes factors would only apply conditionally on a single covariate set. A more nuanced challenge involves treating $\sigma^2$ as a parameter subject to model uncertainty (either zero or positive) while simultaneously considering uncertainty in inclusion patterns of $\bm{\beta}$. For certain models, one could combine the proposed MCMC algorithm with the AutoRJMCMC algorithm of \cite{lamnisos2009transdimensional} to explore a joint model space of models with $\sigma^2=0$ and models with $\sigma^2>0$. Another possible line of research could attempt to exploit ideas from the literature on variance selection in random effects and state-space models (\citealp{bitto2019achieving}; \citealp{cadonna2020triple}), or sampling strategies for stochastic volatility models (\citealp{kim1998stochastic}), both of which rely on Gaussian priors on transformed variance parameters. However, the adaptation of such strategies faces several obstacles and requires the development of fundamentally different posterior simulation algorithms than those considered in this paper, as well as a detailed evaluation of useful prior specifications for $\sigma^2$. We therefore leave a formal treatment of this issue to future research. Instead, we will take a largely empirical approach, comparing log predictive scores between models with $\sigma^2>0$ and $\sigma^2=0$ to assess the evidence in favor of overdispersion in the specific applications we consider. Finally, it is worth mentioning that our empirical results indicate that the adoption of our framework does not lead the inference astray, even if the data are generated by the  simpler model with $\sigma^2=0$. Thus, we generally recommend the adoption of our ULLGM framework in applied research. 

\section{Applications to Simulated Data}
\label{sec:simulations}

To assess the effectiveness of the PLN and BiL model averaging algorithms, we used simulated data, varying both the number of observations ($n=150$ and $1,000$) and the number of regressors ($p=50$, $100$ and $250$), while using $N_i=30$ for the BiL model. In all scenarios, the linear predictor was defined as $1.5 + \bm{x}_i'\bm{\beta}^{\star}$, where the first ten regression coefficients were non-zero. The coefficients were specified as:
\begin{equation*}
    \bm{\beta}^{\star} = \frac{\log(p)}{\sqrt{n}} (2,-3,2,2,-3,3,-2,3,-2,3,0,\ldots,0)' \in \mathbb{R}^P.
\end{equation*}
The regressors $\bm{x}_i$ were drawn from a normal distribution with mean 0 and covariance matrix $\Sigma$, where $\Sigma_{jk}=\rho^{\vert j - k\vert}$, determined by a correlation coefficient $\rho$. We used $\rho=0.6$ in our examples, representing a challenging setting with relatively high correlation among the regressors.

To test the %framework's
resilience against misspecification, we utilized three different data generating processes to generate the latent outcomes $z_i$. First, we added noise terms from $\mathcal{N}(0, \sigma^2)$ with $\sigma^2 = 0.2$ to the linear predictor (the ULLGM case). In the second setting, we used a noise-free linear predictor ($\sigma^2=0$), corresponding to a GLM setting.  %to generate the data. Lastly, 
Finally, we added logarithmic samples from $\mathcal{G}(5.5, 5.5)$ (a gamma distribution with mean one and variance $\frac{1}{5.5}$) as noise to the linear predictor. The implied error distribution has a variance of approximately 0.2, but is  skewed and has a non-zero mean. 

Regarding the prior setup, we choose $m = 5$ to favor sparse models. % over complex models. 
We compared two settings for $g$. Firstly, a `unit information prior' that fixes $g = n$, a popular and empirically successful default in many BMA applications (\citealp{kass1995reference}). The second setting accounts for theoretical shortcomings of deterministic $g$ (\citealp{liang_etal_08}) by letting $g$ be random using a hyper-$g/n$ prior with $a=3$, as favored in \cite{leysteel2012}. For each of the 36 settings per model, we simulated 100 replicate data sets, collected several measures of accuracy -- such as the Brier score, false positive and negative rates and posterior mean model size -- and averaged the results. In each model run, we collected 300,000 posterior samples after an initial burn-in of 250,000 iterations. 

Detailed results of the simulation study are summarized in Tables A1 and A2 in Supplement A11. The findings indicate a reasonably high level of accuracy across all settings. For the PLN (BiL) model, the average Brier score is 0.008 (0.009), suggesting high-quality posterior inclusion probability estimates. The average model size of 13.79 (13.94) slightly exceeds the true model size of ten. This is expected given the challenging scenario with highly correlated regressors. The computational burden to obtain these estimates is quite modest. On a single core of an AMD Ryzen 5 5500U, it takes an average of 4.96 (5.47) minutes for 550,000 MCMC iterations, well above the number of iterations required for chain convergence and accurate posterior results.

Across simulation settings, the results are mostly in line with what one would expect. For instance, accuracy measures mostly improve with $n$. %and decrease in
However, the effect of $p$ is less clear-cut. 
Overall, the simulation runs indicate that the ULLGM framework performs well, even in scenarios with misspecification of the sampling model, implying a certain level of robustness of the ULLGM framework for variable selection and model averaging. Importantly, results do not deteriorate in the $p = 250 > n =150$ setting we consider. In terms of prior choices, we find the unit information prior (UIP) to be slightly preferable in the settings we investigate, with the hyper-$g/n$ prior showing a tendency to favor slightly larger models, accompanied by a posterior on $g$ centered on substantially larger values than $n$. Accordingly, the fraction of model visits to the true model is also somewhat higher for the UIP, especially when $n=1000$.  Interestingly, these advantages of the UIP (the preference for model size closer to the true value of 10 and the higher fraction of true models in the sampler for large $n$) are not observed when the data are generated by a GLM, which leads to very large posterior values for $g$ under the hyper-$g/n$ prior. Inference on $\sigma^2$ is in line with expectations under both priors: close to zero for the GLM and around 0.2 for the cases with overdispersion. In general, a certain sensitivity of model averaging outcomes to prior settings is well documented in the literature and warrants a careful comparison of results based on a range of priors in applied contexts.

\section{Real Data Applications}
\label{sec:real_data_apps}

\subsection{Measles Vaccination Coverage in Ethiopia}
\label{sec:vaccination_application}

Vaccination coverage rates are a key metric for assessing the performance of national health and immunization systems. Such performance indicators are, however, generally measured using national statistics or at the scale of large regions. This is often due to the design of surveys, administrative convenience, or operational constraints. This approach can obscure subnational variations and `coldspots' of low coverage, potentially allowing diseases to persist even when overall coverage rates are high. Hence, to reduce health inequalities and make steps towards disease elimination targets, it is crucial to more accurately characterize fine-scale variations in coverage. Growing demand for subnational health metrics has led to significant interest in empirical models that provide regional vaccination coverage estimates, along with the uncertainties associated with these estimates (\citealp{utazi2018high}). These efforts often rely on Binomial models, which forms the basis of the BiL model in Table \ref{tab:RRSPM}. These models typically incorporate a regression function and spatial smoothing mechanisms, but usually do not address model uncertainty.

Given the potential effectiveness of BMA as a predictive tool, we employ the BiL model to analyze data on vaccination coverage in Ethiopia. Specifically, we utilize data from the 2019 Demographic and Health Survey (DHS) in Ethiopia.\footnote{The DHS Program provides comprehensive, nationally representative survey data on population, health, and nutrition in over 90 countries worldwide.} The data is collected in survey \textit{clusters}, with a cluster typically consisting of 25-30 households, representing for example a rural settlement or an urban neighborhood. The dataset includes a total of $n = 305$ clusters. For each cluster, we record $y_i$, the number of children born in the three years before the survey who have received the first dose of a measles (or measles-containing) vaccine, out of $N_i$, the total number of children born within the same period whose vaccination status is known. The observed
vaccination rates $y_i/N_i$ within these clusters vary from 0\% to 100\%, with an average rate of 44.8\% across all clusters. A map illustrating the distribution of survey clusters across Ethiopia and their respective sample estimates of vaccination coverage rates is presented in Supplementary Figure A6. On this map, clusters with lighter coloring indicate a smaller local sample size $N_i$, implying a smaller influence of cluster $i$ on the BiL parameter estimates.

We gather a set of $p=63$ potentially relevant predictors of vaccination rates and apply BMA to identify a robust subset of determinants. The covariates include a variety of factors that are related to health outcomes, such as regional sociodemographic characteristics, household living standards, and proxies for local economic development like a satellite-based nightlight intensity measure. Additionally, the covariates cover climatic conditions, measures of accessibility of different regions, and several nutritional scores based on anthropometric measurements of children in the survey clusters. To account for latent spatial variation in vaccination rates, we also incorporate indicators for Ethiopia's 11 regions and GPS-based data on the clusters' latitude, longitude, and altitude. This set of covariates encompasses a range of variables that can be found in most DHS surveys or can be collected from publicly accessible sources. Therefore, this analysis might hold broader interest beyond the Ethiopian case study presented here. Detailed information on the full set of covariates, along with summary statistics, is provided in Supplementary Table A3.

\begin{figure*}[t]
\centering
\begin{subfigure}{0.49\textwidth}
    \includegraphics[width=\textwidth]{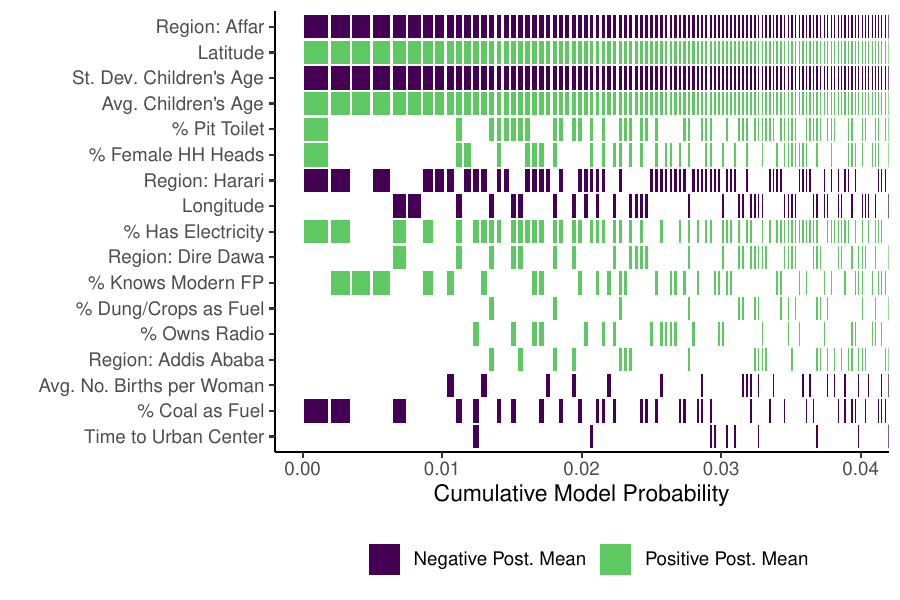}
    \caption{Highest Probability Models.}
    \label{fig:highest_measles}
\end{subfigure}
\hfill
\begin{subfigure}{0.49\textwidth}
    \includegraphics[width=\textwidth]{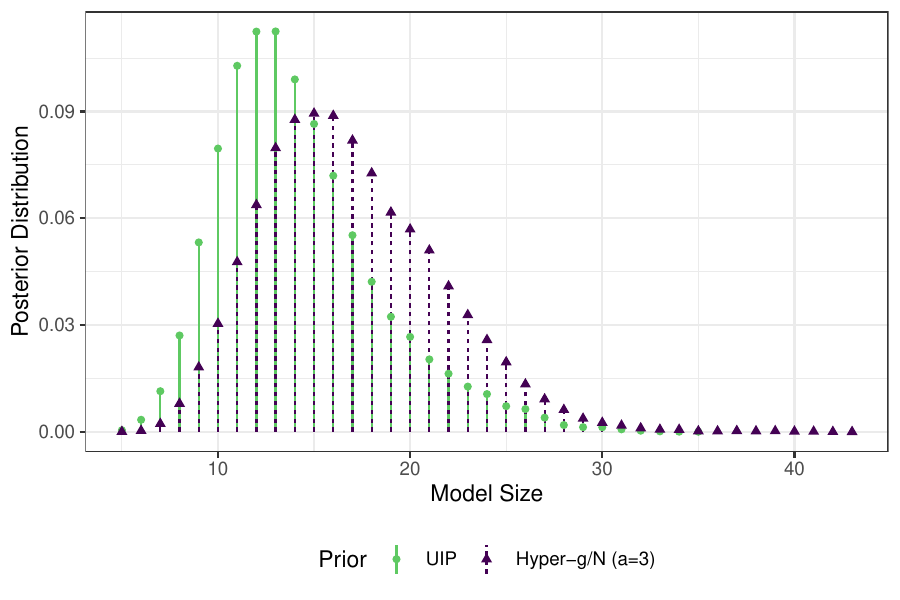}
    \caption{Posterior Distribution of Model Size.}
    \label{fig:size_measles}
\end{subfigure}
        
\caption{Estimation Results (Measles Vaccination Data). Panel (a) shows the highest posterior probability models under the unit information prior. Panel (b) illustrates the posterior distributions of model size. The highest probability models plot only includes variables with estimated $\text{PIP}>0.3$ under the unit information prior.}
\label{fig:results_measles}
\end{figure*}

We implement the BiL model using the algorithm described in Section \ref{sec:computation}, under a UIP prior ($g=n$) and a hyper-$g/n$ ($a=3$) prior, %as suggested in \cite{leysteel2012}
 alongside an agnostic uniform prior on model size ($m=p/2$). The analysis is based on $300,000$ posterior draws, collected following a burn-in phase of $250,000$ iterations. Posterior simulation takes between 15 and 20 minutes on a single core of an AMD Ryzen 5 5500U. We provide the estimated posterior inclusion probabilities and posterior means of $\bm{\beta}$ in Supplementary Figure A7. Under both priors, the posterior means of the intercept are $\mathbb{E}({\alpha}|\bm{y}) = -0.28$ while $\mathbb{E}({\sigma}^2|\bm{y}) = 0.16$ under the unit information prior and $\mathbb{E}({\sigma}^2|\bm{y}) = 0.23$ under the hyper-$g/n$ prior; with estimated posterior standard deviations of $\sigma^2$ around 0.07 this indicates that there is substantial posterior mass away from zero, informally suggesting that the data support the addition of the dispersion parameter $\sigma^2$. Figure~\ref{fig:results_measles}(b) indicates a slight preference for larger models under the hyper-$g/n$ prior (which tends to lead to somewhat smaller values for $g$). The median probability models, which include those covariates with a posterior inclusion probability (PIP) greater than $0.5$, agree on eight influential variables between the two priors used. The average age of children in a cluster is strongly positively associated with vaccination rates (see also Figure~\ref{fig:results_measles}(a)), likely due to increased interactions with healthcare systems over time and the fact that vaccines are typically not scheduled for administration directly after birth, decreasing the likelihood of very young children being vaccinated. Conversely, a larger standard deviation in children's ages within a cluster, indicating a more dispersed age distribution, is significantly associated with lower  vaccination rates, suggesting that age homogeneity can enhance the effectiveness of health interventions and vaccination campaigns. Such uniformity may support more targeted health education and vaccination efforts, encourage communal sharing of health information, and enable healthcare providers to better plan and deliver vaccination services to the predominant age group, thereby boosting overall coverage. The significant positive relationship between latitude and vaccination rates suggests higher coverage in northern clusters, while the pronounced negative impact of being in the \textit{Affar} region—characterized by remoteness, pastoralist communities and regional political tensions — indicates unobserved factors affecting spatial variations in vaccination rates. Model probabilities are in general relatively spread out (as can be seen in Figure~\ref{fig:results_measles}(a)), reflecting a rather high amount of collinearity among the covariates. The highest probability models are detailed in Supplementary Tables A5 and A6. Numerical results on estimated posterior means, standard deviations, and inclusion probabilities are available in Supplementary Table A4.

\subsection{Bilateral Migration Flows Between OECD Countries}
\label{sec:migration_application_paper}

We use the PLN model to examine migration flows between the 38 OECD countries from 2015 to 2020. 
This challenging dataset comprises $n=1,406$ bilateral migration flows, ranging from zero to 1.6 million migrants, leading to a dispersion index of 345,000. BMA is conducted with a set of $p=54$ potentially important covariates and results are presented in Supplement A12.

\subsection{Comparative Predictive Performance}
\label{sec:predictive}

To understand the predictive capabilities of ULLGMs, we carried out a predictive exercise based on the measles vaccination data and the migration data. Each data set was randomly split into test (prediction) and training sets 100 times, with 15\% allocated to the test set and the remaining 85\% to the training set. Then, we estimated models using the training data and evaluated their prediction accuracy on the test data. For the ULLGMs, a unit information prior and a hyper-$g/n$ prior were used. Non-BMA versions of ULLGMs were estimated as well, including the full, null, median probability, and highest probability models, based on the unit information prior BMA results. For the bilateral migration data, we also performed a Poisson regression BMA analysis using adaptations of the AutoRJMCMC algorithm of \cite{lamnisos2009transdimensional}. In addition, the full, null, median probability, and highest probability models, based on the AutoRJMCMC results were included in the analysis. For the vaccination data, we added  Binomial logistic regression models without overdispersion using BMA and also considering the full, null, median, and highest posterior models. For BMA methods, we set $m=p/2$ to stay agnostic about model size a priori. In the case of the GLM models, we used $\alpha \sim \mathcal{N}(0, 1000)$ and $\bm{\beta_k} | M_k \sim \mathcal{N}(0 \bm{\iota}_{p_k}, g(\bm{X}_{k}'\bm{X}_{k})^{-1})$ as priors on the regression parameters. For each model, we collected 300,000 posterior samples after an initial burn-in period of 250,000 iterations. For posterior simulation under Binomial and Poisson GLMs, we employed a multivariate Gaussian posterior approximation to the regression coefficients, based on a Bayesian IWLS algorithm (\citealp{gamerman1997sampling}).

\begin{table*}[t!]
\centering
\renewcommand{\arraystretch}{0.7} % Default value: 1
\resizebox{\textwidth}{!}{%
\begin{tabular}{lrrrrrrrrrr}
\toprule
 & Avg. LPS & Med. LPS & Min. LPS & Max. LPS & \% Best & \% Worst & Avg. Rank & $\sigma^2$ & ${M}$ Size & Time\\
\midrule
\textbf{Full Vaccination Data (n=305)} & & & & & & & & &\\
ULLGM-BMA-HYPER-$g/n$ & \textbf{1.95} & \textbf{1.95} & 1.73 & \textbf{2.27} & \textbf{61} & \textbf{0} & \textbf{1.63} & 0.22 & 16.59 & 1.99\\
ULLGM-BMA-UIP & 1.96 & 1.97 & 1.73 & 2.30 & 11 & \textbf{0} & 2.30 & 0.16 & 13.80 & 1.94\\
ULLGM-FULL-UIP & 2.01 & 2.00 & \textbf{1.71} & 2.39 & 10 & \textbf{0} & 4.36 & 0.01 & 63 & 2.25\\
ULLGM-HP-UIP & 2.01 & 1.99 & 1.76 & 2.37 & 3 & \textbf{0} & 4.55 & 0.25 & 7.34 & 1.74\\
ULLGM-MP-UIP & 2.06 & 2.06 & 1.83 & 2.43 & 1 & \textbf{0} & 6.25 & 0.28 & 7.85 & 1.77\\
ULLGM-NULL & 2.22 & 2.21 & 2.07 & 2.45 & 0 & \textbf{0} & 9.11 & 0.98 & 0 & 1.59\\
BINOM-BMA-UIP & 2.01 & 2.01 & 1.75 & 2.48 & 9 & \textbf{0} & 4.44 & - & 14.25 & 1.57\\
BINOM-FULL-UIP & 2.06 & 2.06 & 1.75 & 2.48 & 4 & \textbf{0} & 6.43 & - & 63 & -\\
BINOM-HP-UIP & 2.09 & 2.08 & 1.81 & 2.48 & 1 & \textbf{0} & 7.38 & - & 12.14 & -\\
BINOM-MP-UIP & 2.14 & 2.16 & 1.83 & 2.51 & 0 & 2 & 8.57 & - & 11.41 & -\\
BINOM-NULL & 2.56 & 2.54 & 2.14 & 3.24 & 0 & 98 & 10.98 & - & 0 & -\\
\addlinespace
\addlinespace
\textbf{Subset Vaccination Data (n=85)} & & & & & & & & &\\
ULLGM-BMA-HYPER-$g/n$ & \textbf{2.11} & 1.90 & 1.13 & 4.60 & 1 & \textbf{0} & 4.67 & 0.29 & 5.70 & 0.46\\
ULLGM-BMA-UIP & 2.12 & 1.91 & 1.11 & 4.63 & 5 & 2 & \textbf{4.39} & 0.25 & 4.72 & 0.43\\
ULLGM-FULL-UIP & 2.27 & 2.11 & 0.41 & 5.30 & 14 & \textbf{0} & 5.14 & 0.01 & 54 & 0.46\\
ULLGM-HP-UIP & 2.13 & 1.91 & 1.09 & 5.63 & 13 & 5 & 4.61 & 0.39 & 1.04 & 0.26\\
ULLGM-MP-UIP & 2.15 & 1.95 & 1.14 & 3.91 & 5 & 1 & 5.53 & 0.45 & 1.02 & 0.26\\
ULLGM-NULL & 2.13 & 2.00 & 0.82 & \textbf{3.84} & 9 & 8 & 5.93 & 0.63 & 0 & 0.20\\
BINOM-BMA-UIP & 2.69 & 2.45 & \textbf{0.33} & 7.05 & 6 & 11 & 7.73 & - & 53.99 & 1.89\\
BINOM-FULL-UIP & 2.69 & 2.44 & \textbf{0.33} & 7.05 & 2 & 15 & 7.67 & - & 54 & -\\
BINOM-HP-UIP & 2.69 & 2.44 & \textbf{0.33} & 7.04 & 4 & 19 & 7.89 & - & 54.00 & -\\
BINOM-MP-UIP & 2.69 & 2.44 & \textbf{0.33} & 7.07 & 2 & 15 & 7.84 & - & 54.00 & -\\
BINOM-NULL & 2.29 & \textbf{1.76} & 0.87 & 6.16 & \textbf{41} & 24 & 4.60 & - & 0 & -\\
\addlinespace
\addlinespace
\textbf{Full Migration Data (n=1,406)} & & & & & & & & &\\
ULLGM-BMA-HYPER-$g/n$ & \textbf{8.18} & 8.18 & \textbf{7.80} & 8.49 & 16 & \textbf{0} & 1.98 & 0.73 & 31.31 & 7.61\\
ULLGM-BMA-UIP & 8.19 & 8.19 & 7.81 & 8.50 & 3 & \textbf{0} & 2.90 & 0.72 & 28.13 & 7.52\\
ULLGM-FULL-UIP & \textbf{8.18} & \textbf{8.17} & 7.79 & \textbf{8.48} & \textbf{72} & \textbf{0} & \textbf{1.64} & 0.69 & 54 & 8.68\\
ULLGM-HP-UIP & 8.20 & 8.19 & 7.82 & 8.51 & 0 & \textbf{0} & 4.47 & 0.72 & 25.97 & 7.85\\
ULLGM-MP-UIP & 8.20 & 8.20 & 7.82 & 8.50 & 9 & \textbf{0} & 4.01 & 0.72 & 26.84 & 7.60\\
ULLGM-NULL & 9.25 & 9.25 & 8.90 & 9.61 & 0 & \textbf{0} & 6.00 & 6.82 & 0 & 7.21\\
POISS-BMA-UIP & $1.4 \times 10^3$ & $1.3  \times 10^3$ & $6.38 \times 10^2$ & $3.7 \times 10^3$ & 0 & {\bf 0} & 8.36 & - & 54.00 & 8.44\\
POISS-FULL-UIP & $1.4 \times 10^3$ & $1.3  \times 10^3$ & $6.39 \times 10^2$ & $3.7 \times 10^3$ & 0 & {\bf 0} & 8.59 & - & 54 & -\\
POISS-HP-UIP & $1.4 \times 10^3$ & $1.3  \times 10^3$ & $6.39 \times 10^2$ & $3.7 \times 10^3$ & 0 & {\bf 0} & 8.61 & - & 54.00 & -\\
POISS-MP-UIP & $1.4 \times 10^3$ & $1.3 \times 10^3$ & $6.38 \times 10^2$ & $3.7 \times 10^3$ & 0 & {\bf 0} & 8.44 & - & 54.00 & -\\
POISS-NULL & $3  \times 10^4$ & $2  \times 10^4$ & $1 \times 10^4$ & $8 \times 10^4$ & 0 & 100 & 11.00 & - & 0 & - \\
\addlinespace
\addlinespace
\textbf{Subset Migration Data (n=38)} & & & & & & & & &\\
ULLGM-BMA-HYPER-$g/n$ & \textbf{8.09} & \textbf{7.77} & 4.18 & \textbf{11.96} & \textbf{49} & \textbf{0} & \textbf{2.22} & 0.16 & 5.37 & 0.32\\
ULLGM-BMA-UIP & 8.13 & 7.93 & 4.15 & 12.18 & 11 & \textbf{0} & 2.95 & 0.27 & 5.50 & 0.29\\
ULLGM-FULL-UIP & 9.19 & 8.55 & 4.88 & 15.83 & 5 & \textbf{0} & 4.81 & 0.16 & 28 & 0.22\\
ULLGM-HP-UIP & 8.33 & 8.32 & 3.87 & 12.11 & 24 & \textbf{0} & 2.92 & 0.26 & 3.95 & 0.19\\
ULLGM-MP-UIP & 8.45 & 7.99 & 4.16 & 13.71 & 3 & \textbf{0} & 3.97 & 0.36 & 3.38 & 0.17\\
ULLGM-NULL & 9.48 & 8.59 & 7.15 & 16.23 & 3 & \textbf{0} & 5.54 & 4.06 & 0.00 & 0.12\\
POISS-BMA-UIP & $2 \times 10^5$ & $2.0 \times 10^3$ & {\bf 3.19} & $7 \times 10^6$ & 5 & 5 & 8.38 & - & 28.00 & 0.86\\
POISS-FULL-UIP & $2 \times 10^5$ & $1.9 \times 10^3$ & {\bf 3.19} & $7 \times 10^6$ & 0 & 19 & 8.54 & - & 28 & -\\
POISS-HP-UIP & $2 \times 10^5$ & $2.0 \times 10^3$ & {\bf 3.19} & $7 \times 10^6$ & 0 & 8 & 8.62 & - & 28.00& - \\
POISS-MP-UIP & $2 \times 10^5$ & $2.0 \times 10^3$ & {\bf 3.19} & $6 \times 10^6$ & 0 & 5 & 8.57 & - & 28.00  & -\\
POISS-NULL & $1 \times 10^4$ & $6.3 \times 10^3$ & $1.54 \times 10^2$ & $3 \times 10^5$ & 0 & 62 & 9.49 & - & 0 & -\\
\bottomrule
\end{tabular}%
}
\caption{Results of predictive exercise (best performance in \textbf{bold}). Smaller values of LPS indicate better predictive performance. Mean, median, minimum, and maximum LPS scores across 100 partitions, the share of replications where a model ranked as the best or worst, and its average ranking are reported. For ULLGMs, average posterior means for $\sigma^2$ are reported. `${M}$ Size' is the posterior mean number of included regressors. `Time' is run time in ms per MCMC iteration.}
\label{tab:results_predictions}
\end{table*}

In addition to analyzing the full samples, we performed leave-one-out cross-validation on subsamples of the data sets to gain insights into predictive performance in smaller samples. For the migration data, we examined the $n=38$ migration flows from OECD countries to Austria, excluding all destination-specific and three multicollinear covariates (resulting in $p=28$). For the vaccination data, we considered the data for the three regions with the lowest vaccination rates ($n=85$, $p=54$). Given the small sample size, we expect sparser models to be relevant a priori, and set $m=5$ to slightly favor smaller models. Graphical summaries of the BMA results for these subsamples, comparable to those presented in Section~\ref{sec:vaccination_application} and Supplement A12, are provided in Supplementary Figures A10 to A13. 

For predictive evaluation, we employ the logarithmic score or log predictive score (LPS, see {\it e.g.}~\citealp{FLS01}), which is a proper scoring rule for counts (see \citealp{Czado_etal_09}). Denoting the training data as $\bm{y}^{\text{t}}$ and the holdout data to be predicted as $\bm{y}^{\text{p}}$ with $n_p$ elements $y^{\text{p}}_i$ and associated covariate values $\bm{x}^{\text{p}}_i$, LPS is defined as 
\begin{equation}\label{eq:LPS}
\text{LPS}=-\frac{1}{n_p} 
\sum_{i=1}^{n_p}
\log P(y^{\text{p}}_i ~|~ 
\bm{x}^{\text{p}}_i, 
\bm{y}^{\text{t}}).
\end{equation}
The required posterior predictive probabilities evaluated at the holdout counts are approximated as detailed in Supplement A13. 

The results are presented in Table~\ref{tab:results_predictions}, where smaller values of LPS indicate better predictive performance. Various statistics of LPS scores for each model are reported across 100 partitions, as well as the share of replications where a model ranked as the best or worst, along with its average ranking. %Additionally, for the ULLGMs, average posterior means for $\sigma^2$ are reported. The model size, indicating the posterior mean number of included regressors, is also documented, averaged over all partitions. `
Time refers to milliseconds per MCMC iteration. GLM results are based on direct sampling from Gaussian posterior approximations and hence no runtime is reported for the cases without BMA. Boxplots of the LPS across the 100 replications are provided in Supplementary Figures A14 and A15.

In the real data applications examined, ULLGMs outperform their counterparts that do not account for overdispersion. In all cases, there is significant evidence that $\sigma^2 > 0$ based on comparing the LPS. For the vaccination data, overdispersion is moderate, resulting in relatively comparable outcomes for ULLGMs and non-ULLGMs. On average, the ULLGMs perform better. At the same time, they tend to select smaller models. We attribute this to the inability to accommodate the variability in the data without the overdispersion parameter: the standard models have to compensate by including more covariates. Similarly, posterior variability is significantly smaller in the GLM cases. This effect is particularly pronounced in the migration data analysis, where substantial overdispersion in the data causes all coefficients in a standard Poisson regression model to appear as highly important predictors. Consequently, the RJMCMC-BMA algorithm for Poisson regression predominantly visits the full model. This still does not adequately capture the data dispersion, which results in overly concentrated predictive distributions and very suboptimal predictive performance. In contrast, ULLGMs can accommodate overdispersion through $\sigma^2$ and produce dramatically better predictive scores. Note that estimates for $\sigma^2$ are substantially higher for the null models (where all overdispersion has to be accommodated through $\sigma^2$) and lower for the full models. Among the ULLGMs, the hyper-$g/n$ prior tends to favor larger models, but provides similar or slightly better predictive performance compared to the unit information prior framework in both data sets. Irrespective of whether we use a ULLGM structure or not, the null models tend to predict badly, especially for the complete data sets. Thus, covariate information substantially improves prediction, empirically justifying the regression framework. The best overall performance is shown by the ULLGM-BMA model with a hyper-$g/n$ prior which has the best average LPS in all cases, never predicts worst and predicts best in roughly half of the holdout samples for two of the four datasets considered.

\section{Concluding Remarks}
\label{sec:conclusion}
In this article, we present a formal and general framework for BMA in non-Gaussian regression models, based on the class of ULLGMs. We provide full characterisations of posterior existence and provide mild and intuitive conditions for model selection consistency within this class. In addition, we  develop a simple and adaptable MCMC algorithm to handle posterior simulation under model uncertainty. Our empirical investigations focus on PLN and BiL regression models for overdispersed count data. A simulation study suggests high accuracy and robustness to likelihood misspecification, making the framework potentially useful in a wide range of settings. Finally, we apply the models to two real data applications and conduct a comparative predictive exercise, further illustrating the advantages of the proposed framework.

For the measles vaccination rate application, we deal with data that are often modeled using spatial methods. The migration data are essentially network data, models for which often include latent variables to capture similarities between the nodes. Here, we used simple regression models for both applications. The ability to use BMA allows us to include many potential predictors, which helps to explicitly capture structures that are usually treated as latent. This approach not only aids in interpretation and simplifies modeling but also enables us to predict  observables using only the covariates. For some applications, combining BMA with latent variable modeling could provide an even more powerful framework. Adapting existing MCMC algorithms for latent variable models to incorporate model uncertainty is a natural extension of the algorithms developed here. %in this article.

Several additional research directions are attractive avenues for future exploration. In terms of substantive applications, the proposed framework is broadly applicable and could be particularly valuable for analyzing model uncertainty in multi-way contingency tables (\citealp{ntzoufras2000stochastic}) and related problems, such as multiple systems analysis (\citealp{silverman2020multiple}). Furthermore, many practically relevant  applications of regression models involve multivariate outcomes. Combining multivariate latent Gaussian models with multivariate Bayesian variable selection techniques (\citealp{brown1998multivariate}) could yield very interesting  modeling environments. 

%%%%%%%%%%%%%%%%%%%%%%%%%%%%%%%%%%%%%%%%%%%%%%
%% Acknowledgements                         %%
%% should be provided in the                %%
%% Acknowledgements section.                %%
%%%%%%%%%%%%%%%%%%%%%%%%%%%%%%%%%%%%%%%%%%%%%%
\begin{acks}[Acknowledgments]
The authors would like to thank the anonymous referees, an Associate Editor and the Editor for their constructive comments which improved the quality of this paper.
\end{acks}

%%%%%%%%%%%%%%%%%%%%%%%%%%%%%%%%%%%%%%%%%%%%%%
%% Funding information, if any,             %%
%% should be provided in the                %%
%% funding section.                         %%
%%%%%%%%%%%%%%%%%%%%%%%%%%%%%%%%%%%%%%%%%%%%%%
% \begin{funding}
% The first author was supported by NSF Grant DMS-??-??????.

% The second author was supported in part by NIH Grant ???????????.
% \end{funding}

%%%%%%%%%%%%%%%%%%%%%%%%%%%%%%%%%%%%%%%%%%%%%%
%% Supplementary Material, including data   %%
%% sets and code, should be provided in     %%
%% {supplement} environment with title      %%
%% and short description. It cannot be      %%
%% available exclusively as external link.  %%
%% All Supplementary Material must be       %%
%% available to the reader on Project       %%
%% Euclid with the published article.       %%
%%%%%%%%%%%%%%%%%%%%%%%%%%%%%%%%%%%%%%%%%%%%%%
\begin{supplement}
\stitle{Online Supplementary Material}
\sdescription{The supplementary material provides further interpretation of the BeC model (Supplement A1); additional details on the moments of the BiL model (A2); an illustration of the limitations of model averaging without accounting for overdispersion (A3); more information on sampling efficiency for $z_i$ (A4); proofs of Theorems 1-4 in (A5) to (A8); details on posterior gradients (A9) and the ADS proposal (A10); additional results for simulated data (A11); an additional application (A12); information on evaluating the log predictive density of ULLGMs (A13); and extensive additional results for the presented applications (A14). }
\end{supplement}

\bibliographystyle{ba}
\bibliography{bibliography}

\clearpage

\renewcommand{\thesection}{A\arabic{section}}
\setcounter{section}{0}

\renewcommand{\theequation}{A\arabic{equation}}
\setcounter{equation}{0}

\renewcommand{\thefigure}{A\arabic{figure}}
\setcounter{figure}{0}

\renewcommand{\thetable}{A\arabic{table}}
\setcounter{table}{0}

\section{Interpretation of $\sigma$ %and prior 
for BeC models}\label{sec:App_BeC}

The use of the latent variable representation of the BeC models is helpful in getting a better understanding of what this model class represents. Below we focus on link functions that are cdfs of scale mixtures of normals (which is the case for the most popular choices). 

If we take for $Q(\cdot)$ the cdf of a standard Normal, the BeC model is equivalent to the  Probit model with an extra unidentified parameter $\sigma^2$. 

\begin{figure}
    \centering
    \includegraphics[width=\textwidth]{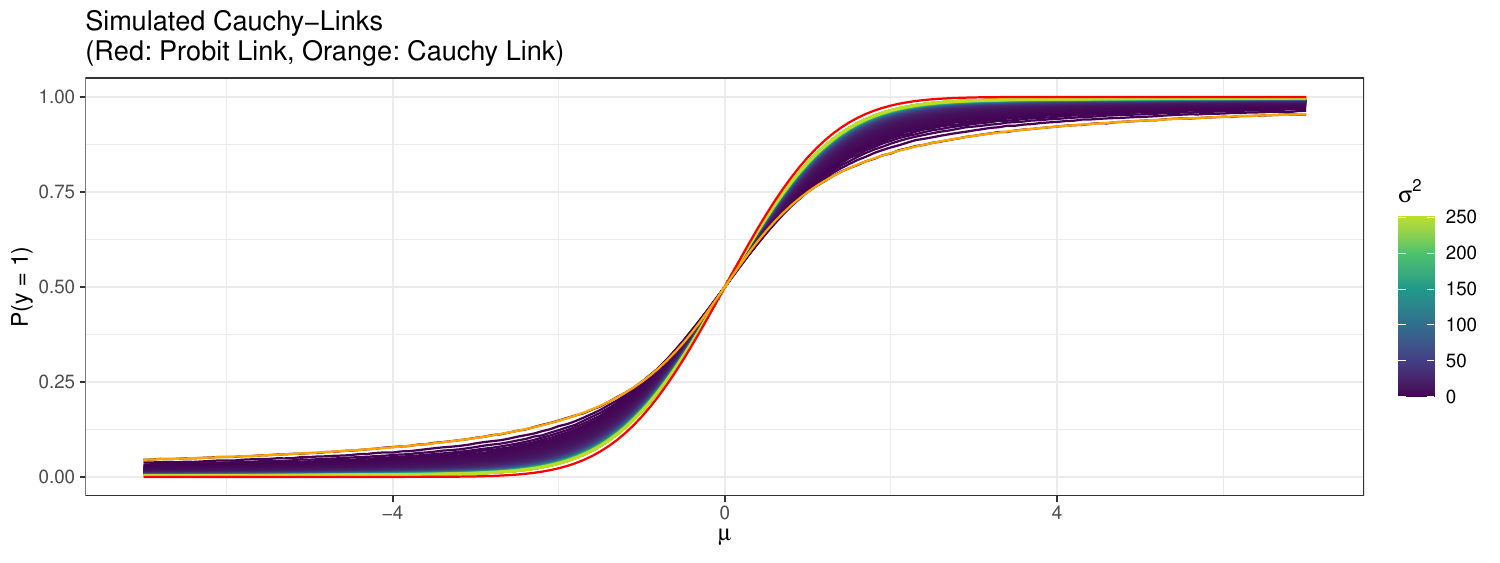}
    \caption{Simulated link functions. Value of $\sigma^2$ determines proximity to probit link and Cauchy link.}
    \label{fig:cauchy_probit_link}
\end{figure}

Choosing alternative $Q(\cdot)$ specifications maps out a class of models with a link function that sits in between that of the corresponding standard binary regression model and that of the Probit BeC model.  For example, taking $Q(\cdot)$ to be a student-$t$ cdf with $\sigma^2=0$ leads to the ``$t$-link'' model of \cite{albert1993bayesian}. 
Let us consider the latent variable representation of this model as follows:
$y_i=1$ for some latent variable $w_i>0$ and $y_i=0$ for $w_i\leq0$ with $w_i\sim N(\alpha+\bm{x}_i'\bm{\beta}, \lambda_i^{-1})$,
where $\lambda_i\sim \mathcal{G}(\nu/2,\nu/2)$. Extending this to the ULLGM setting gives us $w_i|z_i\sim N(z_i, \lambda_i^{-1})$,
and integrating out $z_i$ with (\ref{eq:Gen_z}) leads to $w_i\sim N(\alpha+\bm{x}_i'\bm{\beta},\sigma^2+ \lambda_i^{-1})$. Thus, the probability that $y_i=1$ becomes
\begin{equation}P(y_i=1)=\mathbb{E}_{\lambda_i}\left\{\Phi\left(\frac{\alpha+\bm{x}_i'\bm{\beta}}{\sqrt{\sigma^2+ \lambda_i^{-1}}}\right)\right\},\label{BeC_prob}
\end{equation}
where $\Phi(\cdot)$ denotes the cdf of the standard Normal distribution. Clearly, if $\sigma^2=0$ this simply describes the $t$-link model and as $\sigma^2\to \infty$ we will tend to the overparameterised Probit model with $P(y_i=1)=\Phi(\{\alpha+\bm{x}_i'\bm{\beta}\}/\sigma)$. For nonzero finite values of $\sigma^2$, the probability in (\ref{BeC_prob}) together with $\lambda_i\sim \mathcal{G}(\nu/2,\nu/2)$ describes a hybrid model. That is, in BeC models, the interpretation of $\sigma^2$ is the relative weight of the Probit link version. If we take $Q(\cdot)$ to be the logistic cdf instead, the same kind of argument holds, only changing the distribution for $\lambda_i$. In particular, (\ref{BeC_prob}) applies where now $\lambda_i=(2\psi_i)^2$ and $\psi_i$ has a Kolmogorov-Smirnov distribution \citep{holmesheld2006}. An example for a Cauchy link is shown in Figure~\ref{fig:cauchy_probit_link}. To scale things comparably for different values of $\sigma$, the figure plots simulated values of \begin{equation*}
P(y_i=1)=\mathbb{E}_{\lambda_i}\left\{\Phi\left(\frac{(\alpha+\bm{x}_i'\bm{\beta})\sqrt{\sigma^2 + 1} }{\sqrt{\sigma^2+ \lambda_i^{-1}}}\right)\right\},
\end{equation*}
%(\ref{BeC_prob}) as a function of $(\alpha+\bm{x}_i'\bm{\beta})/\sqrt{\sigma^2+ 1}$
for $\nu=1$ and values of $\sigma^2$ ranging from 0 (Cauchy link) to 250 (close to Probit link).

\begin{figure}
    \centering
    \includegraphics[width=\textwidth]{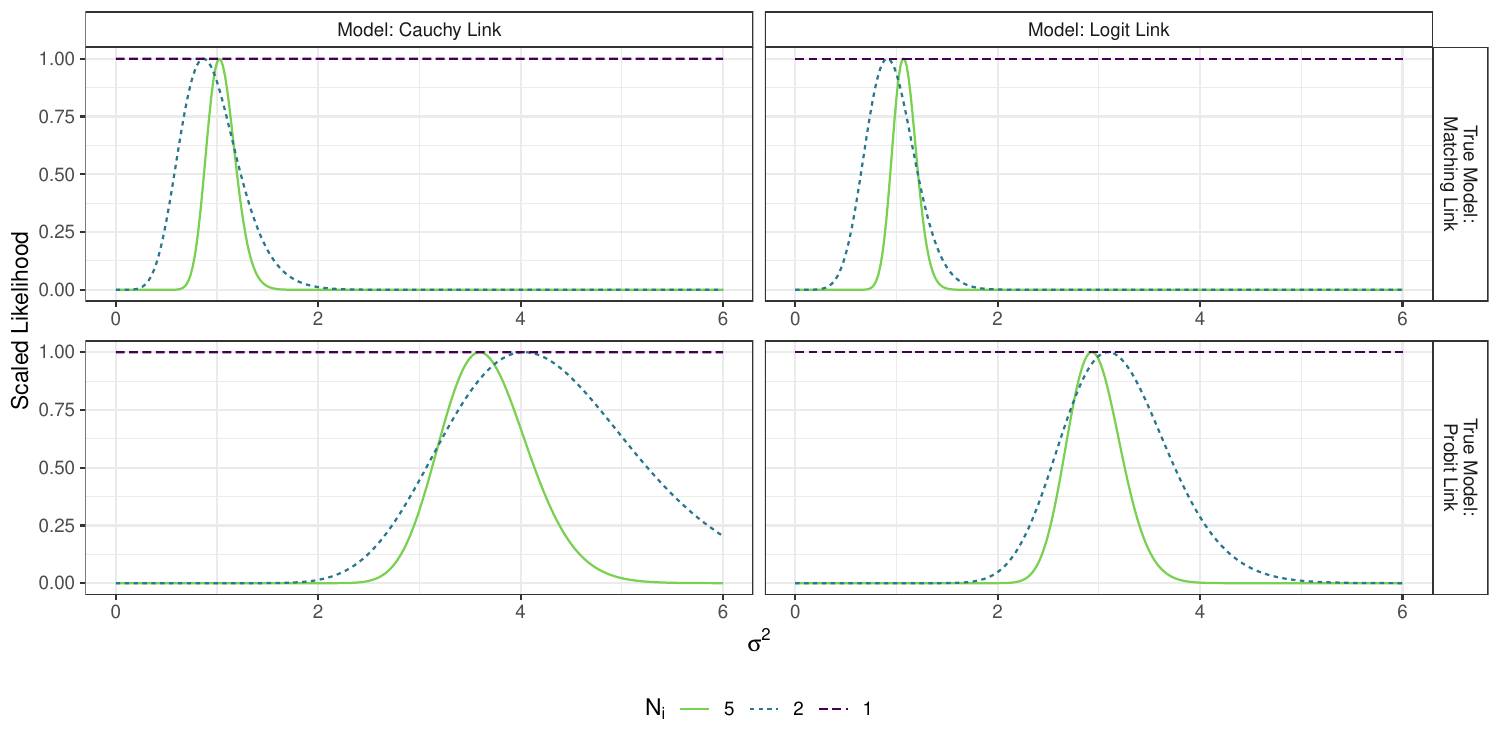}
    \caption{Likelihood evaluations for different values of $\sigma^2$ and $N_i$ under Binomial ULLGM likelihood under a logistic link function (right column) and a Cauchy link function (left column). Data generating processes follow either a misspecified probit model (bottom row) or the correct  Logit or Cauchy link function (top row). In all simulated data sets, $\sigma^2=1$ and $n=1,000$.}
    \label{fig:cauchy_logit_identification}
\end{figure}

Figure~\ref{fig:cauchy_logit_identification} illustrates the likelihood behavior of Binomial ULLGMs for various values of $\sigma^2$ and number of trials $N_i$ and for different specifications of $h(z)$, using logistic and Cauchy link functions. The data are generated from Binomial ULLGMs with $\sigma^2=1$ and three possible link functions: the Cauchy or logistic link functions lead to correct model specifications and the probit link function leads to misspecification.  The figure highlights that while $N_i > 1$ provides  likelihood information about $\sigma^2$,  the likelihood for the Bernoulli case, where $N_i = 1$, is completely flat with respect to $\sigma^2$. This implies that $\sigma^2$ cannot be identified from the likelihood for BeC models. Consequently, the posterior distribution becomes improper under the prior in (\ref{PriorAS})-(\ref{PriorThetaMA}), as discussed in Appendix~\ref{App:proof}. Inference on $\sigma^2$ will be fully determined by the  prior on $\sigma^2$, which needs to be proper.

\section{Moments and Dispersion of $y_i$ under the BiL model}
\label{app:binomial_moments}

In order to approximate the first two moments of $y_i$ under the BiL model, we approximate the logistic cdf with a scaled Gaussian cdf, such that
\begin{equation}\label{eq:BiLinterpret}
\begin{split}
\mathbb{E}(y_i~|~\bm{x}_i, N_i) &= N_i~\mathbb{E}(\pi_i~|~\bm{x}_i)\\
    &= N_i~\int \frac{\exp{(z_i)}}{1+\exp{(z_i)}}~\mathcal{N}(z_i~|~ \alpha+\bm{x}_i'\bm{\beta},  \sigma^2)~dz_i \\&\approx N_i~\int \Phi(b z_i)~\mathcal{N}(z_i~|~\alpha+\bm{x}_i'\bm{\beta}, \sigma^2)~dz_i \\&= N_i~\Phi \left( \frac{\alpha + \bm{x}'_i\bm{\beta}}{\sqrt{b^{-2} + \sigma^2}} \right) = N_i~\Phi \left( \frac{b(\alpha + \bm{x}'_i\bm{\beta})}{\sqrt{1 + b^2\sigma^2}} \right)
\end{split}
\end{equation}
for a suitable value of $b>0$, where $\Phi(\cdot)$ is the cdf of a standard Gaussian random variable. %and where we show that the first equality in line four holds in Supplementary \autoref{app:proof_cdfpdfintegral}. 
To show that the penultimate equality holds we need to verify
\begin{equation}
\int \Phi(\lambda z_i)~\mathcal{N}(z_i~|~\mu,  \sigma^2)~dz_i = \Phi\left(\frac{\mu}{\sqrt{\lambda^{-2} + \sigma^2}}\right)\label{eq:2.5}.
\end{equation}
For this, consider two random variables $X \sim \mathcal{N}(0, \lambda^{-2})$ and $Z \sim \mathcal{N}(\mu, \sigma^2)$. Note that 
\begin{equation}
    P(X \leq Z | Z = z) = P(X \leq z) = \Phi(\lambda z)
\end{equation} 
and, by the law of total probability,
\begin{equation}
P(X \leq Z) = \int P(X \leq Z | Z = z)~\mathcal{N}(z; \mu, \sigma^2)~dz = \int \Phi(\lambda z)~\mathcal{N}(z; \mu, \sigma^2)~dz
\end{equation}
which is equivalent to the left-hand side of (\ref{eq:2.5}). Now note that $P(X \leq Z) = P(X-Z \leq 0)$ and due to Gaussianity, $(X-Z) \sim \mathcal{N}(-\mu, \sigma^2 + \lambda^{-2})$. This implies that $P(X-Z \leq 0) = \Phi\left(\frac{\mu}{\sqrt{\lambda^{-2} + \sigma^2}}\right)$, verifying (\ref{eq:2.5}).

\begin{figure}
    \centering
    \includegraphics[width=0.9\textwidth]{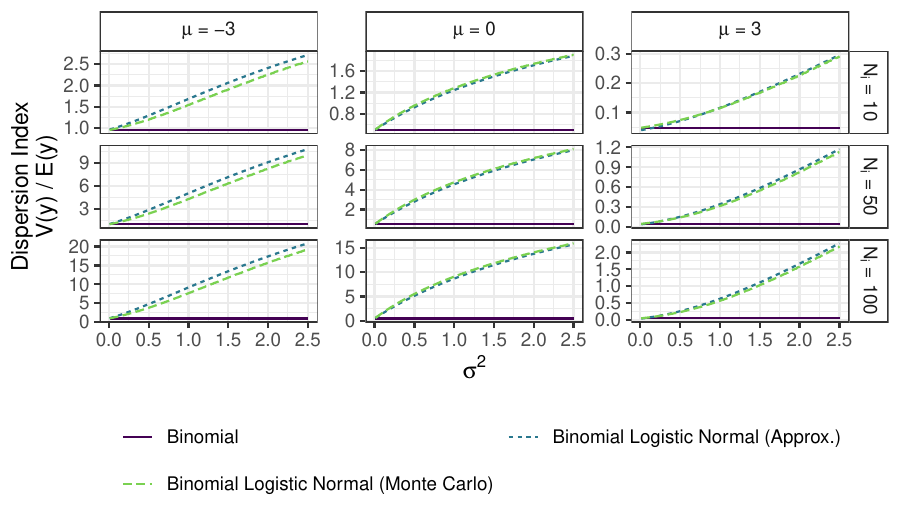}
    \caption{Approximated BiL dispersion index versus Binomial dispersion index.}
    \label{fig:bil_dispersion}
\end{figure}

Approximate variance terms $\mathbb{V}(\pi_i|\mu, \sigma^2)$ and $\mathbb{V}(y_i|N_i, \mu, \sigma^2)$ can be derived based on similar considerations. Consider first the variance of the success probability $\pi_i = [1 + \exp(-z_i)]^{-1}$ for $z_i \sim \mathcal{N}(\mu, \sigma^2)$. Again approximating the logistic cdf with a scaled probit cdf and following \citet{owen1980table}, it can be shown that
\begin{equation}
\begin{split}\label{eq:BiLvariance}
     \mathbb{V}(\pi_i|\mu, \sigma^2) &= \mathbb{E}(\pi^2_i|\mu, \sigma^2) - \mathbb{E}(\pi_i|\mu, \sigma^2)^2\\&\approx \Phi\left(\frac{b \mu}{\sqrt{1+b^2\sigma^2}}\right) - 2T\left(\frac{b\mu}{\sqrt{1+b^2\sigma^2}}, \frac{1}{\sqrt{1 + 2b^2\sigma^2}} \right) - \Phi\left(\frac{b \mu}{\sqrt{1+b^2\sigma^2}}\right)^2
     \end{split}
\end{equation}
for a suitable value of $b>0$ and where $T(h,a)$ is Owen's $T$ function. By the properties of this function, it follows that $\mathbb{V}(\pi_i|\mu, \sigma^2) \rightarrow 0$ as $\sigma^2 \rightarrow 0$ and $\mathbb{V}(\pi_i|\mu, \sigma^2) \rightarrow 0.25$ as $\sigma^2 \rightarrow \infty$. For $\mu \rightarrow \infty$ or $\mu \rightarrow -\infty$, $\mathbb{V}(\pi_i|\mu, \sigma^2) \rightarrow 0$. By the law of total variance, we have 
\begin{equation}
\begin{split}
\label{eq:variance_sp}
    \mathbb{V}(y_i|\mu, \sigma^2) &= \mathbb{E}(\mathbb{V}(y_i|\pi_i)) + \mathbb{V}(\mathbb{E}(y_i|\pi_i))\\
&= \mathbb{E}[N_i \pi_i (1 - \pi_i)] + \mathbb{V}(N_i \pi_i)\\
&= N_i \mathbb{E}[\pi_i (1 - \pi_i)] + N_i^2 \mathbb{V}(\pi_i)\\
&= N_i \mathbb{E}[\pi_i] -  N_i\mathbb{E}[\pi_i^2] + N_i^2 \mathbb{V}(\pi_i),
\end{split}
\end{equation}
which approaches the usual binomial variance $N_i \mathbb{E}[\pi_i (1 - \pi_i)]= N_i\pi_i (1 - \pi_i)$ for $\sigma^2 \rightarrow 0$, as $\sigma^2 \rightarrow 0$ implies $\mathbb{V}(\pi_i) \rightarrow 0$. The dispersion index $\mathbb{V}(y_i|\mu, \sigma^2)/\mathbb{E}(y_i|\mu, \sigma^2)$ is equivalent to
\begin{equation}
    \mathbb{D}(y_i|\mu, \sigma^2, N_i) = \frac{N_i \mathbb{E}[\pi_i] -  N_i\mathbb{E}[\pi_i^2] + N_i^2 \mathbb{V}(\pi_i)}{N_i \mathbb{E}[\pi_i]},
\end{equation}
which tends to the binomial dispersion index $(1-\pi_i)$ for $\sigma^2 \rightarrow 0$ and can be written as 
\begin{equation}
    \mathbb{D}(y_i|\mu, \sigma^2, N_i) = \mathbb{D}_{\text{Binomial}} + N_i\frac{\mathbb{V}(\pi_i)}{\mathbb{E}[\pi_i]},
\end{equation}
where $\mathbb{D}_{\text{Binomial}} =  \frac{\mathbb{E}[\pi_i] -  \mathbb{E}[\pi_i^2]}{\mathbb{E}[\pi_i]}$ stems from the usual binomial specification. The term $N_i \frac{\mathbb{V}(\pi_i)}{\mathbb{E}[\pi_i]}$ accounts for extra-binomial dispersion, and increases in $N_i$ as well as in $\sigma^2$, while it decreases with increasing $\mu$ and vanishes as $\sigma^2 \rightarrow 0$. Figure~\ref{fig:bil_dispersion} shows that the BiL dispersion index is larger than the binomial dispersion index whenever $\sigma^2 > 0$. As $\sigma^2\rightarrow\infty$ the overdispersion term will tend to $N_i/2$, for any finite value of $\mu$. Note, finally, that if we assume a probit link instead of a logistic link, resulting in an overdispersed binomial probit model, then the approximate equalities in (\ref{eq:BiLinterpret}) and (\ref{eq:BiLvariance}) hold exactly with $b=1$.

\section{Effect of Neglecting Overdispersion in Poisson and Binomial Regression}
\label{app:ms_overdispersion}
\begin{figure}
    \centering
    \includegraphics[width=\textwidth]{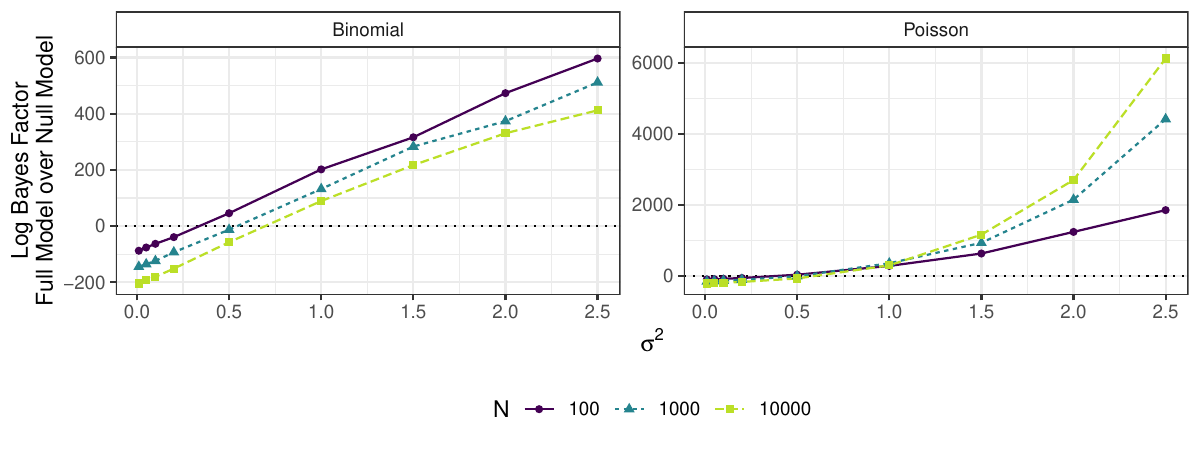}
    \caption{Log Bayes factors of the full model over the (correct) null model for Binomial and Poisson regression for various overdispersion parameters $\sigma^2$. Refer to the text for details.}
    \label{fig:app_overdispersion}
\end{figure}

To illustrate the shortcomings of neglecting overdispersion in model averaging for non-Gaussian models, we conduct a small simulation exercise. Data were simulated from both a BiL model (with 100 trials per observation) and a PLN model, with overdispersion parameter $\sigma^2$ ranging from 0.01 (approximating the GLM case) to 2.5 (indicating clear overdispersion). We vary the sample sizes ($n \in \{100, 1000, 10000\}$) while keeping the number of iid standard Gaussian regressors constant at $p=50$. The linear predictor $z_i$ is simulated from $\mathcal{N}(2, \sigma^2)$, implying no relationship with the regressors. We analysed these data with Poisson and Binomial models and approximated the log Bayes factors of the full model over the null model using the BIC approximation $0.5 \times (\text{BIC}_{\text{Null}} - \text{BIC}_{\text{Full}})$. Each setting was replicated 100 times, and the median Bayes factors across these replications are displayed in Figure \ref{fig:app_overdispersion}.

The results demonstrate that with increasing overdispersion, models without additional variation mechanisms attempt to account for the data variation by adding extra covariates and increasingly favoring larger models. This effect is more pronounced in the Poisson case, which has a more rigid variance structure compared to the binomial case, where variance is influenced by both the number of trials, and the imbalance of the dataset (as reflected in the success probabilities). Nonetheless, in both cases, even moderate amounts of overdispersion strongly favor the full model over the correct null model, even with sample size $n$ growing large. Interestingly, for the Poisson model this effect gets stronger with $n$, while for the Binomial it goes the other way. 

\section{Uncertainty in $z_i$, MCMC efficiency and limiting cases of PLN and BiL models}\label{app:limiting_z}

To develop an understanding of the spread or concentration of the posterior distribution of $z_i$ in the PLN and BiL models, it is helpful to examine the posterior approximations derived in Appendix~\ref{app:integrals}. For the PLN model, the posterior distribution is approximated as follows:
\begin{equation}
\begin{split}
z_i &\sim \mathcal{N}\left(m, s\right)\\
s &= \left(y_i + \sigma^{-2}\right)^{-1}\\
m &= s\left(\log(y_i)y_i + (\alpha + \bm{x}'_i\bm{\beta})\sigma^{-2}\right),
\end{split}
\end{equation}
From this, it becomes evident that larger $y_i$ imply a smaller posterior variance. As $y_i \rightarrow \infty$, the posterior distribution of $z_i$ converges to a point mass at $\log(y_i)$. Conversely, smaller values of $y_i$ result in greater uncertainty in the likelihood contributions of observation $i$. For observations where $y_i = 0$, the Poisson likelihood contribution $\mathcal{P}(e^{z_i})$ provides minimal information beyond $z_i < 2$, and in fact the likelihood function degenerates. Consequently, a certain number of non-zero outcomes is necessary for a proper posterior under improper priors (compare the corresponding proof conditions in Appendix~\ref{App:proof}). This indicates that likelihood identification of $z_i$, and therefore MCMC efficiency, strongly depends on the number of zero outcomes and the size of the remaining counts. For very large counts, the PLN model behaves approximately like a Gaussian regression model with outcome $\log(y_i)$. For small counts, $z_i$ is more strongly informed by prior information, resulting in decreased MCMC efficiency due to increased dependency between $z_i$ and $\alpha$, $\bm{\beta}$, and $\sigma^2$.

Similar considerations apply to the BiL framework, where the posterior approximation of $z_i$ from Appendix~\ref{app:integrals} is given by:
\begin{equation}
\begin{split}
z_i &\sim \mathcal{N}\left(m, s\right)\\
s &= \left([\hat{p}_i(1-\hat{p}_i)N_i] + \sigma^{-2}\right)^{-1}\\
m &= s\left([\text{logit}(\hat{p}_i){N_i\hat{p}_i(1-\hat{p}_i)}] + (\alpha + \bm{x}'_i\bm{\beta})\sigma^{-2}\right).
\end{split}
\end{equation}

From this approximation, it can be seen that likelihood identification is strongest when $N_i \gg 1$ and $y_i \approx 0.5N_i$. For such observations, the BiL model behaves approximately like a Gaussian regression model with outcome $\text{logit}(y_i/N_i)$ as $z_i \rightarrow \text{logit}(y_i/N_i)$ when $N_i \rightarrow \infty$, with the approximation becoming accurate faster when $\hat p_i \approx 0.5$. Conversely, as $N_i$ approaches a single trial (Bernoulli case) and/or outcomes become more imbalanced ($y_i$ close to 0 or $N_i$), likelihood identification weakens and MCMC efficiency decreases. When $y_i = 0$ or $y_i = N_i$, the likelihood contributions become degenerate, even for large $N_i$, which is reflected in the conditions for the proofs in Appendix~\ref{App:proof}.

\section{Proof of Theorem 1}
\label{App:proof}

In this Appendix, we will derive the conditions under which the posterior resulting from  the sampling model in (\ref{eq:Gen_y}) and (\ref{eq:model_k_z}) is well-defined under the improper prior structure in (\ref{PriorAS}) and
(\ref{PriorThetaMA}) for any model in the model space. Theorem 1 considers the case where any additional parameter $r$ is fixed. Thus, in the proof we will not explicitly condition on $r$. 

Denote by $\bm{y}$ the vector of all observations $y_{i}$ and partition $\bm{y}$ as $\bm{y}=(\bm{y}_N',\bm{y}_Z')'$ where $\bm{y}_N$ groups all $n_N$ observations that allow for the integral $\int P(y_i|z_i) d z_i$ to be finite. Now consider the marginal likelihood for model $M_k$
\begin{equation}
P(\bm{y}|M_k)=P(\bm{y}_Z|\bm{y}_N, M_k) P(\bm{y}_N| M_k) 
\label{MLy}
\end{equation}
and we need to show that this marginal likelihood is finite for all values of $\bm{y}$ and for any model $M_k$. First, let us focus on the vector $\bm{y}_N$:
\begin{equation}
P(\bm{y}_N|M_k)=\int P(\bm{y}_N|\bm{z}_N, M_k) p(\bm{z}_N| M_k) d \bm{z}_N,\label{MLyN}
\end{equation}
where $\bm{z}_N$ denotes those $z_{i}$ that correspond to $\bm{y}_N$ and we can write
\begin{equation}
P(\bm{y}_N|\bm{z}_N, M_k)=\prod_{i\in\cal N}P({y}_{i}|{z}_{i}),\label{eq:ygivenzN}
\end{equation}
where $\cal N$ is the set of observation indices of $\bm{y}_N$.
Let us now consider $p(\bm {z}_N| M_k)$. If  the matrix $(\iota:\bm{X_k})$ is of full column rank ({\bf Condition 1}) and if $n_N \ge 2$ ({\bf Condition 2}), we can derive that

   \begin{equation}
p(\bm{z}_N|M_k)\propto g^{-\frac{p_k}{2}}
|\bm{X}_k'
\bm{X}_k|^\frac{1}{2}
|\bm{A}_k|^{-\frac{1}{2}}
[\bm{z}_N'\bm {P}_k\bm{z}_N]^{-\frac{n_N-1}{2}},
\label{MLzN}
\end{equation}
where \begin{equation}
\bm{P}_k=I_{n_N}- (\iota : \bm{X}_{k,N})\left( {\begin{array}{cc}
    n_N^{-1}  & \bm{0}' \\
    \bm{0}  &
    \bm{A}_k^{-1} \\
  \end{array} } \right)  
\left(  \begin{array}{c}
\iota'  \\
\bm{X}_{k,N}' \\
\end{array} \right)
\end{equation}
and $\bm{A}_k=\bm{X}_{k,N}'\bm{X}_{k,N} + g^{-1} \bm{X}_{k}'\bm{X}_{k}$. Under Condition 1, $\bm{A}_k$ is invertible and for fixed choices of $g$, the expression in (\ref{MLzN}) is almost surely bounded from above by a finite number, say $c$. For hyperpriors on $g$ that are proper distributions with pdf $p(g)$,  the relevant marginal likelihood for $\bm{z}_N$ is the expression in  (\ref{MLzN}) integrated with respect to $p(g)$. As $g$ tends to zero,  (\ref{MLzN}) tends to a finite constant in $g$ and as $g$ tends to $\infty$ the expression in (\ref{MLzN}) behaves like $g^{-p_{k}/2}$.  Thus any proper $p(g)$ will lead to a finite value of the marginal likelihood $p(\bm{z}_N|{M}_k)$. For the null model $M_0$ with only the intercept, the prior is simply (\ref{PriorAS}) and the marginal likelihood is \begin{equation}\label{M0A}
p(\bm{z}_N|{M}_0)\propto[(\bm{z}_N-\bar{z}_N \iota)^\prime(\bm{z}_N-\bar{z}_N \iota)]^{-\frac{n_N-1}{2}}\end{equation} 
(with the same proportionality constant as in (\ref{MLzN})), which is also bounded. In the latter expression we have defined \begin{equation}\bar{z}_N=\frac{1}{n_N}\sum_{i\in \cal N} z_{i}.
\end{equation}
Therefore, (\ref{MLyN}) becomes
\begin{equation}
P(\bm{y}_N|M_k)< c\prod_{i\in\cal N}\int P({y}_{i}|{z}_{i}) d {z}_{i},
\label{Int}
\end{equation}
and it is sufficient to show that each of the integrals in the above expression is finite. In the sequel, we will consider the models presented in Table \ref{tab:RRSPM} and drop subscripts for convenience. 

\subsection{PLN model}

Here, we consider 
\begin{equation}
I=\int_\Re P({y}|{z}) d {z}=\int_\Re \frac{\exp[-\exp(z)] (\exp{z})^y}{y!} d z,
\end{equation}
and use the variable transformation $h=\exp(z)$ to obtain
\begin{equation}
I=\frac{1}{y!}\int_{\Re_+} \exp[-h] h^{y-1} d h=\frac{1}{y!} \Gamma(y)=\frac{1}{y},
\end{equation}
which means that  $\bm{y}_N$ consists of all nonzero observations.
This leads directly to 
\begin{equation}
P(\bm{y}_N|M_k)< c\prod_{i\in\cal N} \frac{1}{y_{i}} < \infty. \label{MLyN1}
\end{equation}
 Thus, we have a well-defined posterior distribution after taking into account at least 2 nonzero observations. These observations in $\bm{y}_N$ will then update the improper prior into a proper posterior % similar to (\ref{PostS}) and (\ref{PostTheta}),
which can then be used as the (proper) prior for the analysis of the zero observations in $\bm{y}_Z$. Of course, the latter will lead to a proper posterior with a finite integrating constant. Thus, $P(\bm{y}_Z| \bm{y}_{N}, M_k)<\infty$ and using (\ref{MLy}) and (\ref{MLyN1}) we obtain that
$P(\bm{y}|M_k)<\infty$
which proves the result. Conditions 1 and 2 jointly are thus sufficient for propriety. Condition 1 is also necessary, since we need $\bm{X}_k$ to be of full column rank for the prior specification in (\ref{PriorThetaMA}) and given that the regressors are demeaned this also implies that Condition 1 holds.  In order to prove that Condition 2 is also necessary for propriety, we consider the same line of proof as in Subsection \ref{sec:App1_BeL}. If condition 2 does not hold, we need to rely on observations for which $y_i=0$ to obtain a proper posterior ($\bm{y}_N$ with $n_N<2$ does not lead to a proper posterior). As explained in Subsection \ref{sec:App1_BeL}, the integral in (\ref{IntBer}) then needs to integrate in each $z_i$ which requires that $P(y_i|z_i,M_k)$ tends to zero in the tails for $z_i$. For $y_i=0$ we have \begin{equation}
 P(y_i=0|z_i,M_k)=  \exp[-\exp(z_i)],
\end{equation}
which tends to 1 as $z_i\to -\infty$. Thus, (\ref{IntBer}) will not integrate and condition 2 is necessary for posterior propriety in the PLN model.  

If we change the distribution for the observables $y_i$ or the link function $h(\cdot)$, then all that changes in the proof is the definition of $\bm{y}_N$ and the expression for (\ref{Int}).

\subsection{NBL model}\label{proof:NBL}
 If $y\sim$ Negative Binomial$\left(r, \frac{\exp(z)}{1+\exp{z}}\right)$ then the integrals in (\ref{Int}) are
\begin{equation}
I=\int_\Re P({y}|{z}) d {z}={r+y-1\choose y}\int_\Re \pi^r (1-\pi)^{y} d z,
\end{equation} defining 
$\pi=\frac{\exp(z)}{1+\exp{z}}$. Thus, we obtain
\begin{equation}
I={r+y-1\choose y}\int_{0}^1 \pi^r (1-\pi)^y \left|\frac{d \pi}{d z}\right|^{-1}d\pi
={r+y-1\choose y}\int_{0}^1 \pi^{r-1}(1-\pi)^{y-1} d\pi=\frac{1}{y}.
\end{equation}
The integral above is  finite for all observations where $y>0$. Thus, if we denote by $\bm y_N$ those observations for which $y_i>0$, we have
\begin{equation}
P(\bm{y}_N|M_k)< c < \infty, 
\end{equation}
which means that we have a well-defined posterior distribution after taking into account at least 2 observations in $\bm y_N$. The rest of the proof mirrors that for the PLN model. If we do not have two observations for which $y_i>0$, we can use the same arguments as in Subsection \ref{sec:App1_BeL} to show that the posterior does not exist, so that conditions 1 and 2 are both necessary and sufficient for posterior propriety in the NBL case.

\subsection{BiL model}
 If we use Binomial $y\sim Bin\left(N, \frac{\exp(z)}{1+\exp{z}}\right)$ then we obtain 
\begin{equation}
I=\int_\Re P({y}|{z}) d {z}={N\choose y}\int_\Re \pi^y (1-\pi)^{N-y} d z,
\end{equation} defining 
$\pi=\frac{\exp(z)}{1+\exp{z}}$. Thus, we obtain
\begin{equation}
I={N\choose y}\int_{0}^1 \pi^y (1-\pi)^{N-y} \left|\frac{d \pi}{d z}\right|^{-1}d\pi
={N\choose y}\int_{0}^1 \pi^{y-1} (1-\pi)^{N-y-1} d\pi.
\end{equation}
The integrand above is the kernel of a Beta$(y,N-y)$ distribution. Provided we have $0<y<N$, this leads to 
\begin{equation}
I={N\choose y}\frac{\Gamma(y)\Gamma(N-y)}{\Gamma(N)}=\frac{N}{y(N-y)}
\end{equation}
The latter expression is finite for all observations where $0<y<N$. Thus, if we denote by $\bm y_N$ those observations for which $0<y_i<N_i$, we have
\begin{equation}
P(\bm{y}_N|M_k)< c\prod_{i} \frac{N_i}{y_i(N_i-y_i)} < \infty, \label{MLyN1B}
\end{equation}
which means that we have a well-defined posterior distribution after taking into account at least 2 observations in $\bm y_N$. The rest of the proof mirrors that for the PLN model. If we do not have two observations for which $0<y_i<N_i$, we can use the same arguments as in Subsection \ref{sec:App1_BeL} to show that the posterior does not exist, so that conditions 1 and 2 are both necessary and sufficient for posterior propriety in the case of the BiL model. 

\subsection{ErLN models}\label{proof:erlang}
 The Erlang case where $y_i\sim$ Erlang$(r, \lambda)$ (i.e.~a Gamma distribution with integer shape parameter $r=1,2,\dots$) covers the Exponential model if we take $r=1$. We assume that  $\lambda=\exp(z)$, so that the integrals in (\ref{Int}) are given by
\begin{equation}
I=\int_{\Re_+} 
    p({y}|{z}) d {z}=\frac{y^{r-1}}{\Gamma(r)}\int_{\Re_+}   \exp(rz)\exp\{-y\exp(z)\} d z.
\end{equation} 
%dropping observational indices for notational convenience. 
Using the transformation $\lambda={\exp(z)}$, we obtain 
\begin{equation}
I=\frac{y^{r-1}}{\Gamma(r)}\int_{\Re_+} \lambda^{r-1} \exp\{-y\lambda\} d \lambda
=\frac{1}{y}.
\end{equation}
This integral is finite for all observations where $y$ is different from 0. This is an event of measure zero in the sampling distribution, so the posterior distribution is almost surely well-defined for any value of $r$, taking $\bm{y}_N=\bm{y}$. 
\subsection{LNN model}
 If we use log Normal sampling $y_i\sim$ log-Normal$(\mu,1)$ with $\mu=z$, the integrals in (\ref{Int}) are given by
\begin{equation}
I=\int_{\Re} 
    p({y}|{z}) d {z}={\frac{1}{y\sqrt{2\pi}}}\int_{\Re} \exp{-\frac{1}{2}{(\ln y - \mu)^2}} d \mu.
\end{equation} 
%dropping observational indices for notational convenience. 
This immediately leads to
\begin{equation}
I={\frac{1}{y}},
\end{equation} 
which  is finite for all observations where $y>0$. The  event $y=0$ has zero probability in the sampling distribution, so the posterior distribution is almost surely well-defined, taking $\bm{y}_N=\bm{y}$. 

\subsection{LNLN model}
 If we use log Normal sampling $y_i\sim$ log-Normal$(r,\lambda)$ with $\lambda=\exp(z)$, the integrals in (\ref{Int}) are given by
\begin{equation}
I=\int_{\Re} 
    p({y}|{z}) d {z}={\frac{1}{y\sqrt{2\pi}}}\int_{\Re} \lambda^{-\frac{1}{2}}\exp{-\frac{(\ln y-r)^2}{2 \lambda}} d z.
\end{equation} 
%dropping observational indices for notational convenience. 
Using the transformation $\lambda={\exp(z)}$, we obtain 
\begin{equation}
I={\frac{1}{y\sqrt{2\pi}}}\int_{\Re_+} \lambda^{-\frac{3}{2}}\exp{-\frac{(\ln y-r)^2}{2 \lambda}} d\lambda,
\end{equation} 
which can be solved using the inverse gamma distribution to leave us with
\begin{equation}
I=\frac{1}{y |\ln y-r|}.
\label{eq:ML_LNLN}
\end{equation}
This integral is finite for all observations where $y$ is different from 0 or $\exp(r)$. These are events of measure zero in the sampling distribution, %(whether $r$ is fixed or random)
so the posterior distribution is almost surely well-defined, taking $\bm{y}_N=\bm{y}$. 

\subsection{Bernoulli-based models BeC}\label{sec:App1_BeL}

Consider the marginal likelihood for model $M_k$ based on the entire sample:

\begin{equation}
 P(\bm{y}|M_k)
 =\int P(\bm{y}|\bm{z},M_k) p(\bm{z}|M_k) d\bm{z}=\int\prod_{i=1}^n P({y_i}|z_i,M_k) p(\bm{z}|M_k) d\bm{z}.\label{IntBer}
\end{equation}

As explained in Appendix \ref{App:z}, the marginal density of $\bm{z}$ 
given $M_k$ is  
a quadratic form which does not have sufficiently thin tails to integrate in $\bm{z}$. Thus, for the integral in (\ref{IntBer}) to be finite the tails need to be squeezed by $\prod_i P({y_i}|z_i,M_k)$. In other words, when integrating with respect to $z_i$, the corresponding   $P({y_i}|z_i,M_k)$
needs to go to zero fast enough as $z_i$ tends to $\infty$ and $-\infty$. %This has to hold for any $n=1,\dots,n$. 
Since 
\begin{equation}
P({y_i}|z_i,M_k)=\pi_i^{y_i} (1-\pi_i)^{1-y_i}
\end{equation}
and we have a link function in BeC models that associates $z_i\to\infty$ with $\pi_i\to 1$ 
it is clear that 
an observed $y_i=1$ will not change the right-hand tail of $p(\bm{z}|M_k)$ along dimension $i$ ( $P(y_i=1|z_i,M_k)=\pi_i$, which will be bounded from below for large $z_i$). Similarly, the value $y_i=0$ will leave the left-hand tail untouched. Thus, for any possible observation the marginal likelihood in (\ref{IntBer}) will not be finite and the posterior will not exist. 

\subsection{Marginal prior distribution of $z$}\label{App:z}

As stated in (\ref{eq:ml_a}), the marginal likelihood under fixed $g$ is
\begin{equation}
%p(\bm{z}|M_k)\propto 
%(1+g)^{\frac{n_N-1-p_k}
%{2}}
%\left[\delta\bm{z}'Q_{(\iota:X_k)}\bm{z}+(1-\delta) (\bm{z}-{\bar z}'\iota)'(\bm{z}-{\bar z}'\iota)\right]^{-\frac{n-1}{2}},
p(\bm{z}|M_k)\propto (1+g)^{\frac{n-1-p_k}{2}}
\left[\{1+g(1-R_{k}^2)\}(\bm{z}-{\bar z}\iota)'(\bm{z}-{\bar z}\iota)\right]^{-\frac{n-1}{2}},\label{MLZI}
\end{equation}
where $R_{k}^2$ is the coefficient of determination of $\bm z$ regressed on $\bm{X}_k$ (and an intercept). %and the proportionality constant is the same for all models, including the null model. 
Thus, the pdf of $\bm{z}$ can be written as 
\begin{align}
p(\bm{z}|M_k)&\propto 
\left[g\bm{z}'Q_{\bm{W}_k}\bm{z}+ (\bm{z}-{\bar z}'\iota)'(\bm{z}-{\bar z}'\iota)\right]^{-\frac{n-1}{2}}\\
    &= \left[g\bm{z}'Q_{\bm{W}_k}\bm{z}+ \bm{z}'(I-\frac{1}{n}\iota \iota')\bm{z}\right]^{-\frac{n-1}{2}}\\
    & = \left[\bm{z}'{\bm V}_k\bm{z}\right]^{-\frac{n-1}{2}},\label{MLZ}
\end{align}
where we have defined  $\bm{W}_k=(\iota : \bm{X}_k)$ and 
\begin{equation}
 \bm{V}_k=(1+g)I-\frac{1}{n}\iota\iota'-g\bm{W}_k(\bm{W}_k'\bm{W}_k)^{-1}\bm{W}_k',  
\end{equation}

The $n \times n$ matrix $\bm{V}_k$ is positive definite as it is the sum of two positive definite matrices. The distribution of $\bm{z}$ for each model is reminiscent of a multivariate Student-$t$ but is not  a proper distribution as it would correspond to negative degrees of freedom and an unbounded density at zero. 
As expected, the expression for $p(\bm{z}_N|M_k)$ in (\ref{MLzN}) simplifies to (\ref{MLZ}) if we take $\bm{z}_N=\bm{z}$, barring a proportionality constant $(1+g)^{\frac{n-1-p_k}{2}}$, which appears in (\ref{MLZI}) but is immaterial to the  considerations in this section.

\section{Proof of Theorem 2} \label{App:proof2}

This theorem applies to models with additional parameters and its proof has a similar structure as that for Theorem 1. 

Again, we focus on the marginal likelihood for a subsample $\bm{y}_N$ which groups the observations corresponding to a  finite $\int P(y_i|z_i,r)dz_i$. We can write for the marginal likelihood of $\bm{y}_N$ in model $M_k$:
\begin{equation}
P(\bm{y}_N|M_k)=\int P(\bm{y}_N|\bm{z}_N,r,M_k)p(\bm{z}_N|M_k) P(r|\bm{z}_N, M_k) d\bm{z}_N dr.  
\end{equation}
Using the fact that $p(\bm{z}_N|M_k)$ is bounded by some finite constant $c$ under conditions 1 and 2 (see the proof of Theorem 1), and applying (\ref{eq:r_indep}) along with (\ref{eq:ygivenzN}), we obtain 
\begin{equation*}
P(\bm{y}_N|M_k)<c\int P(\bm{y}_N|\bm{z}_N,r,M_k) P(r| M_k) d\bm{z}_N dr
=c\int \prod_{i\in\cal N}P({y}_i|{z}_i,r) d{z}_i P(r| M_k)  dr.
\end{equation*}
Substituting the definition of $f(r)$ in (\ref{eq:def_f(r)}), we obtain directly
\begin{equation}
P(\bm{y}_N|M_k)<c\int 
 f(r) P(r| M_k) dr,
\end{equation}
so that the condition in (\ref{eq:Th2}) is sufficient for posterior existence. The rest of the proof follows a similar reasoning to the proof of Theorem 1. 

For the LNLN model we have 
$$f(r)=\prod_{i=1}^n \frac{1}{y_i|\ln y_i-r|}$$
from (\ref{eq:ML_LNLN}), which means that the condition in (\ref{eq:Th2}) would require the prior on $r$ to compensate for $f(r)$ behaving like $1/|r|$ in $n$ neighbourhoods around $\ln(y_i)$. This would need the prior on $r$ to have vanishing mass in these neighbourhoods, but of course their location depends on the observations.  Thus, there is no (non-data based) prior that can satisfy (\ref{eq:Th2}) for the LNLN model, so we can not conclude  that posterior inference on $r$ can be conducted with the overall prior structure assumed here for the LNLN model.

\section{Proof of Theorem 3}

In the consistency proof for Theorem 3, we are assuming that the conditions for existence in Theorem 1 hold. In view of the fact that here we are considering the behaviour for large $n$, these conditions are trivially satisfied. Consider the marginal likelihood for model $M_k$
\begin{equation}
P(\bm{y}|M_k)=\int P(\bm{y}|\bm{z}, M_k) p(\bm{z}| M_k) d \bm{z},\label{MLyTrue}
\end{equation}
where $P(\bm{y}|\bm{z}, M_k)=P(\bm{y}|\bm{z})$. Without loss of generality, assume that $M_k$ is the model that generated the data. Then, we can write
\begin{equation}
P(\bm{y}|M_k)=\int P(\bm{y}|\bm{z}) p(\bm{z}| M_l) {\frac{p(\bm{z}| M_k)}{p(\bm{z}| M_l)}} d \bm{z},\label{MLyTrue2}
\end{equation}
where $M_l, l\ne k$, is any other model and thus misspecified.  For finite $n\ge 2$ and if the matrix $(\iota : \bm{X}_i)$ is of full column rank ({\bf Condition 1} of Theorem 1, which is imposed by the prior in (\ref{eq:PM_hyper})), we know from (\ref{MLZI}) that $p(\bm{z}| M_i), i=1,\dots,K$ is almost surely positive and finite, so that the ratio $r_{kl}(\bm{z})=\frac{p(\bm{z}| M_k)}{p(\bm{z}| M_l)}$ is 
bounded from below by some positive number $B_{kl}$ almost surely in $\bm{z}$. This implies that 
\begin{equation}
P(\bm{y}|M_k)>B_{kl}P(\bm{y}|M_l).
\end{equation}
Now $B_{kl}$ is the minimum Bayes factor of $M_k$ versus $M_l$ for $\bm{z}$, and we know that the underlying Gaussian model for $\bm{z}$ with the prior in Section 3 is model-selection consistent for the unit information prior \citep{FLS01} and the hyper-$g/n$ prior \citep{li2018mixtures}. 
As a consequence, under these choices for $g$ (or any other choice that leads to consistency in the Gaussian model, see e.g.~Table 1 in \citealp{leysteel2012}), $\lim_{n\to \infty} r_{kl}(\bm{z})=\infty$
almost surely in $\bm{z}$. Thus, it must be the case that 
$\lim_{n\to \infty} B_{kl}=\infty$, which immediately  leads to 
\begin{equation}
\lim_{n\to \infty}{\frac{P(\bm{y}|M_k)}  {P(\bm{y}|M_l)}}=\infty,  
\end{equation}
proving model-selection consistency for ULLGMs. 

\section{Proof of Theorem 4}

We assume that the posterior exists for each model we consider; in particular, we assume that the sufficient conditions for existence in Theorem 2 hold, which rules out the LNLN model. In addition, we now take the prior on $r$ to also be independent of $M_k$, so that (\ref{eq:r_indep}) is replaced by 
\begin{equation}
r\amalg \bm{z}, M_k. \label{eq:r_indep3}
\end{equation} The marginal likelihood for the model $M_k$ which is assumed to have generated the data is then 
%\begin{equation}
%P(\bm{y}|M_k)=\int P(\bm{y}|\bm{z}, r) p(\bm{z}| M_k) p(r|M_k)d \bm{z} dr,\label{MLyrTrue}
%\end{equation} 
%Then, we can write
\begin{equation}
P(\bm{y}|M_k)=\int P(\bm{y}|\bm{z}, r) p(r) dr \, p(\bm{z}| M_k) d \bm{z}, \label{MLyrTrue2}
\end{equation}
where we can replace the integral in $r$ by
\begin{equation}\label{eq:def_h(z)}
P(\bm{y}|\bm{z})= \int  P(\bm{y}| \bm{z}, r) p(r) dr , 
\end{equation}
which has to be finite a.s.~since the posterior exists. Thus, (\ref{MLyrTrue2}) becomes 
\begin{equation}
P(\bm{y}|M_k)=\int P(\bm{y}|\bm{z}) p(\bm{z}| M_k) d \bm{z}. \label{MLyrTrue3}
\end{equation}
We can then write 
\begin{equation}
P(\bm{y}|M_k)=\int P(\bm{y}|\bm{z}) p(\bm{z}| M_l) {\frac{p(\bm{z}| M_k)}{p(\bm{z}| M_l)}} d \bm{z}, \label{MLyrTrue4}
\end{equation}
which is identical to (\ref{MLyTrue2}) and repeating the same arguments as in the proof of Theorem 3 proves model selection consistency.

\section{Log Posterior Gradients for PLN and BiL models}
\label{sec:gradients}
Note that in any ULLGMs, the gradient of the conditional log posterior of $z_i$ additively decomposes into two parts. The first part is the gradient of the log of the Gaussian prior $p(z_i|\alpha, \bm{\beta}, \sigma^2, \bm{x}_i)$. Regardless of which likelihood is chosen as the basis for a ULLGM, this gradient is given by
\begin{equation*}
    \frac{\partial \log p(z_i|\alpha, \bm{\beta}, \sigma^2, \bm{x}_i)}{\partial z_{i}} = - \frac{z_{i} - \alpha - \bm{x}_i\bm{\beta}}{\sigma^2}.
\end{equation*}

The second part is the gradient of the log of the likelihood term $P(y_i|h(z_i), r)$, which depends on the type of model. For the PLN model, we have 

\begin{equation*}
P(y_i|z_i) = \frac{e^{y_i z_i}e^{-e^{z_i}}}{y_i!}
\end{equation*}
and hence 
\begin{equation*}
\frac{\partial \log P(y_i|z_i)}{\partial z_{i}} = y_{i} - e^{z_{i}}.
\end{equation*}

For the BiL model, we have that 
\begin{equation*}
P(y_i|z_i, N_i) = \left(N_i \atop y_i \right) p_i^{y_i} (1 - p_i)^{N_i - y_i}
\end{equation*}
and hence 
\begin{equation*}
\frac{\partial \log P(y_i|z_i)}{\partial z_{i}} = \frac{y_i - (N_i - y_i) e^{z_i}}{1+e^{z_i}}.
\end{equation*}

\section{Details on Add-Delete-Swap Proposal}
\label{app:correction_term}

Let \(p\) be the total number of predictors and \(p_k\) the number of predictors in the current model \(M_k\). An add-delete-swap algorithm proposes a new model \(M^*\) using one of three moves:
\begin{itemize}
    \item \textbf{Addition:} Add a predictor that is not currently in the model.
    \item \textbf{Deletion:} Remove a predictor from the model.
    \item \textbf{Swap:} Exchange one predictor in the model with one that is not.
\end{itemize}

The selection probabilities for the move types depend on the current model size:
\begin{equation*}
\begin{aligned}
\pi_A(M_k)=
\begin{cases}
1, & \text{if } p_k=0,\\[1mm]
1/3, & \text{if } 0<p_k<p,
\end{cases}
\\
\pi_D(M_k)=
\begin{cases}
1, & \text{if } p_k=p,\\[1mm]
1/3, & \text{if } 0<p_k<p,
\end{cases}
\\
\pi_S(M_k)=
\begin{cases}
1/3, & \text{if } 0<p_k<p,\\[1mm]
0, & \text{otherwise},
\end{cases}
\end{aligned}
\end{equation*}

where the subscripts $A$, $D$, and $S$ refer to add, delete and swap, respectively.

\paragraph{Addition Move.}  
Suppose we are in a model \(M_k\) with \(p_k<p\) and propose to add one predictor. Given that addition is chosen with probability \(\pi_A(M_k)\), we then select one of the \(p-p_k\) predictors (not in \(M_k\)) uniformly. Thus, the forward proposal probability is
\[
q(M^*\mid M_k)=\pi_A(M_k)\cdot\frac{1}{p-p_k}.
\]

In the reverse move from \(M^*\) (which now contains \(p^*=p_k+1\) predictors) the reverse action is a deletion. Note that if \(p_k+1<p\), then deletion is one of three possible moves, whereas if \(p_k+1=p\) (i.e. when the proposed model is full), deletion is forced. Hence, we have:
\[
\pi_D(M^*)=
\begin{cases}
1/3, & \text{if } p_k+1<p,\\[1mm]
1, & \text{if } p_k+1=p,
\end{cases}
\]
and the probability of choosing the specific predictor to remove (from the \(p_k+1\) included) is \(1/(p_k+1)\). Therefore,
\[
q(M_k\mid M^*)=\pi_D(M^*)\cdot\frac{1}{p_k+1}.
\]

The corresponding correction factor in the Metropolis–Hastings ratio is then
\[
\frac{q(M_k\mid M^*)}{q(M^*\mid M_k)}=\frac{\pi_D(M^*)/(p_k+1)}{\pi_A(M_k)/(p-p_k)}.
\]

\paragraph{Deletion Move.}  
Now, assume that \(M_k\) has \(p_k>0\) and we propose to delete one predictor. The deletion move is selected with probability \(\pi_D(M_k)\) and, given this choice, one predictor is removed uniformly among the \(p_k\) predictors in the model. Thus,
\[
q(M^*\mid M_k)=\pi_D(M_k)\cdot\frac{1}{p_k}.
\]

In the reverse move from \(M^*\) (which now contains \(p^*=p_k-1\) predictors), the reverse action is an addition. If \(p_k-1>0\) (i.e. the proposed model is not empty), addition is one of three moves, while if \(p_k-1=0\) then addition is forced. Hence,
\[
\pi_A(M^*)=
\begin{cases}
1/3, & \text{if } p_k-1>0,\\[1mm]
1, & \text{if } p_k-1=0,
\end{cases}
\]
and the probability of selecting the specific predictor to add (from the \(p-p^*\) predictors not in \(M^*\)) is \(1/(p-p^*)\); note that \(p-p^*=p-(p_k-1)=p-p_k+1\). Thus,
\[
q(M_k\mid M^*)=\pi_A(M^*)\cdot\frac{1}{p-p_k+1}.
\]

The correction factor for the deletion move becomes
\[
\frac{q(M_k\mid M^*)}{q(M^*\mid M_k)}=\frac{\pi_A(M^*)/(p-p_k+1)}{\pi_D(M_k)/(p_k)}.
\]

\paragraph{Swap Move.}  
The swap move is only available when \(0<p_k<p\). When selected (with probability \(\pi_S(M_k)=1/3\)), the move involves two steps:
\begin{enumerate}
    \item Delete one predictor from \(M_k\) (chosen uniformly from the \(p_k\) predictors).
    \item Add one predictor from the \(p-p_k\) not in \(M_k\) (chosen uniformly).
\end{enumerate}
Thus, the forward proposal probability is
\[
q(M^*\mid M_k)=\frac{1}{3}\cdot\frac{1}{p_k}\cdot\frac{1}{p-p_k}.
\]
Since the swap move does not change the model size, the reverse move (swapping back) is performed in an analogous manner from \(M^*\) and yields
\[
q(M_k\mid M^*)=\frac{1}{3}\cdot\frac{1}{p_k}\cdot\frac{1}{p-p_k}.
\]
Hence, the correction factor is
\[
\frac{q(M_k\mid M^*)}{q(M^*\mid M_k)}=1.
\]

\section{Results for Simulated Data}
\label{app:SimulatedData}

\begin{table}[ht!]
\renewcommand{\arraystretch}{1} % Default value: 1
\resizebox{\textwidth}{!}{%
  \begin{threeparttable}
  \singlespacing
\caption{Results of Simulation Study (Poisson Log-Normal).}
\label{tab:results_simulations_pln}
\centering
\begin{tabular}{llrrrrrrrrrr}
\toprule
Prior & DGP & $n$ & $p$ & $M$ Size & Frac. True & Brier & FNR & FPR & $\ln(g)$ & $\sigma^2$ & Time\\
\midrule
Hyper-g/n ($a=3$) & ULLGM & 150 & 50 & 14.823 & 0.005 & 0.026 & 0.046 & 0.134 & 3.980 & 0.185 & 222\\
Hyper-g/n ($a=3$) & ULLGM & 150 & 100 & 15.466 & 0.003 & 0.011 & 0.030 & 0.064 & 4.226 & 0.166 & 227\\
Hyper-g/n ($a=3$) & ULLGM & 150 & 250 & 16.023 & 0.004 & 0.003 & 0.041 & 0.027 & 4.477 & 0.168 & 229\\
Hyper-g/n ($a=3$) & ULLGM & 1000 & 50 & 14.891 & 0.011 & 0.021 & 0.018 & 0.128 & 3.893 & 0.193 & 413\\
Hyper-g/n ($a=3$) & ULLGM & 1000 & 100 & 15.048 & 0.011 & 0.007 & 0.012 & 0.058 & 4.087 & 0.194 & 393\\
Hyper-g/n ($a=3$) & ULLGM & 1000 & 250 & 15.349 & 0.008 & 0.002 & 0.007 & 0.024 & 4.210 & 0.197 & 429\\
\addlinespace
Hyper-g/n ($a=3$) & GLM & 150 & 50 & 13.630 & 0.042 & 0.010 & 0.000 & 0.091 & 10.823 & 0.006 & 217\\
Hyper-g/n ($a=3$) & GLM & 150 & 100 & 13.261 & 0.065 & 0.002 & 0.000 & 0.036 & 10.897 & 0.006 & 221\\
Hyper-g/n ($a=3$) & GLM & 150 & 250 & 12.958 & 0.083 & \textbf{0.000} & 0.000 & 0.012 & 11.184 & 0.006 & 227\\
Hyper-g/n ($a=3$) & GLM & 1000 & 50 & 13.588 & 0.030 & 0.011 & 0.000 & 0.090 & 10.558 & 0.003 & 395\\
Hyper-g/n ($a=3$) & GLM & 1000 & 100 & 13.932 & 0.037 & 0.003 & 0.000 & 0.044 & 10.448 & 0.003 & 380\\
Hyper-g/n ($a=3$) & GLM & 1000 & 250 & 14.098 & 0.030 & 0.001 & 0.000 & 0.017 & 10.513 & 0.003 & 372\\
\addlinespace
Hyper-g/n ($a=3$) & Log-Gamma & 150 & 50 & 15.626 & 0.003 & 0.029 & 0.028 & 0.152 & 4.136 & 0.170 & 227\\
Hyper-g/n ($a=3$) & Log-Gamma & 150 & 100 & 15.030 & 0.005 & 0.009 & 0.041 & 0.062 & 4.336 & 0.136 & 226\\
Hyper-g/n ($a=3$) & Log-Gamma & 150 & 250 & 16.528 & 0.002 & 0.005 & 0.021 & 0.029 & 4.744 & 0.133 & 240\\
Hyper-g/n ($a=3$) & Log-Gamma & 1000 & 50 & 14.790 & 0.011 & 0.020 & 0.023 & 0.123 & 4.022 & 0.172 & 390\\
Hyper-g/n ($a=3$) & Log-Gamma & 1000 & 100 & 14.858 & 0.010 & 0.007 & 0.018 & 0.056 & 4.214 & 0.167 & 390\\
Hyper-g/n ($a=3$) & Log-Gamma & 1000 & 250 & 15.519 & 0.007 & 0.002 & 0.006 & 0.024 & 4.317 & 0.172 & 395\\
\addlinespace
Unit Information ($g=n$) & ULLGM & 150 & 50 & 12.926 & 0.019 & 0.019 & 0.030 & 0.081 & 5.011 & 0.154 & \textbf{167}\\
Unit Information ($g=n$) & ULLGM & 150 & 100 & 14.052 & 0.012 & 0.009 & 0.011 & 0.048 & 5.011 & 0.119 & \textbf{190}\\
Unit Information ($g=n$) & ULLGM & 150 & 250 & 15.607 & 0.009 & 0.003 & 0.021 & 0.024 & 5.011 & 0.114 & 225\\
Unit Information ($g=n$) & ULLGM & 1000 & 50 & \textbf{10.129} & 0.080 & 0.010 & 0.078 & 0.023 & 6.908 & 0.198 & 351\\
Unit Information ($g=n$) & ULLGM & 1000 & 100 & 10.403 & 0.131 & 0.004 & 0.069 & 0.011 & 6.908 & 0.197 & 340\\
Unit Information ($g=n$) & ULLGM & 1000 & 250 & 10.718 & \textbf{0.162} & 0.001 & 0.040 & 0.005 & 6.908 & 0.189 & 357\\
\addlinespace
Unit Information ($g=n$) & GLM & 150 & 50 & 13.773 & 0.039 & 0.009 & 0.000 & 0.095 & 5.011 & 0.036 & 214\\
Unit Information ($g=n$) & GLM & 150 & 100 & 13.802 & 0.041 & 0.002 & 0.000 & 0.042 & 5.011 & 0.040 & 242\\
Unit Information ($g=n$) & GLM & 150 & 250 & 13.777 & 0.043 & \textbf{0.000} & 0.000 & 0.016 & 5.011 & 0.045 & 234\\
Unit Information ($g=n$) & GLM & 1000 & 50 & 13.808 & 0.036 & 0.011 & 0.000 & 0.095 & 6.908 & 0.007 & 368\\
Unit Information ($g=n$) & GLM & 1000 & 100 & 14.091 & 0.029 & 0.004 & 0.000 & 0.045 & 6.908 & 0.008 & 366\\
Unit Information ($g=n$) & GLM & 1000 & 250 & 13.872 & 0.039 & 0.001 & 0.000 & 0.016 & 6.908 & 0.009 & 382\\
\addlinespace
Unit Information ($g=n$) & Log-Gamma & 150 & 50 & 13.026 & 0.013 & 0.020 & 0.048 & 0.090 & 5.011 & 0.138 & 212\\
Unit Information ($g=n$) & Log-Gamma & 150 & 100 & 13.948 & 0.010 & 0.008 & 0.023 & 0.047 & 5.011 & 0.120 & 223\\
Unit Information ($g=n$) & Log-Gamma & 150 & 250 & 15.079 & 0.005 & 0.004 & 0.012 & 0.023 & 5.011 & 0.110 & 232\\
Unit Information ($g=n$) & Log-Gamma & 1000 & 50 & \textbf{10.053} & 0.125 & 0.010 & 0.086 & 0.021 & 6.908 & 0.176 & 322\\
Unit Information ($g=n$) & Log-Gamma & 1000 & 100 & 10.851 & 0.158 & 0.003 & 0.025 & 0.013 & 6.908 & 0.165 & 344\\
Unit Information ($g=n$) & Log-Gamma & 1000 & 250 & 11.063 & \textbf{0.191} & 0.001 & 0.017 & 0.005 & 6.908 & 0.164 & 354\\
\bottomrule
\end{tabular}
    \begin{tablenotes}
      \small
      \item \textit{Note:} `DGP' = data generating process; `${M}$ size' = posterior expected model size (DGP has model size 10); `Frac. True' = Fraction of MCMC iterations where true model is visited; `Brier' = Brier score; `FNR' = False negative rate; `FPR' = False positive rate. The column `$\ln g$' reports the log of the posterior mean of $g$ (or $\ln (n)$ for the UIP),  `$\sigma^2$' states the posterior mean of $\sigma^2$, while `Time' is reported in seconds. Results are averages across 50 replications per simulation setting. For metrics that are comparable across settings, the two best results are printed in \textbf{bold}.
    \end{tablenotes}
  \end{threeparttable}%
}
\end{table}

\begin{table}[ht!]
\renewcommand{\arraystretch}{1} % Default value: 1
\resizebox{\textwidth}{!}{%
  \begin{threeparttable}
  \singlespacing
\caption{Results of Simulation Study (Binomial Logistic-Normal).}
\label{tab:results_simulations_bil}
\centering
\begin{tabular}{llrrrrrrrrrr}
\toprule
Prior & DGP & $n$ & $p$ & $M$ Size & Frac. True & Brier & FNR & FPR & $\ln(g)$ & $\sigma^2$ & Time\\
\midrule
Hyper-g/n ($a=3$) & ULLGM & 150 & 50 & 14.665 & 0.005 & 0.025 & 0.057 & 0.127 & 4.147 & 0.164 & 230\\
Hyper-g/n ($a=3$) & ULLGM & 150 & 100 & 15.710 & 0.003 & 0.011 & 0.054 & 0.066 & 4.177 & 0.176 & 236\\
Hyper-g/n ($a=3$) & ULLGM & 150 & 250 & 15.841 & 0.002 & 0.004 & 0.028 & 0.028 & 4.648 & 0.149 & 239\\
Hyper-g/n ($a=3$) & ULLGM & 1000 & 50 & 15.240 & 0.005 & 0.028 & 0.020 & 0.140 & 3.874 & 0.198 & 411\\
Hyper-g/n ($a=3$) & ULLGM & 1000 & 100 & 15.355 & 0.009 & 0.007 & 0.013 & 0.061 & 4.099 & 0.197 & 423\\
Hyper-g/n ($a=3$) & ULLGM & 1000 & 250 & 15.753 & 0.006 & 0.002 & 0.013 & 0.025 & 4.251 & 0.184 & 430\\
\addlinespace
Hyper-g/n ($a=3$) & GLM & 150 & 50 & 13.886 & 0.021 & 0.014 & 0.000 & 0.097 & 10.748 & 0.008 & 294\\
Hyper-g/n ($a=3$) & GLM & 150 & 100 & 13.880 & 0.033 & 0.004 & 0.000 & 0.043 & 10.725 & 0.009 & 227\\
Hyper-g/n ($a=3$) & GLM & 150 & 250 & 13.882 & 0.042 & \textbf{0.001} & 0.000 & 0.016 & 10.691 & 0.009 & 233\\
Hyper-g/n ($a=3$) & GLM & 1000 & 50 & 13.574 & 0.049 & 0.010 & 0.000 & 0.089 & 10.750 & 0.003 & 563\\
Hyper-g/n ($a=3$) & GLM & 1000 & 100 & 13.710 & 0.035 & 0.003 & 0.000 & 0.041 & 10.566 & 0.003 & 432\\
Hyper-g/n ($a=3$) & GLM & 1000 & 250 & 14.373 & 0.026 & \textbf{0.001} & 0.000 & 0.018 & 10.600 & 0.003 & 427\\
\addlinespace
Hyper-g/n ($a=3$) & Log-Gamma & 150 & 50 & 15.335 & 0.003 & 0.031 & 0.033 & 0.148 & 3.967 & 0.206 & 224\\
Hyper-g/n ($a=3$) & Log-Gamma & 150 & 100 & 15.925 & 0.002 & 0.015 & 0.061 & 0.073 & 4.213 & 0.182 & 232\\
Hyper-g/n ($a=3$) & Log-Gamma & 150 & 250 & 16.091 & 0.002 & 0.004 & 0.053 & 0.028 & 4.508 & 0.162 & 238\\
Hyper-g/n ($a=3$) & Log-Gamma & 1000 & 50 & 14.683 & 0.007 & 0.022 & 0.018 & 0.127 & 3.929 & 0.211 & 431\\
Hyper-g/n ($a=3$) & Log-Gamma & 1000 & 100 & 14.705 & 0.008 & 0.008 & 0.027 & 0.056 & 4.047 & 0.209 & 435\\
Hyper-g/n ($a=3$) & Log-Gamma & 1000 & 250 & 15.368 & 0.005 & 0.003 & 0.016 & 0.024 & 4.236 & 0.206 & 432\\
\addlinespace
Unit Information ($g=n$) & ULLGM & 150 & 50 & 13.332 & 0.015 & 0.019 & 0.041 & 0.103 & 5.011 & 0.118 & 224\\
Unit Information ($g=n$) & ULLGM & 150 & 100 & 14.249 & 0.005 & 0.012 & 0.047 & 0.053 & 5.011 & 0.111 & \textbf{216}\\
Unit Information ($g=n$) & ULLGM & 150 & 250 & 15.307 & 0.004 & 0.004 & 0.027 & 0.024 & 5.011 & 0.113 & 228\\
Unit Information ($g=n$) & ULLGM & 1000 & 50 & \textbf{10.170} & 0.087 & 0.010 & 0.080 & 0.023 & 6.908 & 0.201 & 370\\
Unit Information ($g=n$) & ULLGM & 1000 & 100 & 10.533 & 0.113 & 0.004 & 0.054 & 0.012 & 6.908 & 0.188 & 410\\
Unit Information ($g=n$) & ULLGM & 1000 & 250 & 10.886 & \textbf{0.167} & \textbf{0.001} & 0.035 & 0.005 & 6.908 & 0.191 & 420\\
\addlinespace
Unit Information ($g=n$) & GLM & 150 & 50 & 14.333 & 0.022 & 0.013 & 0.003 & 0.109 & 5.011 & 0.041 & 217\\
Unit Information ($g=n$) & GLM & 150 & 100 & 14.505 & 0.019 & 0.004 & {0.001} & 0.050 & 5.011 & 0.044 & 221\\
Unit Information ($g=n$) & GLM & 150 & 250 & 14.520 & 0.023 & \textbf{0.001} & 0.000 & 0.019 & 5.011 & 0.051 & 227\\
Unit Information ($g=n$) & GLM & 1000 & 50 & 13.560 & 0.037 & 0.011 & 0.000 & 0.089 & 6.908 & 0.007 & 429\\
Unit Information ($g=n$) & GLM & 1000 & 100 & 14.465 & 0.028 & 0.005 & 0.000 & 0.050 & 6.908 & 0.009 & 427\\
Unit Information ($g=n$) & GLM & 1000 & 250 & 14.663 & 0.022 & \textbf{0.001} & 0.000 & 0.019 & 6.908 & 0.010 & 430\\
\addlinespace
Unit Information ($g=n$) & Log-Gamma & 150 & 50 & 12.653 & 0.012 & 0.020 & 0.068 & 0.079 & 5.011 & 0.146 & \textbf{215}\\
Unit Information ($g=n$) & Log-Gamma & 150 & 100 & 13.612 & 0.004 & 0.012 & 0.069 & 0.053 & 5.011 & 0.139 & 224\\
Unit Information ($g=n$) & Log-Gamma & 150 & 250 & 15.672 & 0.003 & 0.005 & 0.031 & 0.025 & 5.011 & 0.119 & 231\\
Unit Information ($g=n$) & Log-Gamma & 1000 & 50 & \textbf{10.050} & 0.145 & 0.008 & 0.079 & 0.021 & 6.908 & 0.212 & 399\\
Unit Information ($g=n$) & Log-Gamma & 1000 & 100 & 10.360 & 0.134 & 0.004 & 0.067 & 0.011 & 6.908 & 0.214 & 373\\
Unit Information ($g=n$) & Log-Gamma & 1000 & 250 & 10.868 & \textbf{0.202} & \textbf{0.000} & 0.020 & 0.005 & 6.908 & 0.205 & 410\\
\bottomrule
\end{tabular}
    \begin{tablenotes}
      \small
      \item \textit{Note:} `DGP' = data generating process; `$M$ size' = posterior expected model size (DGP has model size 10); `Frac. True' = Fraction of MCMC iterations where true model is visited; `Brier' = Brier score; `FNR' = False negative rate; `FPR' = False positive rate. The column `$\ln g$' reports the log of the posterior mean of $g$ (or $\ln (n)$ for the UIP),  `$\sigma^2$' states the posterior mean of $\sigma^2$, while `Time' is reported in seconds. Results are averages across 50 replications per simulation setting. For metrics that are comparable across settings, the two best results are printed in \textbf{bold}.
    \end{tablenotes}
  \end{threeparttable}%
}
\end{table}

Tables \ref{tab:results_simulations_pln} and \ref{tab:results_simulations_bil} present key statistics derived from the simulation outcomes for both the PLN and BiL models, averaged across the replications. Posterior model size as well as the proportion of visits to the true model among all MCMC iterations are reported. In addition, we provide measures of the quality of the variable selection results. We consider the Brier score, which is a strictly proper scoring rule that corrects for the number of available covariates $p$. The Brier score is defined as $\frac{1}{p} \sum_{j=1}^p(\text{PIP}_j-a_j)^2$. Here, $\text{PIP}_j$ is the posterior inclusion probability of covariate $j$ and $a_j = 0$ if covariate $j$ is truly excluded while $a_j=1$ otherwise. The closer the Brier score is to zero, the more accurate the variable selection results are. In addition, the tables present the average fractions of false positives and false negatives across all MCMC samples. We provide the fixed value or estimated posterior mean of $g$ on the log scale, as well as the posterior mean of $\sigma^2$. Finally, the tables report the time in seconds to run a chain of 550,000 MCMC iterations
on a single core of an AMD Ryzen 5 5500U processor.

\section{Bilateral Migration Flows Between OECD Countries}
\label{sec:migration_application}

Quantitative models of human migration advance our understanding of migration behavior, can be used to improve existing migration estimates and to inform policy. Dyadic regression models are commonly used to analyze the spatial allocation of migrants, modeling migration flows based on the characteristics of the origin, destination, and the relationship between country pairs. Such models are extensively applied not just in migration studies but also to understand trade flows (\citealp{carrere2006revisiting}) or tourism patterns (\citealp{morley2014gravity}). The preferred frequentist estimation method is the Pseudo Poisson maximum likelihood approach (\citealp{silva2006log}), which simultaneously accounts for the count nature of the outcome data and potential %ly present
overdispersion. Recent studies highlight a growing interest in applying probabilistic modeling to migration data (\citealp{bijak2010forecasting}; \citealp{welch2022probabilistic}), but the issue of model uncertainty has received limited attention in this field. \citet{mitchell2011drivers} explore migration to the UK using model averaging techniques, albeit within a Gaussian regression framework.

\begin{figure}[!h]
    \centering
    \includegraphics[width=\textwidth]{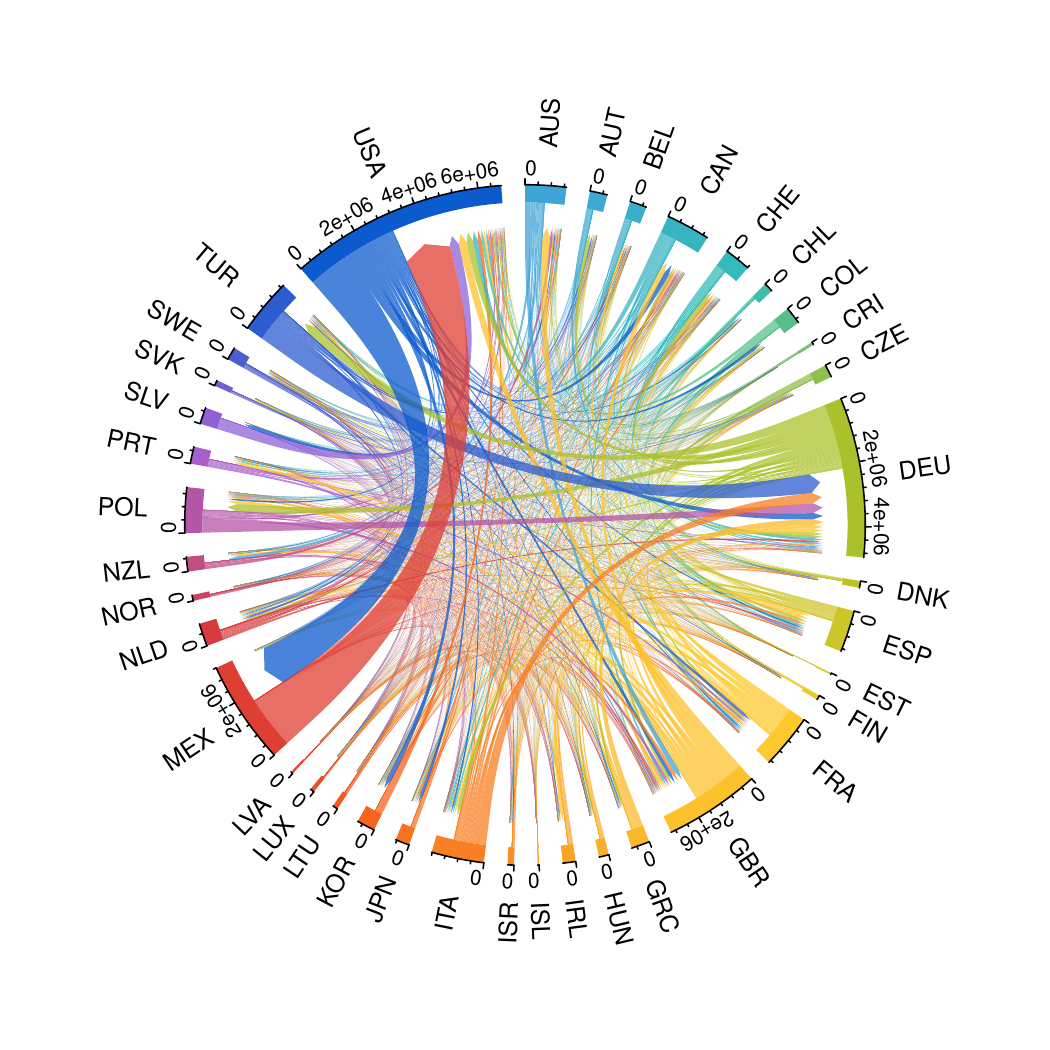}
    \caption{Bilateral Migration Flows between OECD countries. Arrow size is proportional to size of migration flow.}
    \label{fig:oecd_flows_circular}
\end{figure}

In this context, we use the PLN model to examine international migration flows among the 38 OECD countries from 2015 to 2020. These flows are estimates of \citet{abel2019bilateral} following the methodology described in \citet{azose2019estimation}, based on migrant stock data compiled by the United Nations. This challenging dataset comprises $n=38^2-38=1,406$ bilateral migration flows $y_i$, ranging from zero to over 1.6 million migrants (between Mexico and the United States). Dispersion in the data is very high, with the usual dispersion index over 345,000. The flows are depicted in the form of a circular plot in Figure~\ref{fig:oecd_flows_circular}. An initial analysis of the data highlights distinctive features of bilateral migration in this timeframe, such as the prominent migration corridor between Mexico and the United States and the central roles of Germany and the UK as migration hubs in Europe. 

We have compiled a set of $p=54$ variables that hold potential predictive power for bilateral migration flows. This dataset encompasses a variety of country-specific factors for both origin and destination countries, including demographic measures like population size, population age distribution, and educational attainment rates, alongside economic indicators such as GDP per capita and employment rates. It also includes measures of social infrastructure, such as healthcare expenditure as a percentage of GDP, and indices of social and political stability, like the number of battle-related deaths, homicide rates, or the Gini coefficient measuring income inequality. These covariates collectively address a broad spectrum of theories that seek to explain international migration patterns, often highlighting the significance of labor market demands in destination countries, and the availability of knowledge, financial, and social capital in origin countries (\citealp{de2019age}). The dataset also features country-pair variables, like the distance between capitals, to acknowledge the tendency for increased migratory activity between geographically proximate countries. Furthermore, existing bilateral migration stocks and indicator variables for historical colonial ties or a common official language are included to capture the influence of non-geographical distance proxies such as existing migrant networks or cultural similarity on migration dynamics. More details, including summary statistics and the complete list of covariates are provided in Table~\ref{tab:summary_mig}.

\begin{figure}
\centering
\begin{subfigure}{0.49\textwidth}
    \includegraphics[width=\textwidth]{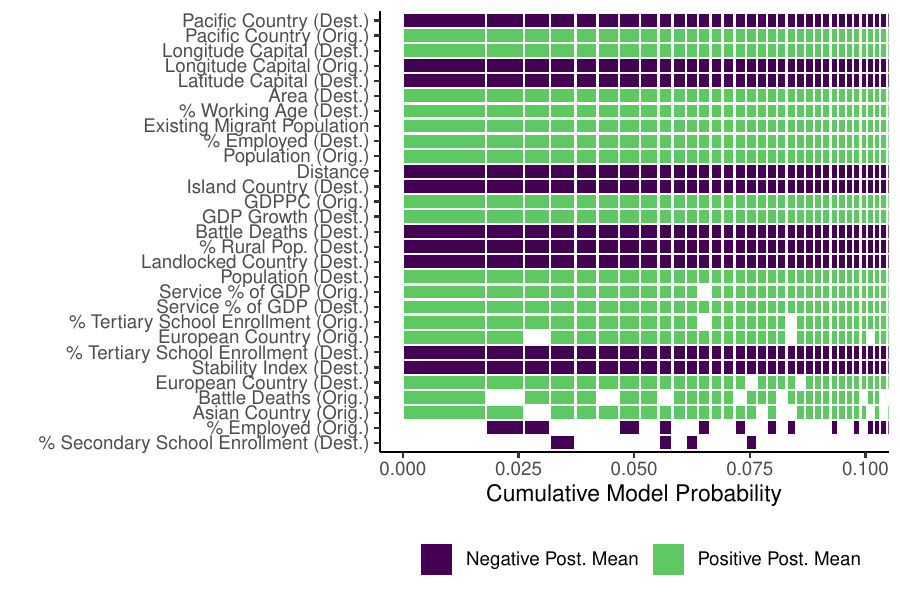}
    \caption{Highest Probability Models.}
    %\label{fig:highest_mig}
\end{subfigure}
\hfill
\begin{subfigure}{0.49\textwidth}
    \includegraphics[width=\textwidth]{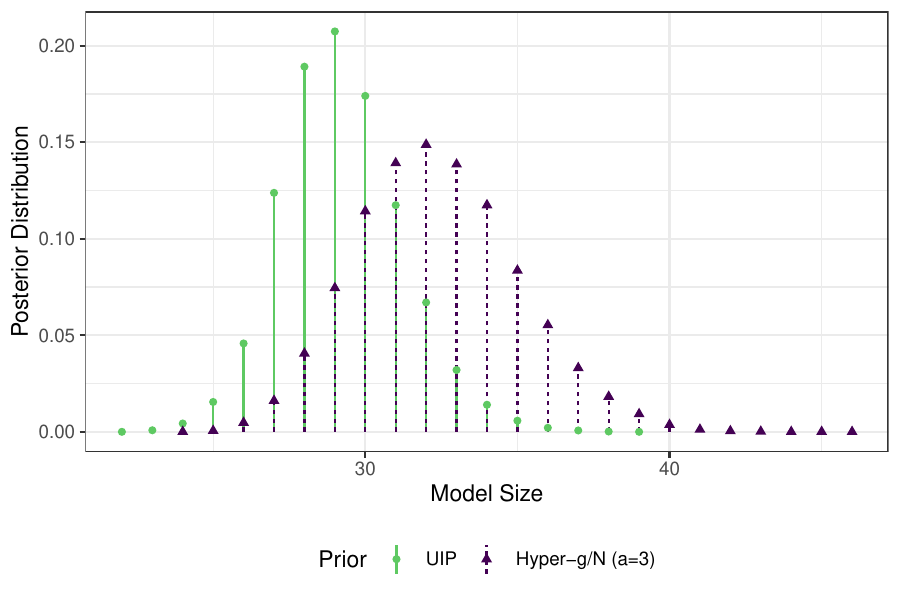}
    \caption{Posterior Distribution of Model Size.}
   % \label{fig:size_mig}
\end{subfigure}
        
\caption{Estimation Results (Bilateral Migration Flow Data). Highest probability models plot includes variables with estimated $\text{PIP}>0.3$ under unit information prior.}
\label{fig:results_migration}
\end{figure}

\begin{figure}[h!]
    \centering
    \begin{subfigure}[b]{0.49\textwidth}
        \includegraphics[width=\textwidth]{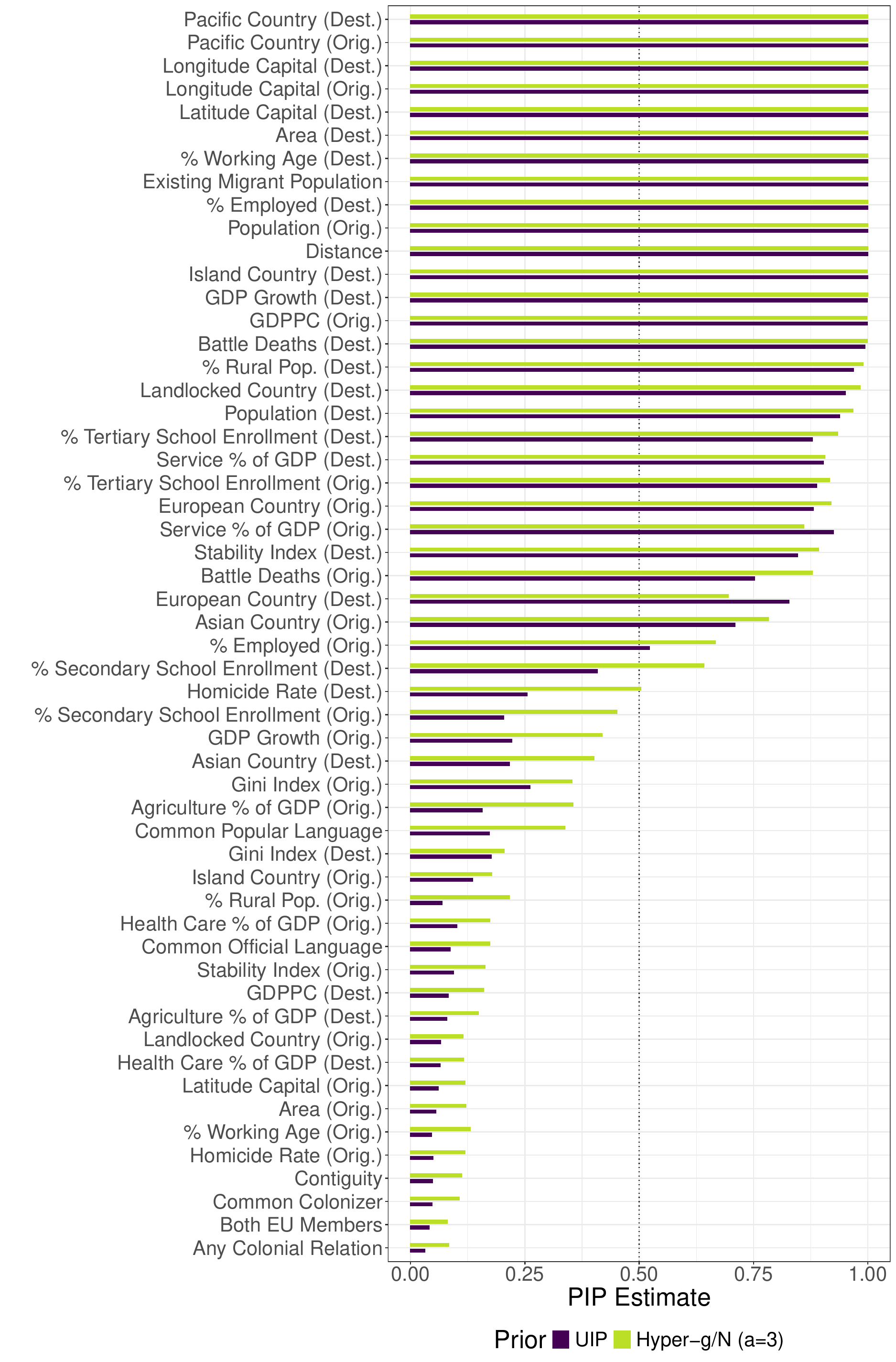}
        \caption{Post. Inclusion Probabilities.}
        \label{fig:image1}
    \end{subfigure}
    \hfill % add some horizontal spacing
    \begin{subfigure}[b]{0.49\textwidth}
        \includegraphics[width=\textwidth]{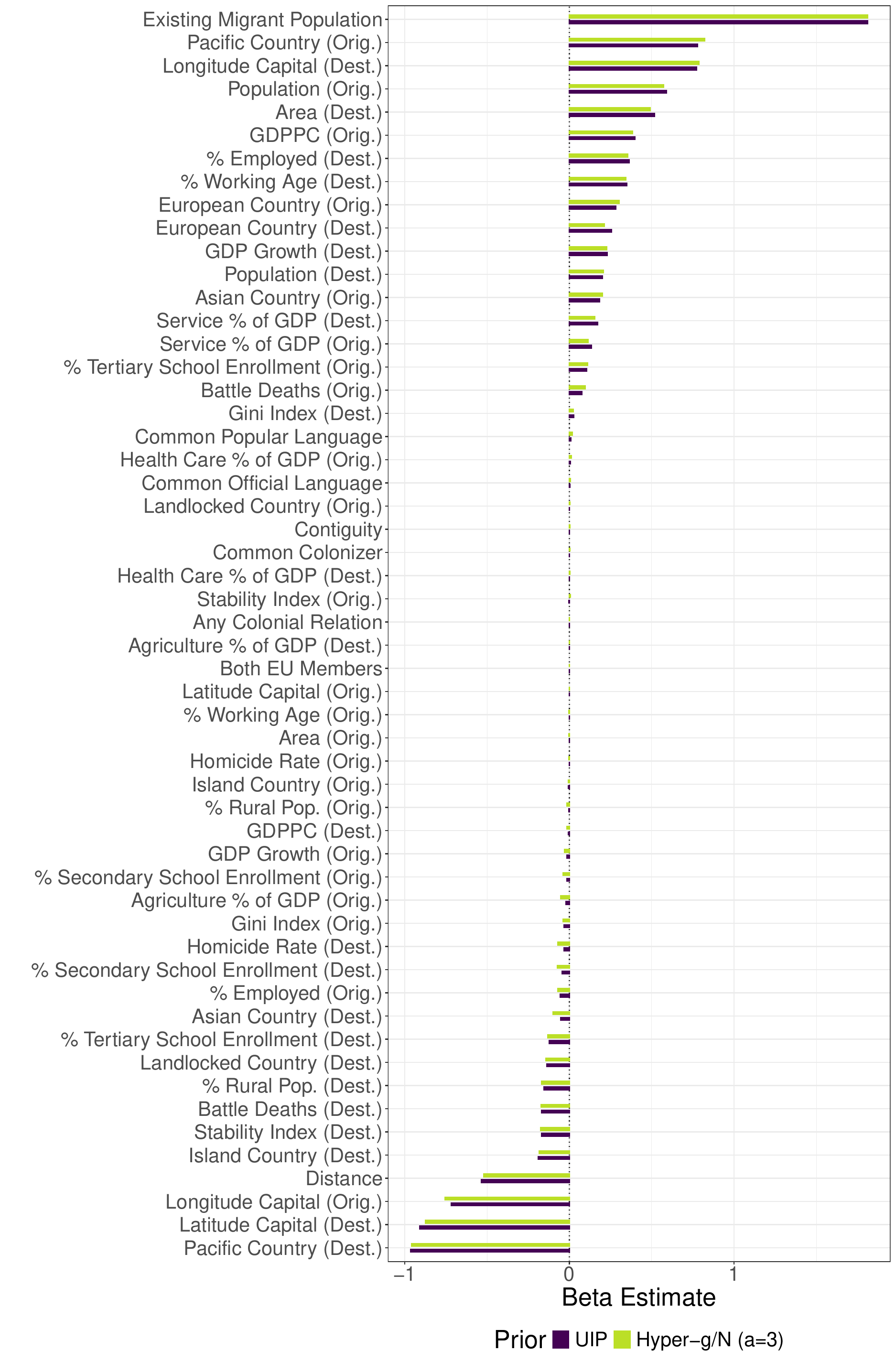}
        \caption{Coefficient Post. Means.}
        \label{fig:image2}
    \end{subfigure}
    \caption{Estimation Results for Bilateral Migration Flow Data.}
    \label{fig:mig_res_beta_pip}
\end{figure}

Based on this data, we conduct a BMA exercise using a Poisson Log-Normal specification. We compare a UIP prior ($g=n$) and a hyper-$g/n$ prior ($a=3$), alongside an agnostic prior on model space ($m=p/2$). The analysis is based on $300,000$ posterior samples after a burn-in period of $250,000$ iterations. Under both UIP and hyper-$g/n$ priors, posterior simulation takes between 26 and 32 minutes on a single core of an AMD Ryzen 5 5500U. The estimated posterior inclusion probabilities and the posterior means of $\bm\beta$ are presented in Figure~\ref{fig:mig_res_beta_pip}. Under both priors, the posterior mean estimates for the intercept and the overdispersion parameter are $\mathbb{E}({\alpha}|\bm{y}) = 6.9$ and $\mathbb{E}({\sigma}^2|\bm{y}) = 0.7$, the latter indicating the presence of significant overdispersion in the model. The highest probability models under the unit information prior are shown in Figure~\ref{fig:results_migration}(a). The posterior model size distribution is depicted in Figure~\ref{fig:results_migration}(b), where the hyper-$g/n$ prior tends to support slightly larger models, with a posterior mean model size of $32.8$, compared to $29.5$ under the unit information prior.

The median probability models under the two priors (including those variables with a PIP estimate greater than 0.5) agree on a set of 29 covariates. The selected variables align with theoretical expectations about migration determinants, highlighting factors like distance (which is negatively correlated to migration flows), or the presence of existing bilateral migrant stocks (positive effect). Combined with the positive effects of population sizes, these findings underscore the roles of both mechanical factors and social networks in predicting migration flows. Additionally, the positive correlation between employment rates in destination countries and migration flows emphasizes the significance of labor markets and economic opportunity for international migration. Positive coefficients of indicator variables for Pacific countries reflect the high migration rates observed between Australia and New Zealand. Posterior model probabilities are relatively spread out, as detailed in Table~\ref{tab:topmodels_oecd_uip} and Table~\ref{tab:topmodels_oecd_hypergn}. Comprehensive details of the posterior means, standard deviations, and inclusion probabilities under both priors are available in Table~\ref{tab:results_oecd}.  

\section{Evaluation of  Predictive Mass Function for PLN and BiL Models}
\label{app:integrals}

To evaluate the quality of predictions in Section~\ref{sec:predictive}, we use LPS which require evaluations of the predictive pmf $P(y^{\text{p}}_i ~|~ 
\bm{x}^{\text{p}}_i, 
\bm{y}^{\text{t}})$ in (\ref{eq:LPS}). In general, for the ULLGMs defined via (\ref{eq:Gen_y}) and  (\ref{eq:model_k_z}), the likelihood contribution of a single data point $P(y_i|\bm{x}_i, \alpha, \bm{\beta}, \sigma^2, M_k)$ is given by 
\begin{equation}
\label{LikGen}      p(y_i|\bm{x}_i,  \bm{\theta},  M_k) = \int_{-\infty}^{\infty}F(y_i | h(z_i))~\mathcal{N}(z_i | \alpha + \bm{x}'_{i,k}\bm{\beta}_k, \sigma^2) dz_i,
\end{equation}
where $\bm{x}_{i,k}'$ is the $i$th row of $\bm{X}_k$ and $\bm{\theta}=(\alpha, \bm{\beta}, \sigma^2)\in \Theta$. 
Under the PLN model, this becomes
\begin{equation}\label{LikPLN}
\begin{split}  P(y_i | &\bm{x}_i,\bm{\theta}, M_k) =\\
    &(2\pi\sigma^2)^{-\frac{1}{2}}\int_0^\infty \exp(-\lambda_i)\frac{\lambda_i^{y_i-1}}{y_i!} \exp \left( -\frac{1}{2\sigma^2}(\log \lambda_i - \alpha - \bm{x}'_{i,k}\bm{\beta}_k)^2  \right) d\lambda_i.
   \end{split}
\end{equation}

Similarly, for the BiL model, we have
\begin{equation}\label{LikBiL}
\begin{split}
    P(y_i | &\bm{x}_i,\bm{\theta},  N_i, M_k) =\\ &(2\pi\sigma^2)^{-\frac{1}{2}} {N_i \choose y_i} \int_0^1 \pi_i^{y_i-1} (1-\pi_i)^{(N_i-y_i - 1)}  \exp \left( -\frac{1}{2\sigma^2}(\text{logit}(\pi_i) - \alpha - \bm{x}'_{i,k}\bm{\beta}_k)^2  \right) d\pi_i,
   \end{split}
\end{equation}
where $\text{logit}(x) = \log(x) - \log(1-x)$. Various ways of evaluating these integrals are available. In principle, both (\ref{LikPLN}) and (\ref{LikBiL}) can directly be evaluated numerically using quadrature rules. However, achieving sufficient numerical stability can be an issue, especially for $\sigma^2$ small or large outcomes $y_i$ or $N_i$. In both cases, the posterior density of $z_i$ increasingly behaves like a point mass, as indicated below. Monte Carlo approximation can be used, but typically requires a large number of samples of $z_i$ for a single likelihood evaluation. A more computationally efficient and accurate approximation to the integral representations is to consider the following definite integral approximation to the indefinite integral in (\ref{LikGen}):
\begin{equation}\label{eq:approxLik}
    P(y_i|\bm{x}_i,  \bm{\theta}, M_k) \approx \int_{C_0}^{C_1}F(y_i | h(z_i))~\mathcal{N}(z_i |  \alpha + \bm{x}'_{i,k}\bm{\beta}_k, \sigma^2) dz_i.
\end{equation}

Ideally, $C_0$ and $C_1$ are chosen in a way that adequately reflects the location of the posterior mass of $z_i$. We therefore suggest to choose $C_0$ and $C_1$ based on approximate posterior moments of $z_i$. A Gaussian approximation to the PLN regression model is given by $\log(y_i) = z_i + u_i$ where $u_i \sim \mathcal{N}(0, y_i^{-1})$ (compare e.g.~\citet{chan2009bayesian}). Combining this approximate model with the prior $z_i \sim \mathcal{N}(\alpha + \bm{x}'_i\bm{\beta}, \sigma^2)$, while dropping the model index $k$ for simplicity, the posterior of $z_i$ is Gaussian with variance $s = (y_i + \sigma^{-2})^{-1}$ and mean $m = s(\log(y_i)y_i + (\alpha + \bm{x}'_i\bm{\beta})\sigma^{-2})$. 

For the BiL model, a similar approximation can be derived by starting from the Gaussian approximation to the Binomial $y_i \sim \mathcal{N}(N_ip_i, N_ip_i(1-p_i))$ with $p_i = \exp(z_i)/(1+\exp(z_i))$. This implies that $y_i/N_i \sim \mathcal{N}(p_i, p_i(1-p_i)/N_i)$. Applying the Delta method to approximate the distribution of the logit-transformed success fraction $y_i/N_i$ then results in $\text{logit}(y_i / N_i) \sim \mathcal{N}(z_i, [N_ip_i(1-p_i)]^{-1})$. For the variance term, a plug-in estimator of $p_i$ is $\hat{p}_i = y_i/N_i$. Combining this approximate model with the prior $z_i \sim \mathcal{N}(\alpha + \bm{x}'_i\bm{\beta}, \sigma^2)$, the approximate posterior of $z_i$ in the BiL model is Gaussian with variance $s = ([\hat{p}_i(1-\hat{p}_i)N_i] + \sigma^{-2})^{-1}$ and mean $m = s([\text{logit}(\hat{p}_i){N_i\hat{p}_i(1-\hat{p}_i)}] + (\alpha + \bm{x}'_i\bm{\beta})\sigma^{-2})$. In cases where $y_i=0$ or $y_i = N_i$, it is necessary to introduce numerical offsets to compute these approximate moments for the BiL model. The same holds when $y_i=0$ for the PLN model (these are exactly the observations that do not contribute to posterior existence in these models, see sections \ref{sec:exist} and \ref{App:proof}). 

We found that choosing $C_0 = m - 6s$ and $C_1 = m + 6s$ based on these approximate posterior densities is an excellent trade-off in terms of computational efficiency and numerical precision, even in the presence of extremely large counts, where posterior densities of $z_i$ behave like a point mass (note how the approximate posterior variances go to zero as $N_i \rightarrow \infty$ for the BiL model and $y_i \rightarrow \infty$ for the PLN model; the same holds when $\sigma^2$ is very small).

In order to approximate the predictive mass functions \begin{equation}
P(y^{\text{p}}_i ~|~ 
\bm{x}^{\text{p}}_i, 
\bm{y}^{\text{t}})
=\sum_{k=1}^K\int_\Theta P(y^{\text{p}}_i|\bm{x}^{\text{p}}_i,  \bm{\theta}, M_k) p(\bm{\theta}| M_k, \bm{y}^\text{t}) P(M_k| \bm{y}^\text{t}) d \bm{\theta},
\end{equation}
required for computing LPS as in (\ref{eq:LPS}), we will take an average based on MCMC posterior draws of $\bm{\theta}, M_k|\bm{y}^\text{t}$. In addition, we will replace $P(y^{\text{p}}_i|\bm{x}^{\text{p}}_i,  \bm{\theta}, M_k)$ by its approximation in (\ref{eq:approxLik}).  

\section{Additional Materials and Results for the Real Data Applications}

This section provides additional details on the real data applications in the form of tables and visualisations of both results and raw data. For both vaccination and migration data, the presented estimates of $\bm{\beta}$ are posterior means (and standard deviations) of the posterior density of $\bm{\beta}$, marginalized over the inclusion indicators (i.e., a Monte Carlo estimate including the MCMC draws where a given coefficient is exactly zero). In both applications, all covariates are standardized before estimation.  For the measles vaccination data, Table~\ref{tab:summary_vacc} provides summary statistics of the included variables. WAZ, HAZ, WHZ are abbreviations for weight-for-age, height-for-age and weight-for-height z-scores, respectively. These are anthropometric indicators based on children's measurements, evaluated relative to a reference distribution, that provide insight into various chronic and acute forms of malnutrition. FP stands for family planning. All three source files for the variables (the DHS raw files, the DHS GPS files and the DHS geospatial covariate files) are available on the DHS programme website after registration. 

For the migration data, summary statistics can be found in Table~\ref{tab:summary_mig}. GDPPC stands for Gross Domestic Product per capita, a commonly used measure of average income in an economy, which is used as proxy for well-being and living standards. GDP stands for Gross Domestic Product. EU stands for European Union. CEPII Gravity DB refers to the publicly available gravity database maintained by the \textit{Centre d'Etudes Prospectives et d'Information Internationales}. UNDESA PD is the Population Division of the \textit{United Nations Department of Economic and Social Affairs}. GDPPC, population and distance between capitals enter the model after a logarithmic transformation. Gross enrolment ratios are defined as total enrolment in a given level of education divided by the population in a given age group. These variables may exceed 100\% if the total number of students enrolled in a given level of education exceeds the official population in the corresponding age group. This can be due to late enrolments, early enrolments and early leaving. 

\newpage

\subsection{Summary Statistics and Tabulated Results}

\begin{table}[!htbp] \centering \renewcommand*{\arraystretch}{1.1}\caption{Summary Statistics Measles Vaccination Data}
\label{tab:summary_vacc}
\resizebox{!}{0.41\textheight}{
\begin{tabular}{lrrrrrrrr}
\hline
Variable & N & Mean & Std. Dev. & Min & Q1 & Q3 & Max & Data Source \\ 
\hline
Vacc. Children & 305 & 4.203 & 2.785 & 0 & 2 & 6 & 13 & DHS Survey Files\\
Total Children & 305 & 10.413 & 4.829 & 1 & 7 & 14 & 24 & DHS Survey Files\\
\% Vaccinated & 305 & 0.448 & 0.263 & 0 & 0.25 & 0.636 & 1 & DHS Survey Files\\
Avg. WAZ & 305 & -118.71 & 55.019 & -291.053 & -154.577 & -82.654 & 34.333 & DHS Survey Files\\
St. Dev. WAZ & 305 & 109.519 & 28.297 & 23.027 & 92.954 & 127.22 & 190.747 & DHS Survey Files\\
Avg. WHZ & 305 & -56.78 & 46.904 & -210.231 & -88.667 & -24.778 & 82.833 & DHS Survey Files\\
St. Dev. WHZ & 305 & 92.315 & 24.21 & 12.028 & 79.778 & 105.051 & 185.649 & DHS Survey Files\\
Avg. HAZ & 305 & -118.458 & 60.485 & -315.526 & -158.5 & -76.967 & 81.727 & DHS Survey Files\\
St. Dev HAZ & 305 & 127.365 & 40.013 & 12.021 & 100.672 & 148.787 & 253.53 & DHS Survey Files\\
\% Knows Modern FP & 305 & 0.937 & 0.13 & 0.304 & 0.935 & 1 & 1 & DHS Survey Files\\
\% Illiterate & 305 & 0.566 & 0.312 & 0 & 0.375 & 0.857 & 1 & DHS Survey Files\\
\% Primary Educ. & 305 & 0.342 & 0.215 & 0 & 0.188 & 0.5 & 1 & DHS Survey Files\\
\% Secondary Educ. & 305 & 0.109 & 0.149 & 0 & 0 & 0.167 & 0.8 & DHS Survey Files\\
\% Tertiary Educ. & 305 & 0.071 & 0.14 & 0 & 0 & 0.077 & 0.818 & DHS Survey Files\\
\% Health Insured & 305 & 0.202 & 0.261 & 0 & 0 & 0.323 & 1 & DHS Survey Files\\
Avg. Wealth Index & 305 & 0.001 & 0.965 & -1.247 & -0.637 & 0.499 & 2.62 & DHS Survey Files\\
St. Dev. Wealth Index & 305 & 0.284 & 0.187 & 0.005 & 0.151 & 0.386 & 1.172 & DHS Survey Files\\
\% Owns Livestock & 305 & 0.633 & 0.371 & 0 & 0.25 & 0.952 & 1 & DHS Survey Files\\
\% Owns Agric. Land & 305 & 0.497 & 0.374 & 0 & 0.081 & 0.857 & 1 & DHS Survey Files\\
\% Owns Radio & 305 & 0.289 & 0.242 & 0 & 0.086 & 0.455 & 1 & DHS Survey Files\\
\% Owns Car & 305 & 0.033 & 0.109 & 0 & 0 & 0 & 1 & DHS Survey Files\\
\% Owns Phone & 305 & 0.027 & 0.08 & 0 & 0 & 0 & 0.5 & DHS Survey Files\\
\% Has Electricity & 305 & 0.377 & 0.442 & 0 & 0 & 0.935 & 1 & DHS Survey Files\\
Time to Water Source & 305 & 33.618 & 42.645 & 0 & 8.833 & 42.5 & 332.308 & DHS Survey Files\\
\% Piped Water & 305 & 0.499 & 0.385 & 0 & 0.067 & 0.889 & 1 & DHS Survey Files\\
\% Well Water & 305 & 0.138 & 0.235 & 0 & 0 & 0.174 & 1 & DHS Survey Files\\
\% Surface Water & 305 & 0.308 & 0.364 & 0 & 0 & 0.6 & 1 & DHS Survey Files\\
\% Flush Toilet & 305 & 0.073 & 0.17 & 0 & 0 & 0.056 & 1 & DHS Survey Files\\
\% Non-Flush Toilet & 305 & 0.006 & 0.037 & 0 & 0 & 0 & 0.412 & DHS Survey Files\\
\% Pit Toilet & 305 & 0.582 & 0.347 & 0 & 0.294 & 0.9 & 1 & DHS Survey Files\\
\% Petrol as Fuel & 305 & 0.011 & 0.048 & 0 & 0 & 0 & 0.4 & DHS Survey Files\\
\% Coal as Fuel & 305 & 0.848 & 0.271 & 0 & 0.826 & 1 & 1 & DHS Survey Files\\
\% Dung/Crops as Fuel & 305 & 0.022 & 0.073 & 0 & 0 & 0 & 0.5 & DHS Survey Files\\
\% Rudimentary Walls & 305 & 0.632 & 0.37 & 0 & 0.268 & 1 & 1 & DHS Survey Files\\
\% Finished Walls & 305 & 0.2 & 0.322 & 0 & 0 & 0.281 & 1 & DHS Survey Files\\
Avg. No. Bedrooms & 305 & 1.4 & 0.408 & 1 & 1.103 & 1.583 & 3.6 & DHS Survey Files\\
\% Female HH Heads & 305 & 0.199 & 0.213 & 0 & 0.045 & 0.286 & 1 & DHS Survey Files\\
Avg. No. Births per Woman & 305 & 3.706 & 1.236 & 1.333 & 2.714 & 4.591 & 7.231 & DHS Survey Files\\
\% Married Women & 305 & 0.93 & 0.094 & 0.429 & 0.889 & 1 & 1 & DHS Survey Files\\
\% Non-Residents & 305 & 0.011 & 0.031 & 0 & 0 & 0 & 0.222 & DHS Survey Files\\
Urban Cluster & 305 & 0.305 & 0.461 & 0 & 0 & 1 & 1 & DHS Survey Files\\
\% Female Children & 305 & 0.485 & 0.128 & 0.167 & 0.4 & 0.559 & 0.889 & DHS Survey Files\\
Avg. Women's Age & 305 & 28.747 & 2.45 & 20.571 & 27.091 & 30.5 & 36.5 & DHS Survey Files\\
St. Dev. Women's Age & 305 & 5.997 & 1.52 & 1.414 & 4.953 & 7.047 & 12.021 & DHS Survey Files\\
Avg. Children's Age & 305 & 28.659 & 4.293 & 4.333 & 26.588 & 31.152 & 41.75 & DHS Survey Files\\
St. Dev. Children's Age & 305 & 17.074 & 2.74 & 1.414 & 15.959 & 18.462 & 26.41 & DHS Survey Files\\
Under-5 Mortality Rate & 305 & 0.058 & 0.074 & 0 & 0 & 0.091 & 0.5 & DHS Survey Files\\
Latitude & 305 & 9.488 & 2.175 & 4.028 & 8.055 & 10.696 & 14.379 & DHS GPS Data\\
Longitude & 305 & 38.909 & 2.643 & 33.198 & 37.116 & 41.245 & 46.953 & DHS GPS Data\\
Altitude & 305 & 1568.696 & 655.751 & 230.69 & 1121.18 & 2008.7 & 3154.79 & DHS GPS Data\\
Pop. Density & 305 & 1198.24 & 3870.277 & 2.066 & 55.042 & 545.667 & 30101.07 & DHS Geospatial Covariates\\
Time to Urban Center & 305 & 95.444 & 107.793 & 0 & 10.771 & 134.744 & 605.559 & DHS Geospatial Covariates\\
Avg. Temperature & 305 & 21.464 & 3.491 & 14.547 & 18.715 & 23.797 & 29.155 & DHS Geospatial Covariates\\
Avg. Precipitation & 305 & 89.746 & 32.799 & 16.395 & 61.768 & 116.522 & 155.29 & DHS Geospatial Covariates\\
Malaria Prevalence & 305 & 0.035 & 0.052 & 0 & 0.004 & 0.049 & 0.371 & DHS Geospatial Covariates\\
Nightlight Intensity & 305 & 1.214 & 3.41 & 0 & 0 & 0.185 & 21.463 & DHS Geospatial Covariates\\
Region: Affar & 305 & 0.082 & 0.275 & 0 & 0 & 0 & 1 & DHS Survey Files\\
Region: Amhara & 305 & 0.115 & 0.319 & 0 & 0 & 0 & 1 & DHS Survey Files\\
Region: Oromiya & 305 & 0.115 & 0.319 & 0 & 0 & 0 & 1 & DHS Survey Files\\
Region: Somali & 305 & 0.082 & 0.275 & 0 & 0 & 0 & 1 & DHS Survey Files\\
Region: Benishangul-Gumuz & 305 & 0.082 & 0.275 & 0 & 0 & 0 & 1 & DHS Survey Files\\
Region: SNNP & 305 & 0.115 & 0.319 & 0 & 0 & 0 & 1 & DHS Survey Files\\
Region: Gambela & 305 & 0.082 & 0.275 & 0 & 0 & 0 & 1 & DHS Survey Files\\
Region: Harari & 305 & 0.082 & 0.275 & 0 & 0 & 0 & 1 & DHS Survey Files\\
Region: Addis Ababa & 305 & 0.082 & 0.275 & 0 & 0 & 0 & 1 & DHS Survey Files\\
Region: Dire Dawa & 305 & 0.082 & 0.275 & 0 & 0 & 0 & 1 & DHS Survey Files\\
\hline
\end{tabular}
}
\end{table}

\begin{table}[t]
\caption{Estimation Results (Measles Vaccination).}
\label{tab:results_vacc}
\centering
\resizebox{!}{0.45\textheight}{\begin{tabular}{@{\extracolsep{4pt}}lrrrrrr}
\toprule
&\multicolumn{3}{c}{Unit Information Prior}&\multicolumn{3}{c}{Hyper-$g/n$ Prior}\\\cline{2-4}\cline{5-7}\addlinespace
Variable & Post. Mean & Post. SD & PIP & Post. Mean & Post. SD & PIP \\
\midrule
Avg. Children's Age & 0.304 & 0.062 & 1.000 & 0.300 & 0.063 & 1.000\\
St. Dev. Children's Age & -0.339 & 0.061 & 1.000 & -0.336 & 0.063 & 1.000\\
Latitude & 0.374 & 0.071 & 1.000 & 0.369 & 0.077 & 1.000\\
Region: Affar & -0.355 & 0.072 & 1.000 & -0.349 & 0.077 & 0.999\\
\% Pit Toilet & 0.145 & 0.111 & 0.706 & 0.146 & 0.107 & 0.742\\
\% Female HH Heads & 0.095 & 0.095 & 0.570 & 0.101 & 0.096 & 0.616\\
Region: Harari & -0.099 & 0.106 & 0.536 & -0.100 & 0.103 & 0.573\\
Longitude & -0.134 & 0.156 & 0.515 & -0.131 & 0.157 & 0.523\\
\% Has Electricity & 0.119 & 0.141 & 0.493 & 0.099 & 0.129 & 0.458\\
Region: Dire Dawa & 0.090 & 0.104 & 0.486 & 0.086 & 0.101 & 0.498\\
\% Knows Modern FP & 0.080 & 0.099 & 0.468 & 0.092 & 0.101 & 0.545\\
\% Dung/Crops as Fuel & 0.072 & 0.089 & 0.455 & 0.077 & 0.087 & 0.518\\
\% Owns Radio & 0.051 & 0.082 & 0.338 & 0.068 & 0.088 & 0.458\\
Region: Addis Ababa & 0.060 & 0.100 & 0.318 & 0.058 & 0.096 & 0.348\\
Avg. No. Births per Woman & -0.049 & 0.083 & 0.317 & -0.050 & 0.082 & 0.345\\
\% Coal as Fuel & -0.069 & 0.119 & 0.314 & -0.071 & 0.120 & 0.353\\
Time to Urban Center & -0.043 & 0.074 & 0.308 & -0.046 & 0.074 & 0.357\\
Avg. Precipitation & -0.051 & 0.116 & 0.220 & -0.052 & 0.115 & 0.263\\
\% Flush Toilet & 0.029 & 0.066 & 0.214 & 0.029 & 0.066 & 0.232\\
Region: Somali & -0.031 & 0.078 & 0.199 & -0.030 & 0.079 & 0.233\\
Avg. Wealth Index & 0.049 & 0.146 & 0.188 & 0.020 & 0.137 & 0.187\\
Region: Oromiya & -0.015 & 0.042 & 0.155 & -0.017 & 0.044 & 0.194\\
\% Married Women & 0.015 & 0.044 & 0.152 & 0.021 & 0.051 & 0.219\\
Avg. WAZ & 0.020 & 0.075 & 0.146 & 0.032 & 0.115 & 0.214\\
Avg. No. Bedrooms & -0.012 & 0.037 & 0.146 & -0.017 & 0.044 & 0.200\\
\% Tertiary Educ. & 0.019 & 0.055 & 0.145 & 0.022 & 0.058 & 0.186\\
Avg. HAZ & 0.015 & 0.056 & 0.140 & 0.023 & 0.082 & 0.243\\
Urban Cluster & 0.017 & 0.054 & 0.131 & 0.025 & 0.064 & 0.198\\
\% Owns Agric. Land & -0.015 & 0.050 & 0.118 & -0.020 & 0.058 & 0.173\\
Altitude & 0.013 & 0.052 & 0.106 & 0.018 & 0.064 & 0.159\\
St. Dev. Wealth Index & 0.008 & 0.033 & 0.105 & 0.010 & 0.037 & 0.142\\
Avg. Temperature & 0.001 & 0.056 & 0.100 & 0.010 & 0.070 & 0.163\\
Nightlight Intensity & 0.010 & 0.045 & 0.098 & 0.013 & 0.052 & 0.156\\
\% Owns Livestock & -0.013 & 0.057 & 0.094 & -0.016 & 0.066 & 0.156\\
Pop. Density & 0.008 & 0.038 & 0.093 & 0.007 & 0.039 & 0.125\\
St. Dev. WAZ & 0.006 & 0.027 & 0.088 & 0.009 & 0.034 & 0.139\\
\% Surface Water & -0.006 & 0.027 & 0.088 & -0.009 & 0.035 & 0.141\\
St. Dev HAZ & 0.006 & 0.027 & 0.086 & 0.009 & 0.034 & 0.132\\
\% Secondary Educ. & -0.006 & 0.032 & 0.083 & -0.011 & 0.042 & 0.148\\
St. Dev. WHZ & -0.005 & 0.025 & 0.081 & -0.008 & 0.031 & 0.124\\
Under-5 Mortality Rate & -0.004 & 0.022 & 0.079 & -0.007 & 0.028 & 0.132\\
\% Female Children & -0.004 & 0.021 & 0.078 & -0.007 & 0.028 & 0.126\\
\% Finished Walls & 0.006 & 0.036 & 0.071 & 0.013 & 0.053 & 0.136\\
\% Non-Flush Toilet & -0.004 & 0.018 & 0.069 & -0.005 & 0.023 & 0.116\\
\% Health Insured & 0.004 & 0.023 & 0.068 & 0.007 & 0.031 & 0.121\\
Avg. WHZ & -0.003 & 0.043 & 0.067 & -0.007 & 0.071 & 0.110\\
Region: Gambela & -0.001 & 0.020 & 0.064 & -0.001 & 0.026 & 0.098\\
Avg. Women's Age & 0.000 & 0.019 & 0.063 & 0.000 & 0.022 & 0.089\\
\% Owns Car & 0.003 & 0.023 & 0.062 & 0.005 & 0.032 & 0.111\\
\% Owns Phone & -0.002 & 0.021 & 0.061 & -0.006 & 0.030 & 0.121\\
\% Illiterate & -0.003 & 0.025 & 0.058 & -0.003 & 0.032 & 0.103\\
\% Rudimentary Walls & -0.001 & 0.019 & 0.057 & 0.000 & 0.026 & 0.088\\
Region: Amhara & 0.003 & 0.020 & 0.057 & 0.004 & 0.027 & 0.106\\
Region: Benishangul-Gumuz & 0.001 & 0.017 & 0.056 & 0.003 & 0.023 & 0.103\\
Region: SNNP & 0.001 & 0.017 & 0.056 & 0.001 & 0.023 & 0.092\\
\% Piped Water & 0.003 & 0.020 & 0.054 & 0.003 & 0.027 & 0.094\\
\% Non-Residents & -0.001 & 0.014 & 0.053 & -0.002 & 0.019 & 0.099\\
St. Dev. Women's Age & 0.000 & 0.013 & 0.053 & 0.000 & 0.018 & 0.090\\
\% Well Water & 0.002 & 0.016 & 0.052 & 0.004 & 0.024 & 0.098\\
\% Petrol as Fuel & 0.001 & 0.015 & 0.051 & 0.002 & 0.021 & 0.095\\
Time to Water Source & 0.000 & 0.014 & 0.049 & 0.000 & 0.019 & 0.083\\
\% Primary Educ. & 0.000 & 0.013 & 0.048 & -0.001 & 0.019 & 0.085\\
Malaria Prevalence & 0.000 & 0.013 & 0.048 & 0.001 & 0.019 & 0.090\\
\midrule
$\alpha$ & -0.278 & 0.049 & - & -0.283 & 0.051 & -\\
$\sigma^2$ & 0.162 & 0.078 & - & 0.229 & 0.072 & -\\
$g$ & 305.000 & 0.000 & - & 90.762 & 61.736 & -\\
Model Size & 14.173 & 4.141 & - & 16.847 & 4.579 & -\\
\addlinespace
\bottomrule
\end{tabular}}
\end{table}

\begin{table}[t]
\caption{Top Five Highest Probability Models Using UIP Prior (Measles Vaccination Data).}
\label{tab:topmodels_vacc_uip}
\centering
\begin{tabular}{llllll}
\toprule
 & Model \#1 & Model \#2 & Model \#3 & Model \#4 & Model \#5\\
 \midrule
\% Knows Modern FP &  & x & x & x & \\
Avg. Wealth Index &  &  & x & x & \\
\% Has Electricity & x & x &  &  & x\\
\% Pit Toilet & x &  &  &  & \\
\% Coal as Fuel & x & x &  &  & x\\
\% Female HH Heads & x &  &  &  & \\
Avg. Children's Age & x & x & x & x & x\\
St. Dev. Children's Age & x & x & x & x & x\\
Latitude & x & x & x & x & x\\
Longitude &  &  &  &  & x\\
Region: Affar & x & x & x & x & x\\
Region: Somali & x &  &  &  & \\
Region: Harari & x & x &  & x & \\
Region: Dire Dawa &  &  &  &  & x\\
\midrule
Posterior Model Probability & 0.002 & 0.002 & 0.001 & 0.001 & 0.001\\
\bottomrule
\end{tabular}
\end{table}

\begin{table}[t]
\caption{Top Five Highest Probability Models Using Hyper-$g/n$ Prior (Measles Vaccination Data).}
\label{tab:topmodels_vacc_hypergn}
\centering
\begin{tabular}{llllll}
\toprule
 & Model \#1 & Model \#2 & Model \#3 & Model \#4 & Model \#5\\
 \midrule
\% Knows Modern FP &  & x &  &  & \\
Avg. Wealth Index &  &  &  &  & x\\
\% Owns Radio & x &  & x & x & \\
\% Has Electricity & x & x & x & x & \\
\% Pit Toilet &  &  &  &  & x\\
\% Coal as Fuel & x & x &  & x & \\
Avg. No. Births per Woman & x &  & x &  & \\
Avg. Children's Age & x & x & x & x & x\\
St. Dev. Children's Age & x & x & x & x & x\\
Latitude & x & x & x & x & x\\
Longitude &  &  &  & x & x\\
Region: Affar & x & x & x & x & x\\
Region: Harari & x & x & x &  & \\
Region: Dire Dawa &  &  &  & x & \\
\midrule
Posterior Model Probability & $4 \times 10^{-4}$ & $2 \times 10^{-4}$ &
$2 \times 10^{-4}$ &
$2 \times 10^{-4}$ &
$2 \times 10^{-4}$\\
\bottomrule
\end{tabular}
\end{table}

\begin{table}[!htbp] \centering \renewcommand*{\arraystretch}{1.1}\caption{Summary Statistics Bilateral Migration Data}
\label{tab:summary_mig}
\resizebox{!}{0.45\textheight}{
\begin{tabular}{lrrrrrrrr}
\hline
Variable & N & Mean & Std. Dev. & Min & Q1 & Q3 & Max & Data Source \\ 
\hline
Migration Flow & 1406 & 13366.951 & 67942.166 & 0 & 191.25 & 5874 & 1635815 & Abel and Cohen (2019)\\
GDPPC (Orig.) & 1406 & 10.208 & 0.74 & 8.325 & 9.565 & 10.699 & 11.532 & CEPII Gravity DB\\
GDPPC (Dest.) & 1406 & 10.208 & 0.74 & 8.325 & 9.565 & 10.699 & 11.532 & CEPII Gravity DB\\
\% Rural Pop. (Orig.) & 1406 & 21.446 & 10.408 & 2.021 & 13.49 & 28.787 & 46.223 & World Bank\\
\% Rural Pop. (Dest.) & 1406 & 21.446 & 10.408 & 2.021 & 13.49 & 28.787 & 46.223 & World Bank\\
Contiguity & 1406 & 0.053 & 0.223 & 0 & 0 & 0 & 1 & CEPII Gravity DB\\
Distance & 1406 & 8.154 & 1.153 & 4.007 & 7.224 & 9.166 & 9.896 & CEPII Gravity DB\\
Common Colonizer & 1406 & 0.004 & 0.065 & 0 & 0 & 0 & 1 & CEPII Gravity DB\\
Any Colonial Relation & 1406 & 0.003 & 0.053 & 0 & 0 & 0 & 1 & CEPII Gravity DB\\
Common Official Language & 1406 & 0.08 & 0.271 & 0 & 0 & 0 & 1 & CEPII Gravity DB\\
Common Popular Language & 1406 & 0.094 & 0.292 & 0 & 0 & 0 & 1 & CEPII Gravity DB\\
Population (Orig.) & 1406 & 9.5 & 1.493 & 5.802 & 8.599 & 10.784 & 12.679 & CEPII Gravity DB\\
Population (Dest.) & 1406 & 9.5 & 1.493 & 5.802 & 8.599 & 10.784 & 12.679 & CEPII Gravity DB\\
Both EU Members & 1406 & 0.329 & 0.47 & 0 & 0 & 1 & 1 & CEPII Gravity DB\\
Gini Index (Orig.) & 1406 & 34.202 & 6.527 & 24.55 & 30.033 & 36.683 & 51.083 & World Bank\\
Gini Index (Dest.) & 1406 & 34.202 & 6.527 & 24.55 & 30.033 & 36.683 & 51.083 & World Bank\\
\% Employed (Orig.) & 1406 & 69.006 & 6.997 & 50.315 & 65.394 & 73.933 & 84.433 & World Bank\\
\% Employed (Dest.) & 1406 & 69.006 & 6.997 & 50.315 & 65.394 & 73.933 & 84.433 & World Bank\\
\% Tertiary School Enrollment (Orig.) & 1406 & 73.057 & 21.614 & 19.76 & 62.431 & 84.974 & 136.695 & World Bank\\
\% Tertiary School Enrollment (Dest.) & 1406 & 73.057 & 21.614 & 19.76 & 62.431 & 84.974 & 136.695 & World Bank\\
\% Secondary School Enrollment (Orig.) & 1406 & 112.066 & 17.125 & 66.897 & 101.848 & 117.715 & 158.052 & World Bank\\
\% Secondary School Enrollment (Dest.) & 1406 & 112.066 & 17.125 & 66.897 & 101.848 & 117.715 & 158.052 & World Bank\\
Existing Migrant Population & 1406 & 6.949 & 3.406 & 0 & 5.19 & 9.335 & 16.27 & UNDESA PD\\
\% Working Age (Orig.) & 1406 & 65.61 & 2.565 & 59.108 & 64.133 & 66.881 & 72.814 & World Bank\\
\% Working Age (Dest.) & 1406 & 65.61 & 2.565 & 59.108 & 64.133 & 66.881 & 72.814 & World Bank\\
Island Country (Orig.) & 1406 & 0.184 & 0.388 & 0 & 0 & 0 & 1 & Various\\
Island Country (Dest.) & 1406 & 0.184 & 0.388 & 0 & 0 & 0 & 1 & Various\\
Health Care \% of GDP (Orig.) & 1406 & 8.911 & 2.315 & 4.28 & 7.094 & 10.503 & 17.029 & World Bank\\
Health Care \% of GDP (Dest.) & 1406 & 8.911 & 2.315 & 4.28 & 7.094 & 10.503 & 17.029 & World Bank\\
GDP Growth (Orig.) & 1406 & 1.62 & 1.693 & -0.893 & 0.642 & 2.307 & 9.323 & World Bank\\
GDP Growth (Dest.) & 1406 & 1.62 & 1.693 & -0.893 & 0.642 & 2.307 & 9.323 & World Bank\\
Stability Index (Orig.) & 1406 & 66.528 & 22.383 & 8.995 & 57.74 & 81.757 & 98.58 & World Bank\\
Stability Index (Dest.) & 1406 & 66.528 & 22.383 & 8.995 & 57.74 & 81.757 & 98.58 & World Bank\\
Service \% of GDP (Orig.) & 1406 & 64.211 & 6.198 & 54.322 & 58.432 & 69.433 & 79.575 & World Bank\\
Service \% of GDP (Dest.) & 1406 & 64.211 & 6.198 & 54.322 & 58.432 & 69.433 & 79.575 & World Bank\\
Area (Orig.) & 1406 & 12.114 & 1.72 & 7.858 & 10.841 & 13.017 & 16.116 & CEPII\\
Area (Dest.) & 1406 & 12.114 & 1.72 & 7.858 & 10.841 & 13.017 & 16.116 & CEPII\\
Landlocked Country (Orig.) & 1406 & 0.158 & 0.365 & 0 & 0 & 0 & 1 & CEPII\\
Landlocked Country (Dest.) & 1406 & 0.158 & 0.365 & 0 & 0 & 0 & 1 & CEPII\\
Latitude Capital (Orig.) & 1406 & 38.041 & 26.02 & -44.283 & 37.5 & 52.533 & 64.15 & CEPII\\
Latitude Capital (Dest.) & 1406 & 38.041 & 26.02 & -44.283 & 37.5 & 52.533 & 64.15 & CEPII\\
Longitude Capital (Orig.) & 1406 & 9.201 & 60.926 & -99.167 & -6.25 & 24.1 & 174.783 & CEPII\\
Longitude Capital (Dest.) & 1406 & 9.201 & 60.926 & -99.167 & -6.25 & 24.1 & 174.783 & CEPII\\
Asian Country (Orig.) & 1406 & 0.079 & 0.27 & 0 & 0 & 0 & 1 & CEPII\\
European Country (Orig.) & 1406 & 0.684 & 0.465 & 0 & 0 & 1 & 1 & CEPII\\
Pacific Country (Orig.) & 1406 & 0.053 & 0.223 & 0 & 0 & 0 & 1 & CEPII\\
Asian Country (Dest.) & 1406 & 0.079 & 0.27 & 0 & 0 & 0 & 1 & CEPII\\
European Country (Dest.) & 1406 & 0.684 & 0.465 & 0 & 0 & 1 & 1 & CEPII\\
Pacific Country (Dest.) & 1406 & 0.053 & 0.223 & 0 & 0 & 0 & 1 & CEPII\\
Agriculture \% of GDP (Orig.) & 1406 & 2.483 & 1.617 & 0.22 & 1.15 & 3.607 & 6.513 & World Bank\\
Agriculture \% of GDP (Dest.) & 1406 & 2.483 & 1.617 & 0.22 & 1.15 & 3.607 & 6.513 & World Bank\\
Battle Deaths (Orig.) & 1406 & 0.02 & 0.097 & 0 & 0 & 0 & 0.603 & World Bank\\
Battle Deaths (Dest.) & 1406 & 0.02 & 0.097 & 0 & 0 & 0 & 0.603 & World Bank\\
Homicide Rate (Orig.) & 1406 & 4.526 & 10.911 & 0.264 & 0.722 & 2.561 & 61.21 & World Bank\\
Homicide Rate (Dest.) & 1406 & 4.526 & 10.911 & 0.264 & 0.722 & 2.561 & 61.21 & World Bank\\
\hline
\end{tabular}
}
\end{table}

\begin{table}[t]
\caption{Estimation Results (Bilateral Migration Data).}
\label{tab:results_oecd}
\centering
\resizebox{!}{0.45\textheight}{\begin{tabular}{@{\extracolsep{4pt}}lrrrrrr}
\toprule
&\multicolumn{3}{c}{Unit Information Prior}&\multicolumn{3}{c}{Hyper-$g/n$ Prior}\\\cline{2-4}\cline{5-7}\addlinespace
Variable & Post. Mean & Post. SD & PIP & Post. Mean & Post. SD & PIP \\
\midrule
Distance & -0.535 & 0.046 & 1.000 & -0.521 & 0.048 & 1.000\\
Population (Orig.) & 0.591 & 0.050 & 1.000 & 0.572 & 0.057 & 1.000\\
\% Employed (Dest.) & 0.363 & 0.054 & 1.000 & 0.356 & 0.052 & 1.000\\
Existing Migrant Population & 1.811 & 0.049 & 1.000 & 1.811 & 0.048 & 1.000\\
\% Working Age (Dest.) & 0.349 & 0.042 & 1.000 & 0.342 & 0.039 & 1.000\\
Island Country (Dest.) & -0.189 & 0.036 & 1.000 & -0.184 & 0.038 & 0.999\\
Area (Dest.) & 0.518 & 0.099 & 1.000 & 0.491 & 0.097 & 1.000\\
Latitude Capital (Dest.) & -0.908 & 0.131 & 1.000 & -0.876 & 0.120 & 1.000\\
Longitude Capital (Orig.) & -0.719 & 0.146 & 1.000 & -0.755 & 0.144 & 1.000\\
Longitude Capital (Dest.) & 0.773 & 0.136 & 1.000 & 0.787 & 0.141 & 1.000\\
Pacific Country (Orig.) & 0.780 & 0.139 & 1.000 & 0.823 & 0.142 & 1.000\\
Pacific Country (Dest.) & -0.964 & 0.159 & 1.000 & -0.958 & 0.157 & 1.000\\
GDPPC (Orig.) & 0.398 & 0.064 & 0.999 & 0.385 & 0.080 & 0.998\\
GDP Growth (Dest.) & 0.230 & 0.041 & 0.999 & 0.227 & 0.042 & 1.000\\
Battle Deaths (Dest.) & -0.171 & 0.042 & 0.994 & -0.173 & 0.040 & 0.999\\
\% Rural Pop. (Dest.) & -0.154 & 0.066 & 0.969 & -0.171 & 0.063 & 0.990\\
Landlocked Country (Dest.) & -0.136 & 0.049 & 0.950 & -0.145 & 0.043 & 0.983\\
Population (Dest.) & 0.202 & 0.086 & 0.938 & 0.208 & 0.075 & 0.968\\
Service \% of GDP (Orig.) & 0.135 & 0.053 & 0.925 & 0.114 & 0.062 & 0.860\\
Service \% of GDP (Dest.) & 0.172 & 0.076 & 0.903 & 0.155 & 0.073 & 0.906\\
\% Tertiary School Enrollment (Orig.) & 0.106 & 0.050 & 0.888 & 0.111 & 0.050 & 0.917\\
European Country (Orig.) & 0.282 & 0.152 & 0.881 & 0.304 & 0.145 & 0.919\\
\% Tertiary School Enrollment (Dest.) & -0.124 & 0.060 & 0.878 & -0.131 & 0.053 & 0.934\\
Stability Index (Dest.) & -0.169 & 0.093 & 0.847 & -0.176 & 0.088 & 0.892\\
European Country (Dest.) & 0.256 & 0.134 & 0.827 & 0.213 & 0.160 & 0.694\\
Battle Deaths (Orig.) & 0.076 & 0.054 & 0.752 & 0.098 & 0.053 & 0.879\\
Asian Country (Orig.) & 0.183 & 0.138 & 0.709 & 0.202 & 0.137 & 0.783\\
\% Employed (Orig.) & -0.057 & 0.063 & 0.522 & -0.072 & 0.063 & 0.666\\
\% Secondary School Enrollment (Dest.) & -0.046 & 0.063 & 0.409 & -0.072 & 0.066 & 0.642\\
Gini Index (Orig.) & -0.034 & 0.066 & 0.261 & -0.040 & 0.066 & 0.352\\
Homicide Rate (Dest.) & -0.034 & 0.068 & 0.255 & -0.072 & 0.085 & 0.503\\
GDP Growth (Orig.) & -0.016 & 0.035 & 0.221 & -0.031 & 0.044 & 0.419\\
Asian Country (Dest.) & -0.052 & 0.113 & 0.216 & -0.100 & 0.142 & 0.401\\
\% Secondary School Enrollment (Orig.) & -0.015 & 0.034 & 0.203 & -0.038 & 0.053 & 0.451\\
Gini Index (Dest.) & 0.027 & 0.068 & 0.176 & 0.024 & 0.063 & 0.205\\
Common Popular Language & 0.010 & 0.024 & 0.172 & 0.019 & 0.032 & 0.337\\
Agriculture \% of GDP (Orig.) & -0.020 & 0.058 & 0.156 & -0.054 & 0.090 & 0.355\\
Island Country (Orig.) & -0.007 & 0.022 & 0.136 & -0.007 & 0.022 & 0.178\\
Health Care \% of GDP (Orig.) & 0.008 & 0.035 & 0.101 & 0.013 & 0.041 & 0.173\\
Stability Index (Orig.) & -0.003 & 0.030 & 0.094 & 0.006 & 0.043 & 0.163\\
Common Official Language & 0.003 & 0.014 & 0.087 & 0.006 & 0.019 & 0.173\\
GDPPC (Dest.) & -0.008 & 0.036 & 0.082 & -0.015 & 0.054 & 0.160\\
Agriculture \% of GDP (Dest.) & 0.001 & 0.038 & 0.079 & 0.000 & 0.054 & 0.148\\
\% Rural Pop. (Orig.) & -0.004 & 0.024 & 0.069 & -0.017 & 0.046 & 0.216\\
Landlocked Country (Orig.) & 0.002 & 0.012 & 0.066 & 0.003 & 0.015 & 0.115\\
Health Care \% of GDP (Dest.) & 0.000 & 0.025 & 0.065 & 0.003 & 0.030 & 0.115\\
Latitude Capital (Orig.) & -0.002 & 0.022 & 0.060 & -0.001 & 0.034 & 0.119\\
Area (Orig.) & -0.002 & 0.018 & 0.055 & -0.004 & 0.028 & 0.121\\
Homicide Rate (Orig.) & -0.002 & 0.013 & 0.049 & -0.005 & 0.023 & 0.119\\
Contiguity & 0.001 & 0.008 & 0.048 & 0.003 & 0.012 & 0.112\\
Common Colonizer & 0.001 & 0.007 & 0.046 & 0.002 & 0.011 & 0.106\\
\% Working Age (Orig.) & -0.001 & 0.010 & 0.045 & -0.004 & 0.017 & 0.130\\
Both EU Members & 0.000 & 0.010 & 0.040 & 0.000 & 0.013 & 0.080\\
Any Colonial Relation & 0.000 & 0.004 & 0.030 & 0.001 & 0.007 & 0.083\\
\midrule
$\alpha$ & 6.894 & 0.023 & - & 6.894 & 0.023 & -\\
$\sigma^2$ & 0.723 & 0.029 & - & 0.729 & 0.030 & -\\
$g$ & 1406.000 & 0.000 & - & 395.971 & 113.743 & -\\
Model Size & 29.202 & 1.981 & - & 32.364 & 2.642 & -\\
\addlinespace
\bottomrule
\end{tabular}}
\end{table}

\begin{table}[t]
\caption{Top Five Highest Probability Models Using UIP Prior (Bilateral Migration Data).}
\label{tab:topmodels_oecd_uip}
\centering
\resizebox{\textwidth}{!}
{\begin{tabular}{llllll}
\toprule
 & Model \#1 & Model \#2 & Model \#3 & Model \#4 & Model \#5\\
 \midrule
GDPPC (Orig.) & x & x & x & x & x\\
\% Rural Pop. (Dest.) & x & x & x & x & x\\
Distance & x & x & x & x & x\\
Population (Orig.) & x & x & x & x & x\\
Population (Dest.) & x & x & x & x & x\\
Gini Index (Orig.) &  &  & x &  & \\
\% Employed (Orig.) &  & x & x &  & \\
\% Employed (Dest.) & x & x & x & x & x\\
\% Tertiary School Enrollment (Orig.) & x & x & x & x & x\\
\% Tertiary School Enrollment (Dest.) & x & x & x & x & x\\
\% Secondary School Enrollment (Dest.) &  &  &  & x & \\
Existing Migrant Population & x & x & x & x & x\\
\% Working Age (Dest.) & x & x & x & x & x\\
Island Country (Orig.) &  &  &  &  & x\\
Island Country (Dest.) & x & x & x & x & x\\
GDP Growth (Dest.) & x & x & x & x & x\\
Stability Index (Dest.) & x & x & x & x & x\\
Service \% of GDP (Orig.) & x & x & x & x & x\\
Service \% of GDP (Dest.) & x & x & x & x & x\\
Area (Dest.) & x & x & x & x & x\\
Landlocked Country (Dest.) & x & x & x & x & x\\
Latitude Capital (Dest.) & x & x & x & x & x\\
Longitude Capital (Orig.) & x & x & x & x & x\\
Longitude Capital (Dest.) & x & x & x & x & x\\
Asian Country (Orig.) & x & x &  & x & x\\
European Country (Orig.) & x & x &  & x & x\\
Pacific Country (Orig.) & x & x & x & x & x\\
European Country (Dest.) & x & x & x & x & x\\
Pacific Country (Dest.) & x & x & x & x & x\\
Battle Deaths (Orig.) & x &  & x & x & x\\
Battle Deaths (Dest.) & x & x & x & x & x\\
\midrule
Posterior Model Probability & 0.018 & 0.008 & 0.006 & 0.006 & 0.005\\
\bottomrule
\end{tabular}}
\end{table}

\begin{table}[t]
\caption{Top Five Highest Probability Models Using Hyper-$g/n$ Prior (Bilateral Migration Data).}
\label{tab:topmodels_oecd_hypergn}
\centering
\resizebox{\textwidth}{!}
{\begin{tabular}{llllll}
\toprule
 & Model \#1 & Model \#2 & Model \#3 & Model \#4 & Model \#5\\
 \midrule
GDPPC (Orig.) & x & x & x & x & x\\
\% Rural Pop. (Dest.) & x & x & x & x & x\\
Distance & x & x & x & x & x\\
Population (Orig.) & x & x & x & x & x\\
Population (Dest.) & x & x & x & x & x\\
Gini Index (Orig.) & x &  &  &  & \\
\% Employed (Orig.) & x & x &  &  & \\
\% Employed (Dest.) & x & x & x & x & x\\
\% Tertiary School Enrollment (Orig.) & x & x & x & x & x\\
\% Tertiary School Enrollment (Dest.) & x & x & x & x & x\\
\% Secondary School Enrollment (Dest.) &  &  &  & x & \\
Existing Migrant Population & x & x & x & x & x\\
\% Working Age (Dest.) & x & x & x & x & x\\
Island Country (Orig.) &  &  &  &  & x\\
Island Country (Dest.) & x & x & x & x & x\\
GDP Growth (Dest.) & x & x & x & x & x\\
Stability Index (Dest.) & x & x & x & x & x\\
Service \% of GDP (Orig.) & x & x & x & x & x\\
Service \% of GDP (Dest.) & x & x & x & x & x\\
Area (Dest.) & x & x & x & x & x\\
Landlocked Country (Dest.) & x & x & x & x & x\\
Latitude Capital (Dest.) & x & x & x & x & x\\
Longitude Capital (Orig.) & x & x & x & x & x\\
Longitude Capital (Dest.) & x & x & x & x & x\\
Asian Country (Orig.) &  & x & x & x & x\\
European Country (Orig.) &  & x & x & x & x\\
Pacific Country (Orig.) & x & x & x & x & x\\
European Country (Dest.) & x & x & x & x & x\\
Pacific Country (Dest.) & x & x & x & x & x\\
Battle Deaths (Orig.) & x &  & x & x & x\\
Battle Deaths (Dest.) & x & x & x & x & x\\
\midrule
Posterior Model Probability & 0.002 & 0.001 & 0.001 & 0.001 & 0.001\\
\bottomrule
\end{tabular}}
\end{table}

\clearpage
\subsection{Visualizations of the Data and of the Model Averaging Results}

\begin{figure}[!h]
    \centering
    \includegraphics[width=0.75\textwidth]{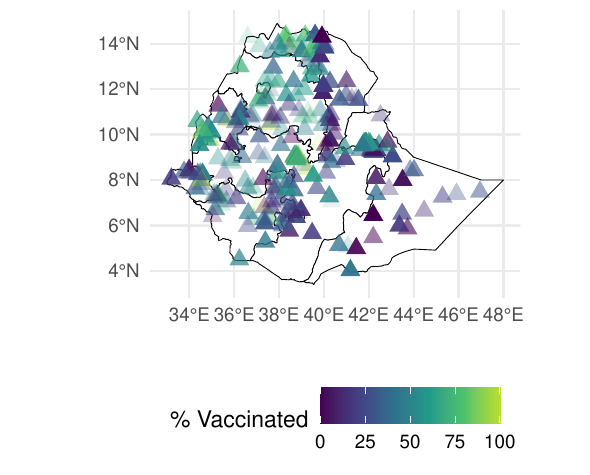}
    \caption{Map of Survey Clusters in Ethiopia DHS Survey 2019. Cluster transparency is inversely proportional to local sample size.}
    \label{fig:ethiopia_map}
\end{figure}

\begin{figure}[h!]
    \centering
    \begin{subfigure}[b]{0.49\textwidth}
        \includegraphics[width=\textwidth]{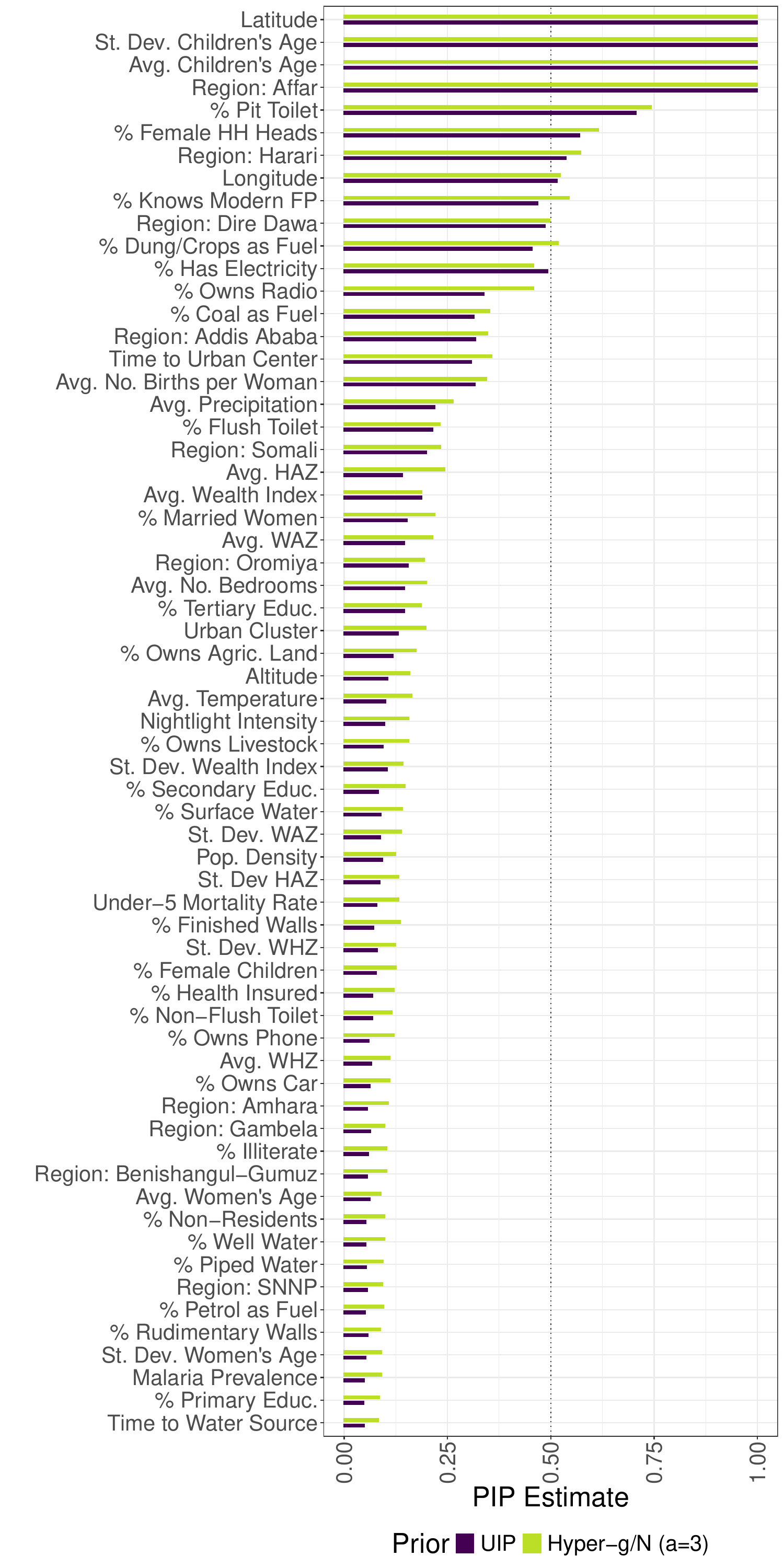}
        \caption{Post. Inclusion Probabilities.}
        \label{fig:image1}
    \end{subfigure}
    \hfill % add some horizontal spacing
    \begin{subfigure}[b]{0.49\textwidth}
        \includegraphics[width=\textwidth]{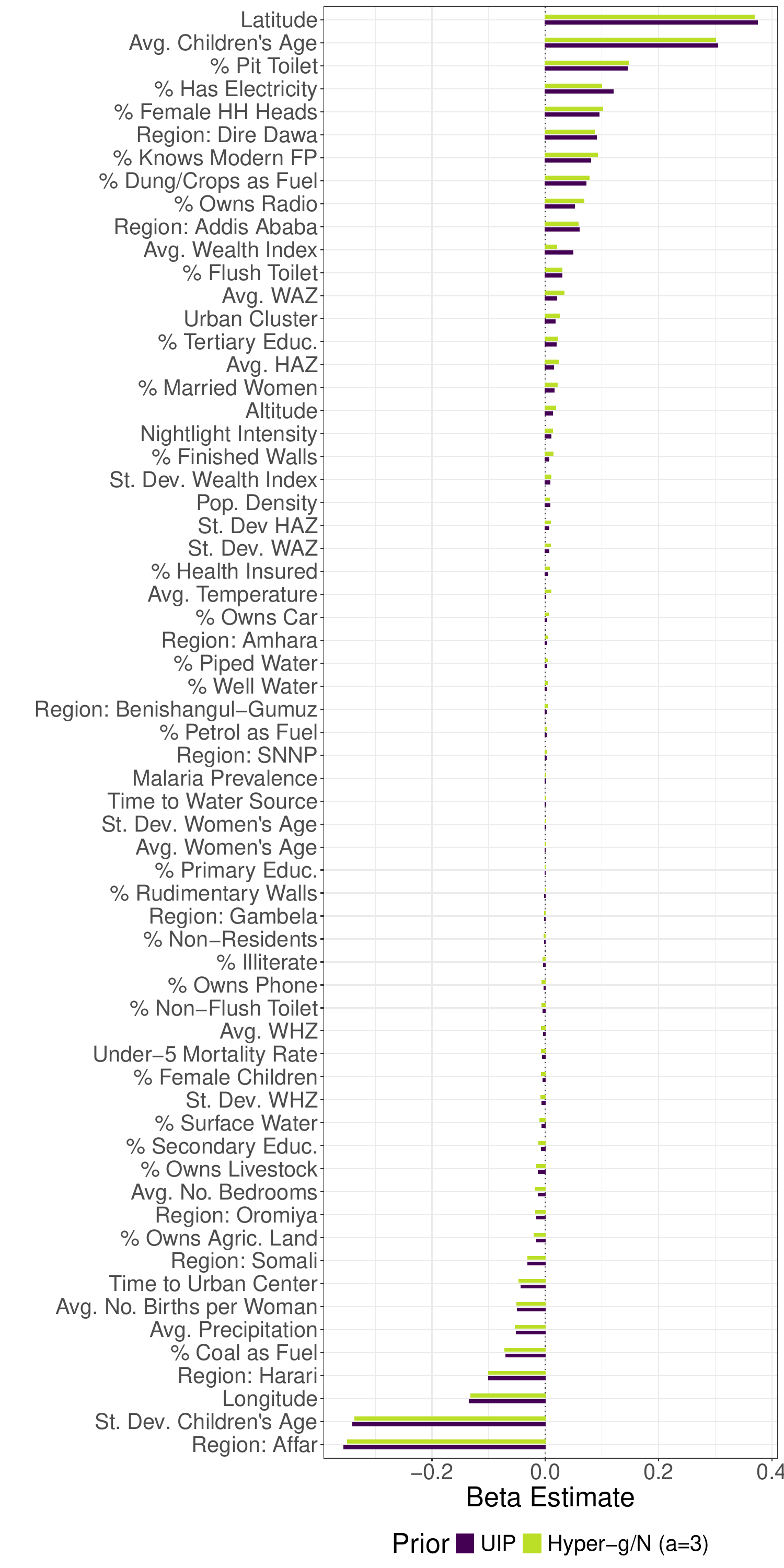}
        \caption{Coefficient Post. Means.}
        \label{fig:image2}
    \end{subfigure}
    \caption{Estimation Results for Measles Vaccination Data.}
    \label{fig:vacc_res_beta_pip}
\end{figure}

\begin{figure}
\centering
\begin{subfigure}{0.49\textwidth}
    \includegraphics[width=\textwidth]{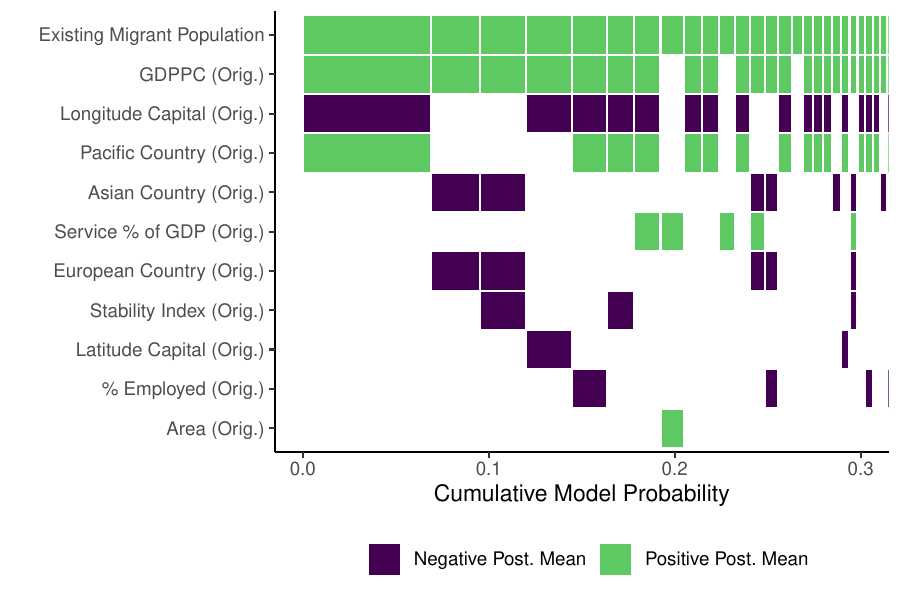}
    \caption{Highest Probability Models.}
    \label{fig:highest_mig}
\end{subfigure}
\hfill
\begin{subfigure}{0.49\textwidth}
    \includegraphics[width=\textwidth]{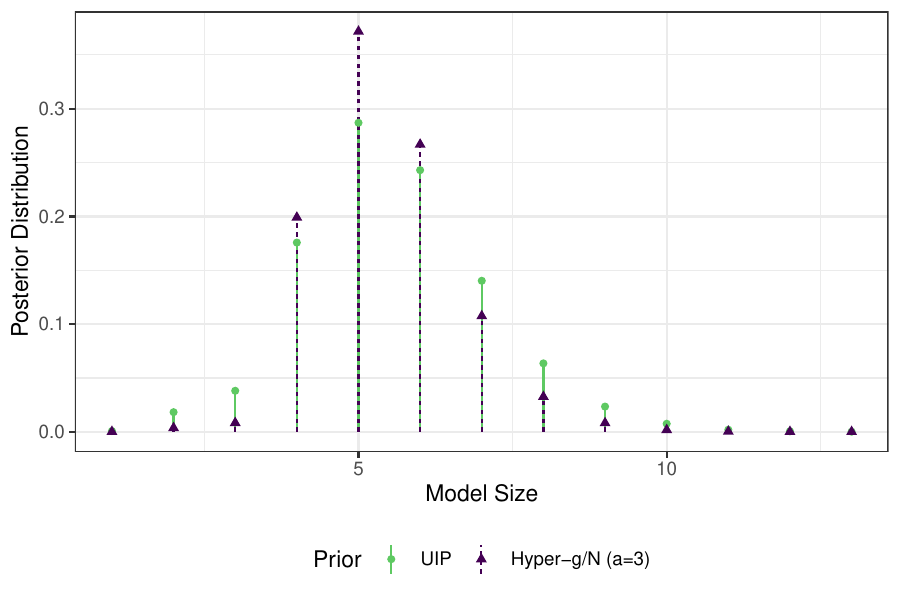}
    \caption{Posterior Distribution of Model Size.}
    \label{fig:size_mig}
\end{subfigure}
        
\caption{Estimation Results (Austrian Migration Flow Data Subset). Highest probability models plot includes variables with estimated $\text{PIP}>0.1$ under unit information prior.}
\label{fig:results_migration_aut}
\end{figure}

\begin{figure}[h!]
    \centering
    \begin{subfigure}[b]{0.49\textwidth}
        \includegraphics[width=\textwidth]{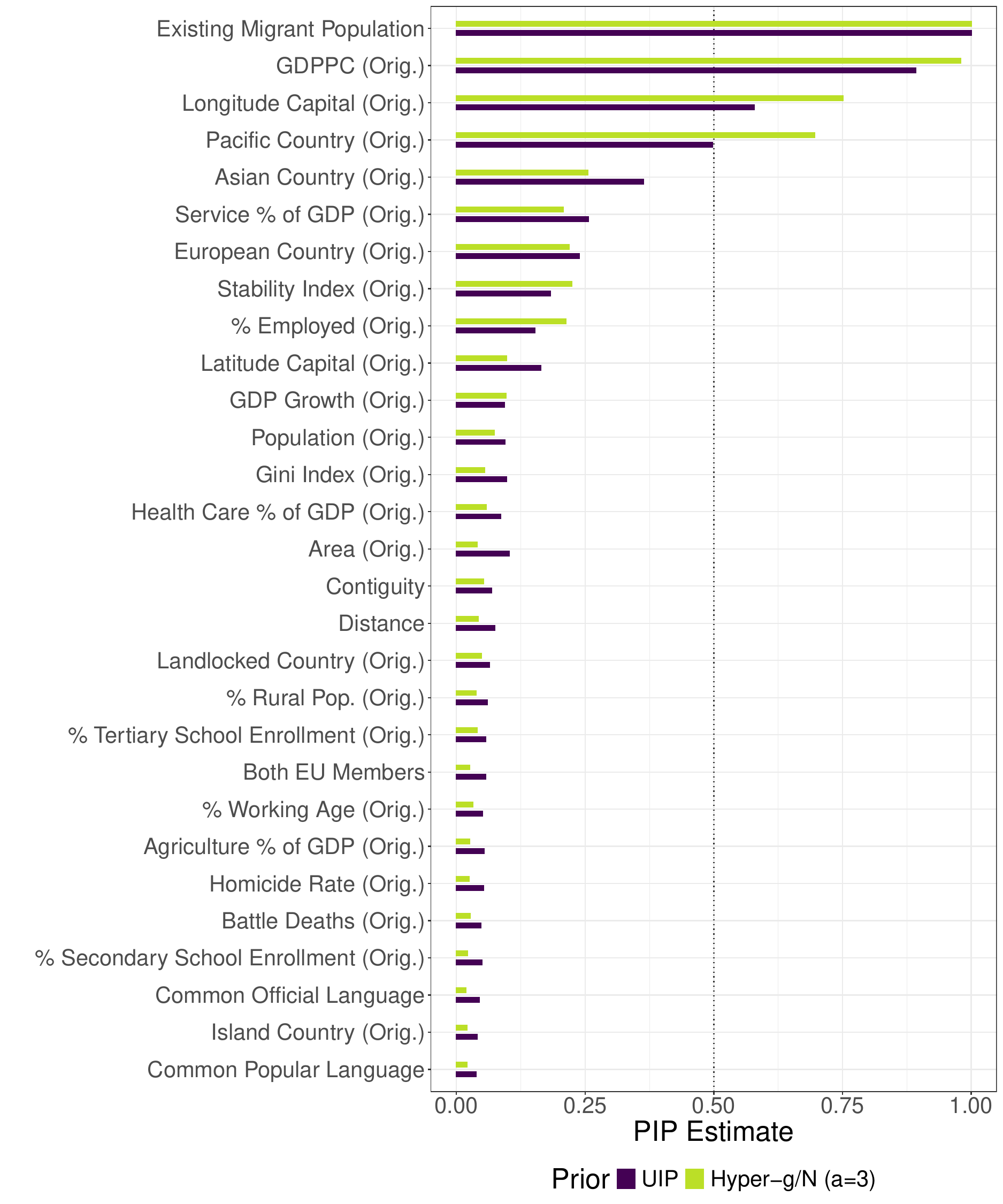}
        \caption{Post. Inclusion Probabilities.}
        \label{fig:image1}
    \end{subfigure}
    \hfill % add some horizontal spacing
    \begin{subfigure}[b]{0.49\textwidth}
        \includegraphics[width=\textwidth]{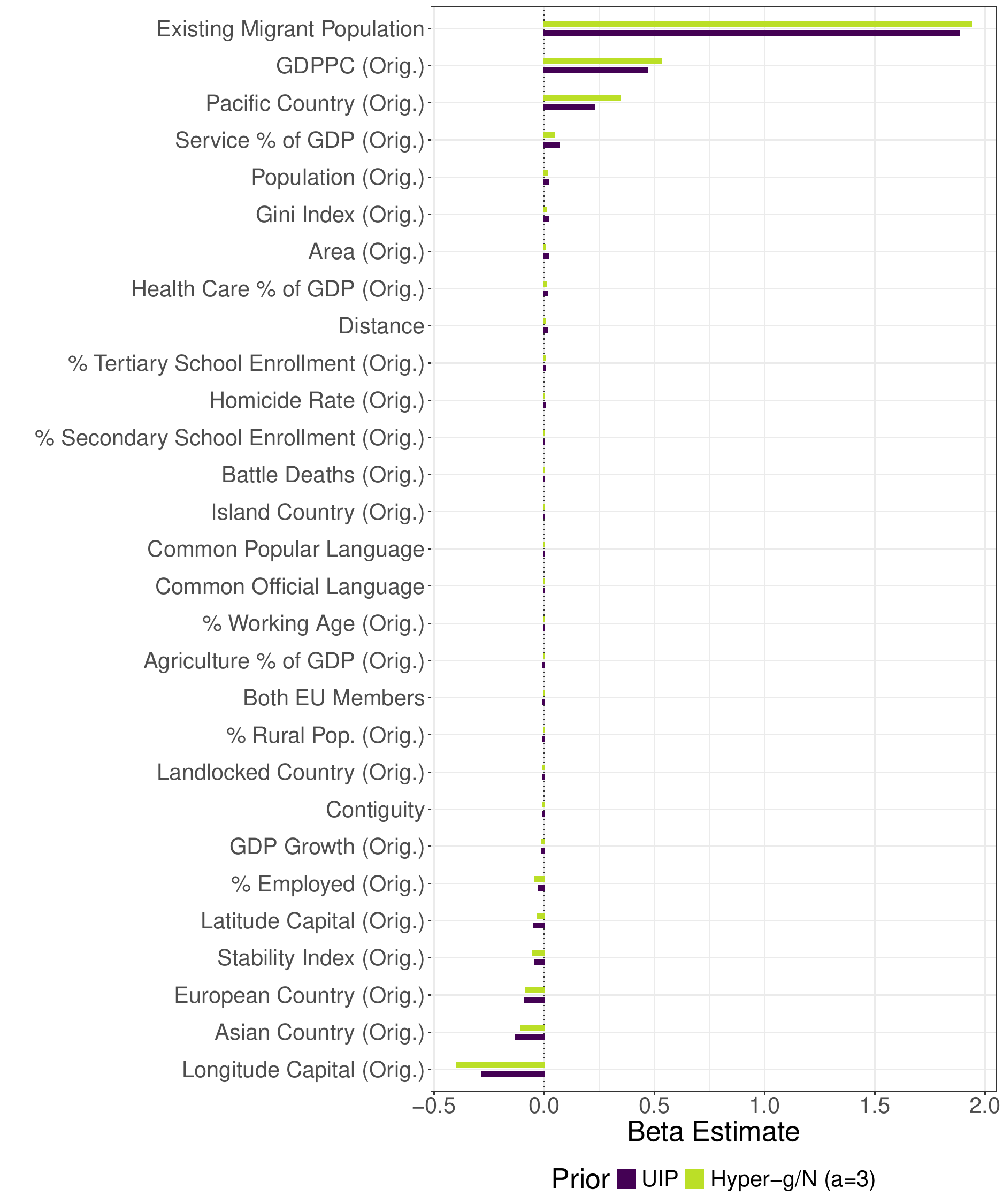}
        \caption{Coefficient Post. Means.}
        \label{fig:image2}
    \end{subfigure}
    \caption{Estimation Results for Austrian Subset of Bilateral Migration Flow Data.}
    \label{fig:mig_res_beta_pip_aut}
\end{figure}

\begin{figure}
\centering
\begin{subfigure}{0.49\textwidth}
    \includegraphics[width=\textwidth]{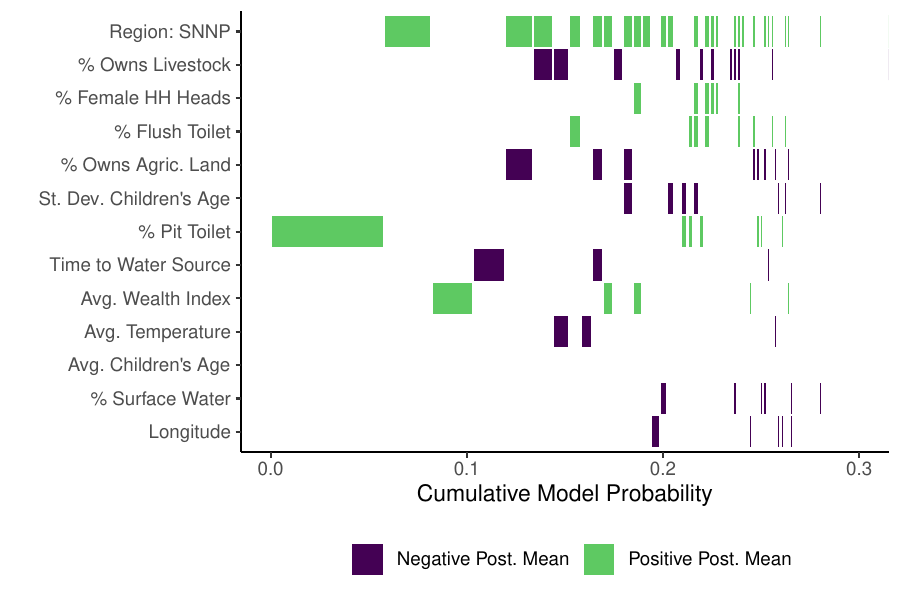}
    \caption{Highest Probability Models.}
    \label{fig:highest_mig}
\end{subfigure}
\hfill
\begin{subfigure}{0.49\textwidth}
    \includegraphics[width=\textwidth]{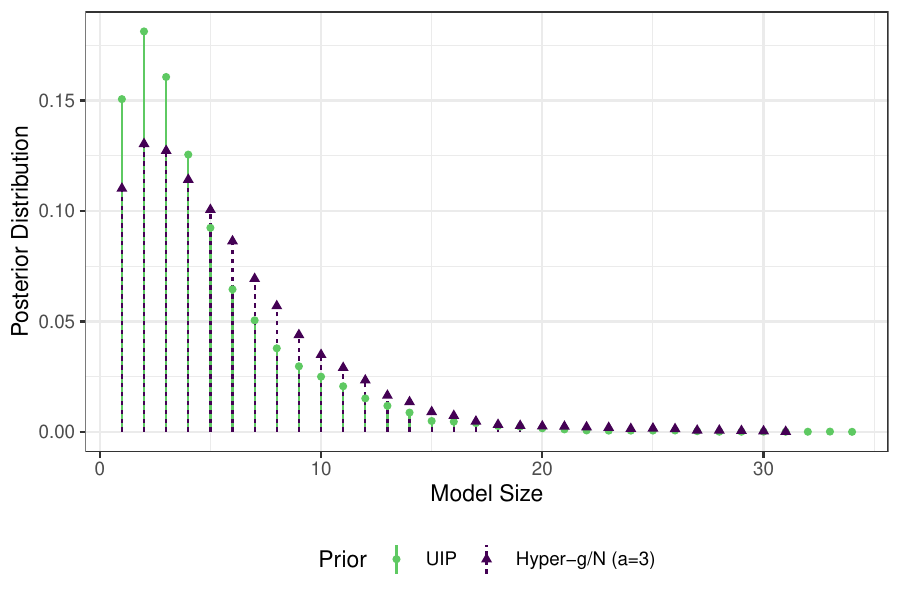}
    \caption{Posterior Distribution of Model Size.}
    \label{fig:size_mig}
\end{subfigure}
        
\caption{Estimation Results (Measles Vaccination Data Subset). Highest probability models plot includes variables with estimated $\text{PIP}>0.1$ under unit information prior.}
\label{fig:results_vaccination_region}
\end{figure}

\begin{figure}[h!]
    \centering
    \begin{subfigure}[b]{0.49\textwidth}
        \includegraphics[width=\textwidth]{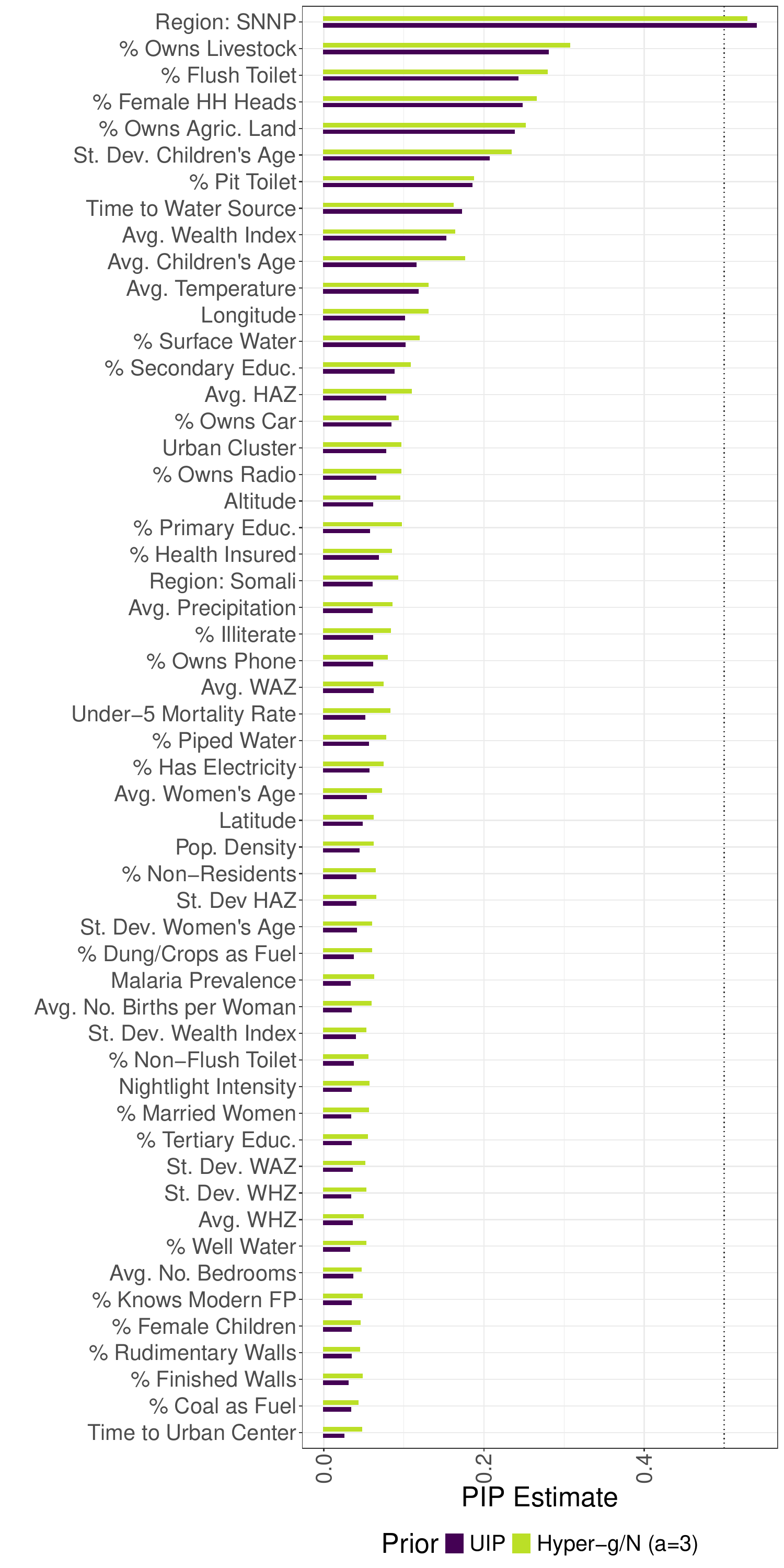}
        \caption{Post. Inclusion Probabilities.}
        \label{fig:image1}
    \end{subfigure}
    \hfill % add some horizontal spacing
    \begin{subfigure}[b]{0.49\textwidth}
        \includegraphics[width=\textwidth]{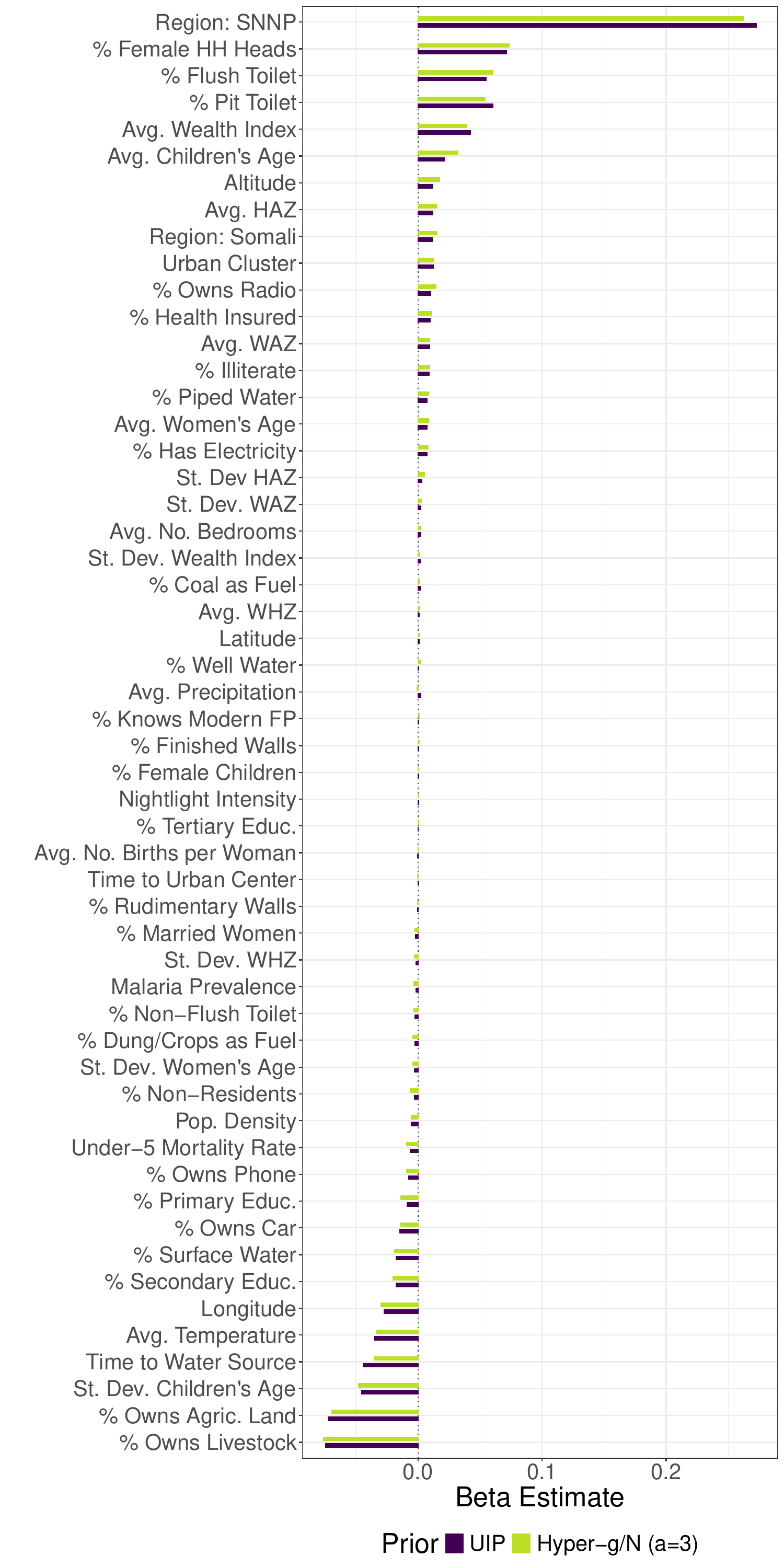}
        \caption{Coefficient Post. Means.}
        \label{fig:image2}
    \end{subfigure}
    \caption{Estimation Results for Low Vaccination Rate Subset of Measles Vaccination Data.}
    \label{fig:vacc_res_beta_pip_regional}
\end{figure}

\begin{figure}[h!]
    \centering
    \begin{subfigure}[b]{0.49\textwidth}
        \includegraphics[width=\textwidth]{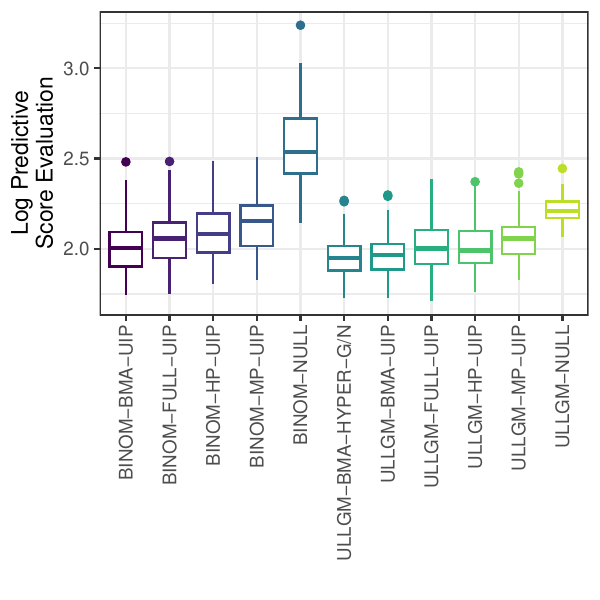}
        \caption{Full Vaccination Data (All Models).}
        \label{fig:image1}
    \end{subfigure}\hfill
    \begin{subfigure}[b]{0.49\textwidth}
        \includegraphics[width=\textwidth]{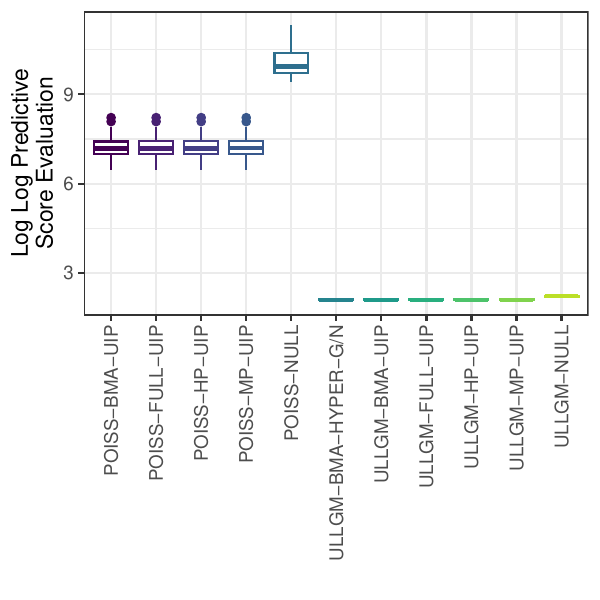}
        \caption{Full Migration Data (All Models).}
        \label{fig:image2}
    \end{subfigure}\\
    \begin{subfigure}[b]{0.49\textwidth}
        \includegraphics[width=\textwidth]{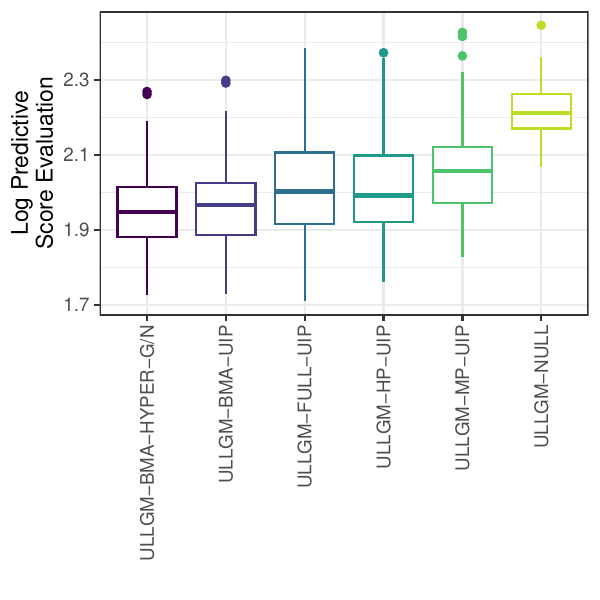}
        \caption{Full Vaccination Data (ULLGMs).}
        \label{fig:image1}
    \end{subfigure}\hfill
    \begin{subfigure}[b]{0.49\textwidth}
        \includegraphics[width=\textwidth]{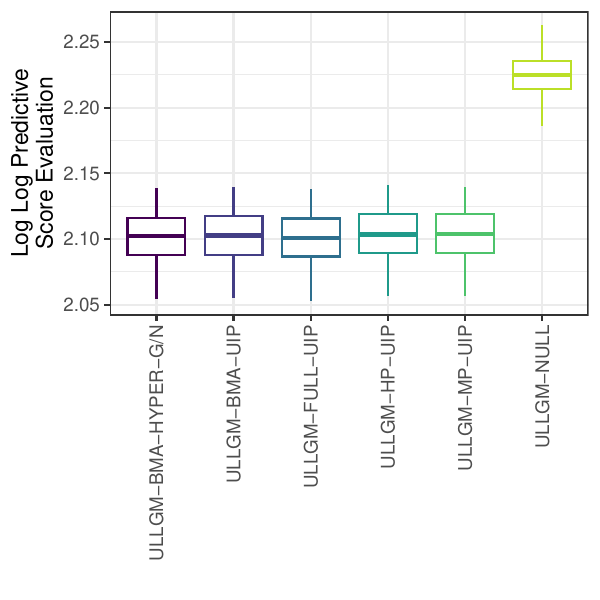}
        \caption{Full Migration Data (ULLGMs).}
        \label{fig:image2}
    \end{subfigure}\\
    \caption{Boxplots of log predictive scores across 100 random training-test partitions for the full data sets. For the migration data, the results are on a double log scale.}
    \label{fig:lps_boxplots}
\end{figure}

\begin{figure}[h!]
    \centering
    \begin{subfigure}[b]{0.49\textwidth}
        \includegraphics[width=\textwidth]{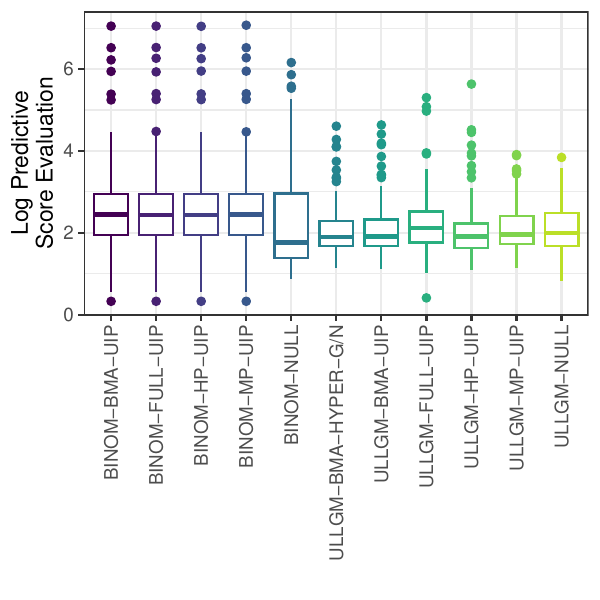}
        \caption{Subset Vaccination Data (All Models).}
        \label{fig:image1}
    \end{subfigure}\hfill
    \begin{subfigure}[b]{0.49\textwidth}
        \includegraphics[width=\textwidth]{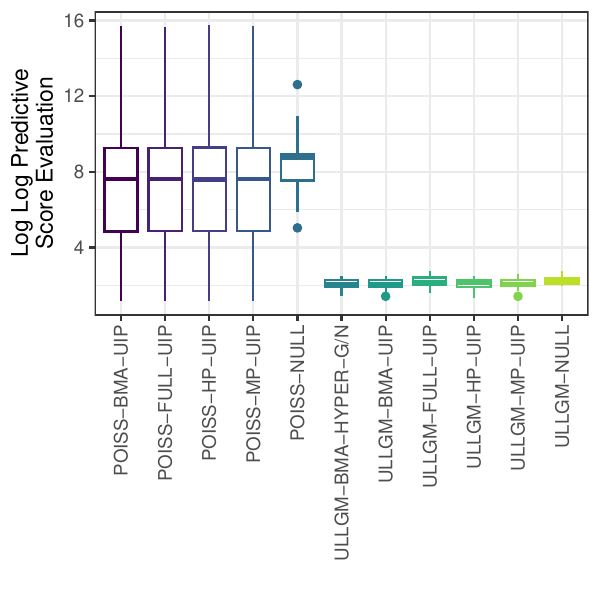}
        \caption{Subset Migration Data (All Models).}
        \label{fig:image2}
    \end{subfigure}\\
    \begin{subfigure}[b]{0.49\textwidth}
        \includegraphics[width=\textwidth]{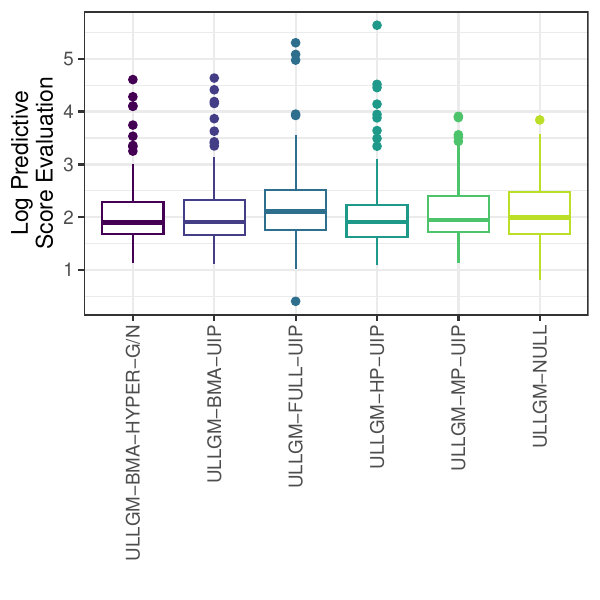}
        \caption{Subset Vaccination Data (ULLGMs).}
        \label{fig:image1}
    \end{subfigure}\hfill
    \begin{subfigure}[b]{0.49\textwidth}
        \includegraphics[width=\textwidth]{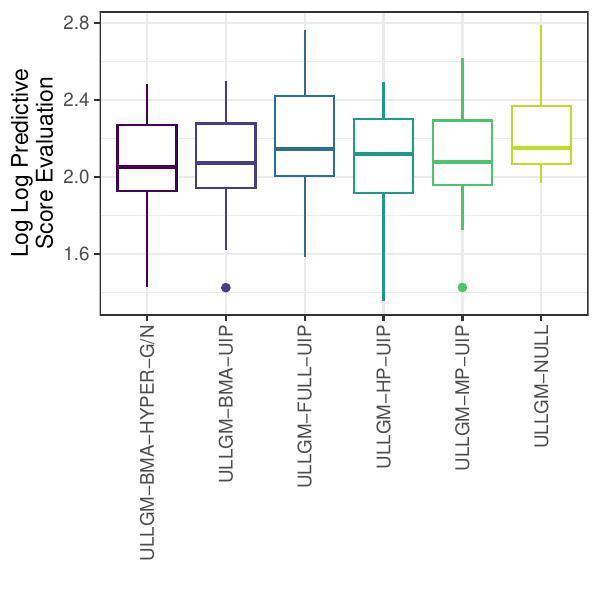}
        \caption{Subset Migration Data (ULLGMs).}
        \label{fig:image2}
    \end{subfigure}\\
    \caption{Boxplots of log predictive scores across hold-out samples in the reduced samples. For the migration data, the results are on a double log scale.}
    \label{fig:lps_boxplots_subset}
\end{figure}

\end{document}